\documentclass[a4paper,11pt]{article}
\usepackage{jheppub}
\usepackage{amsfonts,amsthm,upgreek,bm}
\usepackage{mathrsfs}
\usepackage{float}
\usepackage[mathscr]{eucal}
\usepackage[T1]{fontenc}
\usepackage{pifont}
\usepackage{enumitem}
\usepackage[dvipsnames]{xcolor}
\usepackage{bbold}
\usepackage{braket}
\usepackage{titlesec}
\titleformat{\paragraph}{\normalfont\itshape}{\thesubsubsection}{1em}{}
\numberwithin{equation}{section}

\def\<{\langle}
\def\>{\rangle}

\def\A{\mathcal{A}}
\def\B{\mathcal{B}}

\def\L{\mathcal{L}}
\def\H{\mathcal{H}}

\def\N{\mathcal{N}}

\def\r{\mathcal{R}}

\def\one{\mathbb{1}}
\def\ii{\mathrm{i}}
\def\rmin{r_{\mathrm{min}}}

\newcommand{\be}{\begin{equation}\begin{aligned}}
\newcommand{\ee}{\end{aligned}\end{equation}}

\newcommand{\ms}{\mathscr}

\newcommand{\defn}{\mathrel{\mathop:}=} 

\let\oldsetminus\setminus
\renewcommand{\setminus}{\!\oldsetminus\!}

\let\oldint\int
\renewcommand{\int}{\oldint\limits}
\let\oldlim\lim
\renewcommand{\lim}{\oldlim\limits}
\renewcommand{\bar}{\overline}

\newcommand{\scri}{\ms I}

\definecolor{indigo(dye)}{rgb}{0.0, 0.25, 0.42}

\begin{document}

\title{Holographic algebras at null infinity}

\author[1]{Chang-Han Chen,}
\author[2]{Geoff Penington,}
\author[3]{Gautam Satishchandran,}
\author[2]{and Elisa Tabor}

\affiliation[1]{Leinweber Institute for Theoretical Physics and Department of Physics,\\ University of California, Berkeley, CA 94720, USA}
\affiliation[2]{Leinweber Institute for Theoretical Physics, Stanford University, Stanford, CA 94305, USA}
\affiliation[3]{Princeton Gravity Initiative, Princeton University, Princeton, NJ 08544, USA}

\emailAdd{changhanc@berkeley.edu}
\emailAdd{geoffp@stanford.edu}
\emailAdd{gautam.satish@princeton.edu}
\emailAdd{etabor@stanford.edu}

\abstract{
    We construct an algebra of observables associated to a cut of null infinity in asymptotically flat spacetimes. The Bondi mass associated with a sharp cut of future null infinity is not a well-defined quantum operator: its fluctuations diverge even after smearing in retarded time. We instead introduce a finite-radius, time-smeared version of the Bondi mass and adjoin this operator to the matter and graviton observables in an arbitrarily small asymptotic neighborhood of the cut. We separately analyze spacetimes with and without black holes and find, in both cases, Type III$_1$ von Neumann algebras that satisfy non-trivial nesting relations. In Minkowski spacetime, the resulting algebra reconstructs the spacelike wedge associated with the cut. In a stationary black hole spacetime, it reconstructs the region bounded by the cut and the black hole bifurcation surface, providing an asymptotically flat analogue of an entanglement wedge. In the limit as the cut is moved to past infinity, we recover a Type I$_{\infty}$ algebra for spacetimes without black holes and a Type II$_{\infty}$ algebra for black hole spacetimes. For algebras at finite cuts of retarded time, we construct a Type II$_\infty$ regularization of the black hole algebra whose renormalized von Neumann entropy agrees with the generalized entropy. In appendices, we prove two technical results about quantum fields, in a Schwarzschild spacetime, that asymptote to the Minkowski vacuum at infinity: a split property for unbounded, spacelike separated regions and the construction of a faithful, normal, and semifinite Hartle-Hawking weight whose modular flow is Schwarzschild time evolution.
}

\maketitle

\setcounter{page}{2}
\setcounter{tocdepth}{5}

\section{Introduction}
In asymptotically anti-de Sitter space, quantum gravity is holographic. Asymptotically $AdS_5\times \mathbb{S}^5$ Type IIB string theory is nonperturbatively equivalent to $\mathcal{N}=4$ super-Yang Mills (SYM) \cite{Maldacena:1997re}; more generally, we expect that any consistent, nonperturbative theory of asymptotically AdS quantum gravity is equivalent to some unitary field theory living on its conformal boundary. 

However, the original vision of the holographic principle \cite{tHooft:1993dmi, Susskind:1994vu} went far beyond this. In this vision, any theory of quantum gravity, or at least any theory that could contain black holes, was supposed to be holographic. The part that was unclear prior to AdS/CFT --- and that is still unclear today for spacetimes without asymptotically AdS boundary conditions --- is what ``holographic'' actually means. What, for example, is the analogue of the boundary CFT for an asymptotically flat spacetime?

One way to answer this question would be to find the flat-space analogue of $\mathcal{N}=4$ SYM --- an explicit top-down construction of the holographic dual of some particular asymptotically flat theory. While there has been significant progress in that direction,\footnote{Matrix theory \cite{Banks:1996vh} and the IKKT matrix model \cite{Ishibashi:1996xs} provide important candidate nonperturbative formulations of flat-space M-theory and type IIB string theory, respectively. However, neither proposal takes the form of a codimension-one boundary dual to an asymptotically flat spacetime, which is the type of construction considered here. In the context of celestial holography, explicit top-down holographic dualities have been constructed in special self-dual settings such as a four-dimensional WZW model in Mabuchi gravity on the asymptotically Euclidean Burns space as well as certain self-dual gauge theories on a broader class of four-dimensional, self-dual backgrounds \cite{Costello:2022jpg,Costello:2023hmi,Bittleston:2024efo}. } it is fair to say that we do not, as of yet, have a complete description of the proposed dual theory and its holographic interpretation. We will not attempt to formulate such a top-down theory here.

An alternative approach is to work from the bottom up: to start from perturbative semiclassical gravity in an asymptotically flat background and to try to see the first hints of holographic structure emerging \cite{Marolf:2008mf, Raju:2019qjq, Chowdhury:2020hse, Chowdhury:2021nxw, Bousso:2022hlz, Bousso:2023sya, Bousso:2026btw, Jiang:2017ecm, Apolo:2020bld} (see also the celestial holography program, e.g., \cite{Strominger:2017zoo, Pasterski:2021raf, Donnay:2022aba, Ciambelli:2022vot, Pasterski:2023ikd, Donnay:2023mrd, Chen:2023tvj} and references therein). In retrospect, this provides an alternative history for how AdS/CFT could have been discovered, or at least conjectured: indeed, many of the relevant ingredients for such a development were already in place prior to Maldacena's work \cite{Bekenstein:1973ur, Hawking:1975vcx, Hawking:1982dh, Brown:1986nw}.

A modern version of a bottom-up argument for AdS/CFT goes as follows. In the limit where the gravitational coupling constant $G_N$ goes to zero, quantum gravity reduces to ordinary quantum field theory on a curved background spacetime. In particular, it is a local theory: graviton and matter operators all commute so long as they act in causally separated regions. However, even in this decoupled limit there exist relics of the diffeomorphism gauge constraints that exist in the full nonlinear theory. Just as the total charge of an electromagnetic system can be measured by integrating the electric field surrounding it, there exists a gravitational charge associated to each isometry of the background spacetime that can be measured using (second-order perturbations of) the metric at asymptotic infinity. One of these charges is the ADM Hamiltonian, which generates boundary time evolution via $[H_{\rm ADM},a] = -\ii \partial_t a$.

\begin{figure}
    \centering
    \includegraphics[width=0.9\linewidth]{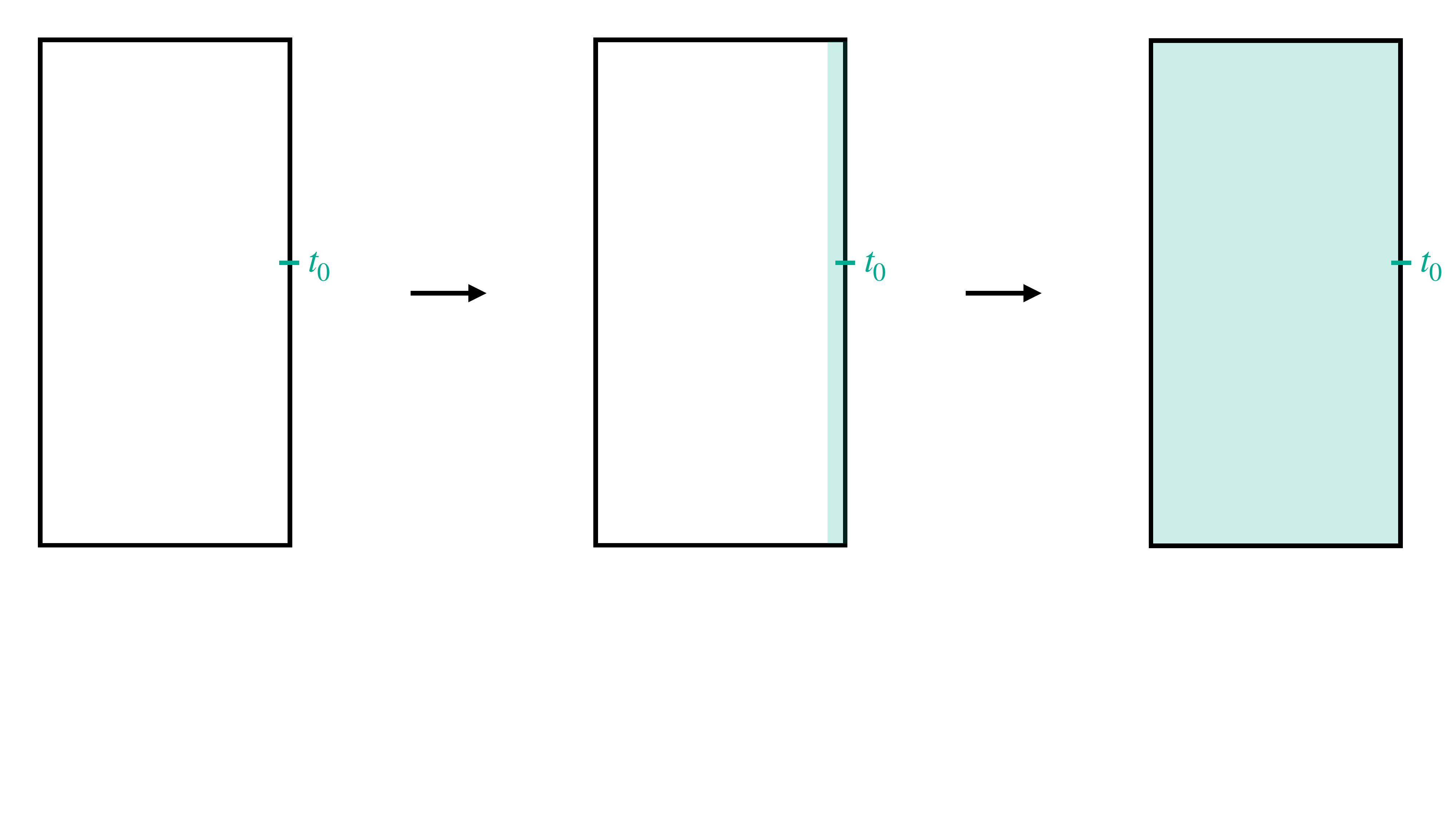}
    \vspace{-2.5cm}
    \caption{The algebra associated to a $t=t_{0}$ cut of the boundary is equivalent to the algebra of operators in the full spacetime of empty AdS. To see this, we first use the ADM Hamiltonian to evolve operators to arbitrary boundary times. Then, in the second step we apply the timelike tube theorem (or HKLL reconstruction) to additionally reconstruct operators deep in the bulk.}
    \label{fig:ads}
\end{figure}

Suppose an external superobserver is able to make arbitrary measurements of bulk fields near asymptotic infinity at some time $t=t_0$, shown in the left panel of Fig.~\ref{fig:ads}. In other words, they can measure operators $a(t_0)$ describing the asymptotic behaviour of matter or gravitons, and, as we just explained, they can measure $H_{\rm ADM}$. Additionally, they can measure sums and products of those operators, along with any operator that can be arbitrarily well approximated by those sums and products. As von Neumann famously showed \cite{Neumann1930ZurAD}, this is sufficient for them to measure any operator in the double commutant algebra $\mathcal{A}_{t_0}=\{ a(t_0), H_{\rm ADM}\}''$. In particular, $\mathcal{A}_{t_0}$ contains the unitaries $\exp(- \ii H_{\rm ADM} t)$ and hence also asymptotic boundary operators $a(t_1) = e^{\ii H_{\rm ADM}(t_1-t_0)}a(t_0) e^{-\ii H_{\rm ADM}(t_1-t_0)}$ at arbitrary time $t_1$.\footnote{Note that the analogous statement in classical gravity is not true, as explained in \cite{Jacobson:2019gnm}. Classically, the Hamiltonian $H_{\rm ADM}$ is still a boundary term. However the classical Hamiltonian flow $e^{-t\{H_{\rm ADM},\;\cdot\;\}}$ is non-analytic in $t$ and so cannot be approximated by the boundary data encoding its Taylor expansion.} So, our observer, who, a priori, could only measure boundary data at $t=t_0$ can in fact measure boundary data at any time. Without appealing to any prior knowledge of AdS/CFT, we have discovered boundary unitarity.

Indeed, something stronger than this is true: in empty AdS, the timelike tube theorem \cite{Borchers:1961, Strohmaier:2023opz} of quantum field theory, or, more prosaically, HKLL reconstruction \cite{Hamilton:2005ju, Hamilton:2006az}, says that any operator can be approximated by operators at asymptotic infinity (so long as those operators are allowed to act at arbitrary times). So, the $t = t_0$ boundary algebra $\mathcal{A}_{t_0}$ actually includes every operator acting on the $G_N \to 0$ QFT Hilbert space. (In settings other than empty AdS, such as an AdS-Schwarzschild black hole, it includes every operator in the boundary's causal wedge.) Within this somewhat narrow limit of quantum gravity, we can already see the emergence of holography.

The goal of this paper is to discover similar realizations of holography in the semiclassical limit of asymptotically flat spacetimes. Previous work on perturbative holography in flat spacetimes \cite{Marolf:2008mf, Laddha:2020kvp} has mostly focused on global questions, such as whether all bulk information is accessible at spatial infinity, or whether the $S$ matrix is unitary.\footnote{Reference \cite{Laddha:2020kvp} also considered the information accessible to a cut of $\mathscr{I}^{+}$. The main differences in our treatment are: (1) a careful treatment of the Bondi mass, which is not a densely defined quantum operator without regularization, and (2) the holographic reconstruction of information deep in the bulk spacetime, rather than merely quantum fields that reach null infinity at some later time.} However, this misses one of the most interesting aspects of holography in asymptotically AdS spacetimes, namely the existence of a unitary boundary time evolution relating the information accessible on different boundary slices. In an alternative-history bottom-up discovery of AdS/CFT, perturbative boundary unitarity would probably have been the single biggest clue to the existence, nonperturbatively, of a local, unitary boundary CFT.

The closest flat-space analogue of a boundary slice in AdS is a cut of either future or past null infinity $\scri^\pm$. Classically, on each such cut we can measure a gravitational charge, known as the Bondi mass, that, like the ADM Hamiltonian, generates time translations. However, unlike in asymptotically AdS spacetimes (with reflecting boundary conditions), this charge is not conserved: instead, energy can leak out of the spacetime via radiation that escapes between different cuts. This fact will have important consequences for the correct holographic interpretation of the information accessible at a cut. Unlike in AdS, the evolution from one cut to another will not be unitary. Instead, holographic time evolution along null infinity is an irreversible process, with all bulk information encoded holographically at spatial infinity, but with information, like energy, leaking out over time via radiation. Mathematically, this process is described by a set of (strictly) nested Type III$_1$ von Neumann algebras
\begin{equation}
    \A(u_1) \subsetneq \mathcal{A}(u_0) \qquad u_1 > u_0
\end{equation}
that are each purified by their already emitted radiation.

 We leave the question of exactly what these observations mean for a putative nonperturbative holographic dual theory living on $\scri^+$ to future work. However, it is clear that the answer cannot be a unitary field theory. Since radiative degrees of freedom decouple at null infinity, we do think it is likely that the basic structure of holographic degrees of freedom leaking out into radiation continues to exist, with both the holographic algebra and the radiation algebra in its commutant described by Type III$_1$ factors. We also note, in passing, the importance of this structure for the black hole information problem in asymptotically flat spacetimes: after regulating local vacuum entanglement near the cut, the entanglement entropy between the emitted Hawking radiation and the holographic algebra describing the black hole and its surrounding is precisely the object that is expected to follow a Page curve \cite{Penington:2019npb, Almheiri:2019hni, Geng:2020fxl, Geng:2021hlu, Raju:2020smc, Antonini:2025sur, Geng:2026asi}.

The basic structure of the paper is as follows. In sec.~\ref{sec:AFS} we review the classical phase space of asymptotically flat perturbations and its quantization. We review how both the ADM mass and the Bondi mass arise as Hamiltonian charges and how the Bondi mass-loss formula relates their difference to the energy radiated through null infinity. We then quantize the radiative degrees of freedom and construct the Hilbert space on which the algebra naturally acts.

In sec.~\ref{sec:bondi}, we meet our first major technical obstruction. The Bondi mass, as conventionally defined, is a perfectly well-behaved classical object. Unfortunately, however, it suffers from quantum divergences that mean it only exists as a sesquilinear form and not as a true (densely defined) quantum operator. In particular, unlike the ADM Hamiltonian, it cannot be exponentiated to time evolve quantum fields. Our solution to this issue is to construct a ``regulated Bondi mass'' at a large but finite sphere of radius $r_0$ and smear it over some small time interval $\delta u$. We show that this process yields a densely defined operator on the perturbative Hilbert space. 

With these preliminaries in hand, we turn in sec.~\ref{sec:algebra} to the main construction of the paper. In sec.~\ref{subsec:i0}, we first review the construction of an algebra associated to spatial infinity, illustrating how this simpler construction connects to previously studied bulk algebras \cite{Marolf:2008mf, Kudler-Flam:2023qfl, Chen:2024rpx, Klinger:2026tws}. Then in sec.~\ref{subsec:construct}, we include the regulated Bondi mass in the algebra defined in sec.~\ref{sec:AFS}, giving a precise definition of what we mean by the algebra $\A(u_0)$ associated to a cut of $\scri^+$. We first ``fatten'' the cut into a thin strip $\r[u_0, \delta u, \rmin]$ containing all points at radii $r > \rmin$ and at times $u_0 < u < u_0 + \delta u$: we define the algebra $ \A_{\delta u,\rmin}(u_0)$ to be the double commutant of all fields in $\r[u_0, \delta u, \rmin]$. This allows $ \A_{\delta u,\rmin}(u_0)$ to include both (a) the regulated Bondi mass described above as well as (b) massive fields whose excitations never actually reach null infinity. To recover an algebra that can be cleanly associated to the original cut, we simply take the limits $\delta u \to 0$ and $\rmin \to \infty$ by taking intersections over all the associated algebras. It turns out that, in all the cases we consider, $\A_{\delta u,\rmin}(u_0)$ is independent of both $\delta u$ and $\rmin$ and so this limit is somewhat trivial.

\begin{figure}
    \centering
    \includegraphics[width=0.75\linewidth]{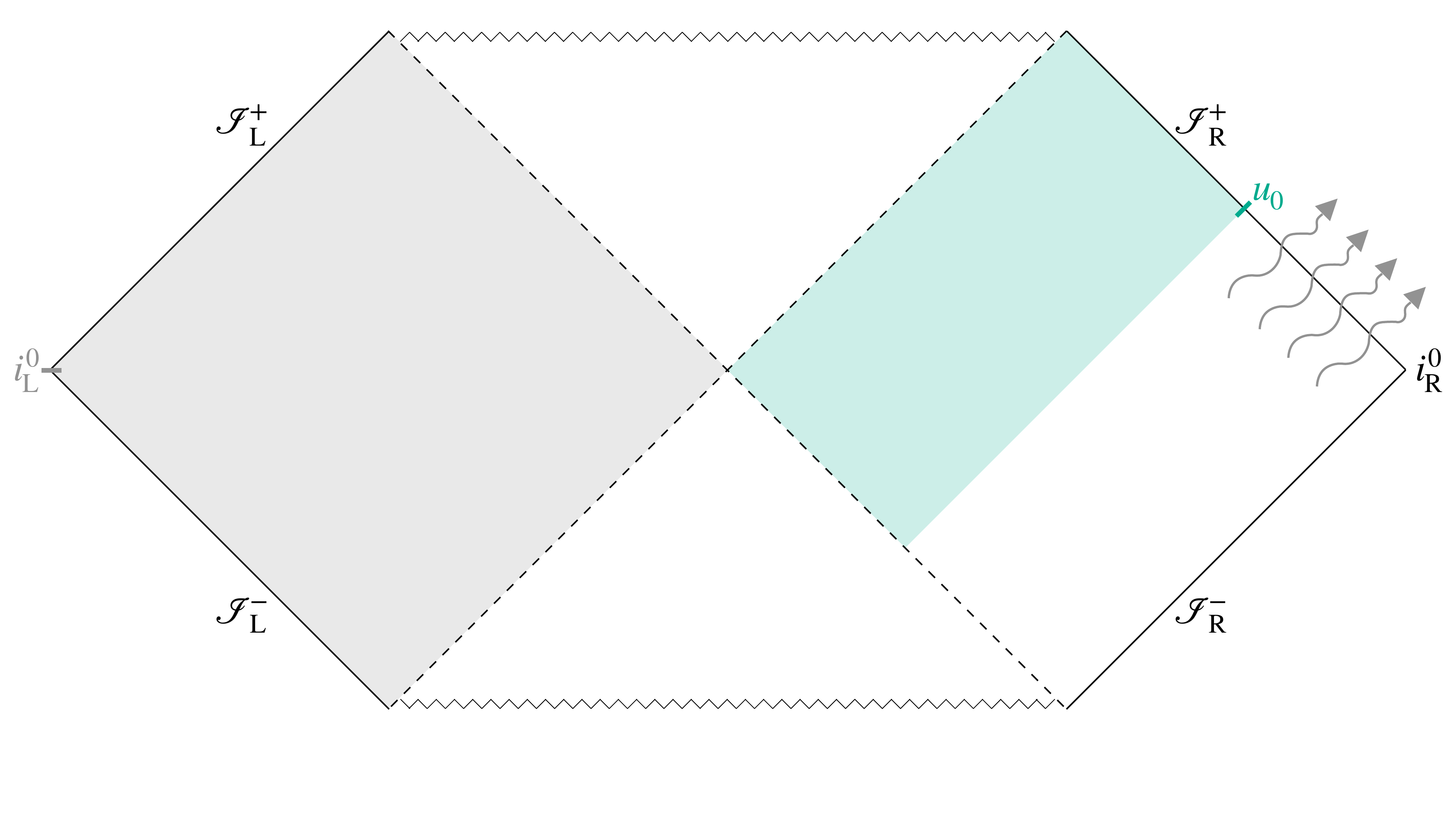}
    \vspace{-.6cm}
    \caption{The entanglement wedge (light green) associated to a $u=u_0$ cut of $\scri^+_{\rm R}$ in a two-sided Schwarzschild black hole is the region bounded by the cut and the black hole bifurcation surface. The commutant of the algebra encoded in this region is shown in gray; it consists of the left exterior (the entanglement wedge of left spatial infinity) along with all radiation to the past of the $u=u_0$ cut.}
    \label{fig:schw_rad}
\end{figure}

We then show that the algebra $\mathcal{A}(u_0)$ is holographic, first for perturbations of Minkowski space in sec.~\ref{subsec:algnoblackholes} and then for Schwarzschild black holes in sec.~\ref{subsec:schw_alg}. The arguments are similar in spirit to the ones given for AdS above; summaries can be found in Fig.~\ref{fig:mink4} and Fig.~\ref{fig:schw4} respectively. The main additional complicating factor is that we cannot use the regulated Bondi mass to evolve operators into the past of the cut on which the Bondi mass was defined. As a result, in Minkowski space, the algebra $\A(u_0)$ only encodes operators that are spacelike separated from the cut $u_0$. The missing degrees of freedom, i.e. the operators in its commutant $\A(u_0)'$, are precisely the radiation that reached $\scri^+$ before $u_0$. For a Schwarzschild black hole, the encoded region, which can be thought of as a flat-space version of an entanglement wedge, is the region spacelike separated from the $u=u_{0}$ cut of $\mathscr{I}^{+}$ and the bifurcation surface (see Fig.~\ref{fig:schw_rad}).\footnote{We note that this is consistent with the proposal for entanglement wedges of general gravitating regions given in \cite{Bousso:2022hlz, Bousso:2023sya, Bousso:2025joj, Bousso:2025fgg, Sahu:2025upe}.}

In empty AdS, the algebra $\A_{t_0}$ includes all operators acting on the Hilbert space, and is therefore a Type I$_{\infty}$ von Neumann algebra. For an AdS-Schwarzschild black hole, on the other hand, it turns out to be Type II$_\infty$, with renormalized entropies that correctly account for the Bekenstein-Hawking entropy \cite{Witten:2021unn, Chandrasekaran:2022eqq}.\footnote{The general relationship between Type II algebras and the entropy of black holes as well as cosmological horizons has been recently investigated by a number of authors \cite{Chandrasekaran:2022cip, Kudler-Flam:2023qfl, Akers:2024bel, Kudler-Flam:2024psh, Chen:2024rpx, Penington:2025hrc, Herderschee:2025nsb, Klinger:2026tws, Cui:2026bcd}.} The asymptotically flat algebras behave differently at any finite cut $u_0$. In both Minkowski and Schwarzschild spacetimes, the algebra $\A(u_0)$ is Type III$_1$, reflecting the ultraviolet entanglement with radiation that has already escaped through $\scri^+$ before $u_0$. In the limit $u_{0}\to -\infty$ the early-radiation factor disappears and we recover the algebra at spatial infinity $\mathcal{A}(i^{0})$. In the case of flat spacetime, $\mathcal{A}(i^{0})$ is Type I$_{\infty}$ and, in the case of a black hole spacetime, it is Type II$_{\infty}$ \cite{Kudler-Flam:2023qfl,Chen:2024rpx,Klinger:2026tws}. Nevertheless, in the case of black holes, the underlying Type II$_{\infty}$ structure is still present even for the holographic algebras at finite cuts of $\mathscr{I}^{+}$. Although $\mathcal A(u_0)$ is Type III$_1$, regulating its QFT divergences produces a Type II$_\infty$ algebra whose renormalized entropy matches the expected generalized entropy of the black hole. Finally, in sec.~\ref{subsec:partial}, we argue that a partial cut reconstructs the future of its generators on null infinity, but no open bulk region, in sharp contrast with subregion reconstruction in AdS/CFT \cite{Ryu:2006bv, Hubeny:2007xt, Engelhardt:2014gca}.

Two technical results used throughout the black-hole analysis are
proved in the appendices. In appendix~\ref{app:split}, we prove the split property for two spacelike separated, unbounded regions in Schwarzschild that extend toward left and right spatial infinity, respectively. This result is an essential ingredient in establishing that the regulated Bondi mass is densely defined in sec.~\ref{subsec:RegularizedBondiMass} and in sec.~\ref{subsec:schw_alg} where we derive several structural properties of the holographic algebra $\A(u_0)$  in black hole spacetimes. In appendix~\ref{app:fns}, we prove that the Hartle-Hawking weight, first conjectured to exist in \cite{Chen:2024rpx}, is faithful, normal, and semifinite on a Hilbert space of states that behave like the Minkowski vacuum at infinity, thereby justifying the crossed-product
construction of the left and right exterior Type II$_\infty$ algebras. While its role in the present paper is largely technical, the existence of this weight also has independent significance as the physical interpretation within that Hilbert space of a Euclidean black hole path integral.

\section{Asymptotically Flat Spacetimes} \label{sec:AFS}

In this section, we review the description of perturbations about asymptotically flat spacetimes. We first discuss the classical phase space of these perturbations in sec.~\ref{subsec:firstorderclass}. Then in sec.~\ref{subsec:quantize} we review their quantization. We refer the reader to, e.g., \cite{Prabhu:2022zcr} for further details.

\subsection{Classical phase space} \label{subsec:firstorderclass}  

The spacetime $(M,g)$ is asymptotically flat  if there exist asymptotically Minkowskian coordinates $x^{\mu}$ such that $g_{\mu \nu}$ differs from $\eta_{\mu \nu}$ by $O(\rho^{-(d-3)})$ as $\rho \to \infty$ --- where $\rho^{2}=\sum_{i=1}^{d-1} x_{i}^{2}$ --- and $N$th derivatives of $g_{\mu \nu}$ are $O(\rho^{-(d-3+N)})$. In four dimensions we also impose Regge-Teitelboim parity conditions on the asymptotic behavior of the metric in spatial directions (see Appendix A of \cite{Regge:1974zd}).  Additionally, we will require that the metric be asymptotically flat at null infinity which imposes analogous conditions on the behavior of the metric in null directions that we will specify shortly. For simplicity, we will restrict attention to metrics $g_{ab}$ that satisfy the vacuum Einstein equations and we will simply state the necessary modifications to the relevant formulae in the presence of matter. 

\subsubsection*{The symplectic structure of the gravitational phase space}

From the Lagrangian of general relativity one can naturally obtain a symplectic structure on the tangent vectors on phase space \cite{Lee:1990nz}. If $g_{ab}(\lambda)$ is any one-parameter family of asymptotically flat metrics then we may define a tangent vector as
\begin{equation}
\delta g_{ab}\defn \frac{d g_{ab}}{d\lambda}\bigg\vert_{\lambda=0} \,.
\end{equation}
The symplectic form on these tangent vectors is given by 
\begin{equation}
W_{\Sigma}(\delta_{1}g,\delta_{2}g) = -\int_{\Sigma}\sqrt{h} d^{d-1}x~ n^{a}\omega_{a}(\delta_{1}g,\delta_{2}g)\,,
\end{equation}
where $h_{ab}$ is the induced metric on any Cauchy surface $\Sigma$ with volume form $\sqrt{h}d^{d-1}x$ and $n^{a}$ is the future-directed unit normal vector to $\Sigma$. The vector field $\omega^{a}$ is 
\begin{equation}
\omega^{a}(\delta_{1}g,\delta_{2}g) = \frac{1}{16\pi G_{\textrm{N}}}P^{abcdef}[\delta_{2}g_{bc}\nabla_{d}\delta_{1}g_{ef}-\delta_{1}g_{bc}\nabla_{d}\delta_{2}g_{ef}]
\end{equation}
where
\begin{equation}
P^{abcdef} \equiv g^{ae}g^{fb}g^{cd}
-\tfrac{1}{2}g^{ad}g^{be}g^{fc}
-\tfrac{1}{2}g^{ab}g^{cd}g^{ef}
-\tfrac{1}{2}g^{bc}g^{ae}g^{fd}
+\tfrac{1}{2}g^{bc}g^{ad}g^{ef}.
\end{equation}
It can be shown that if $g_{ab}$ satisfies Einstein's equation and $\delta g_{ab}$ satisfies the linearized Einstein equation then $\omega_{a}$ is conserved --- i.e., $\nabla^{a}\omega_{a}=0$. Consequently, if the metric and its perturbations are on-shell then $W_{\Sigma}$ is independent of the Cauchy surface $\Sigma$.

We can obtain a more recognizable form for the symplectic form by decomposing the metric $g_{ab}(\lambda)$ on $\Sigma$ into canonically conjugate variables $(h_{ab},p^{ab})$ where $p^{ab}$ is the conjugate momentum relative to the induced metric  $h_{ab}$ on $\Sigma$ given by 
\begin{equation}
p^{ab} = \sqrt{h}(K^{ab} - h^{ab}K)\,.
\end{equation}
In terms of these variables, the symplectic form becomes 
\begin{equation}
W_\Sigma(\delta_{1}g,\delta_{2}g) = -\frac{1}{16\pi G_{\textrm{N}}}\int_{\Sigma}d^{d-1}x~(\delta_{1}h_{ab} \delta_{2}p^{ab} - \delta_{2}h_{ab} \delta_{1}p^{ab})\,,
\end{equation}
which converges if the metric satisfies our asymptotic conditions. 

\subsubsection*{ADM conserved charges from large diffeomorphisms}

In general relativity, diffeomorphisms are a gauge symmetry of the theory. At the level of our phase space analysis this follows from the fact that the symplectic form is degenerate on any pure gauge solution. To see this we note that if we feed $\mathcal{L}_{X}g$ into one of the slots of the integrand of the symplectic form where $X^{a}$ is any $\lambda$-independent, smooth vector field, it  can be shown that $\omega_{a}$ can be expressed as the following identity \cite{Iyer:1994ys} 
\begin{equation}
\label{eq:sympformdecomp}
\omega(\tfrac{d}{d\lambda} g,\mathcal{L}_{X}g) = i_{X}(E(g)\cdot \tfrac{d}{d\lambda}g) + \frac{d}{d\lambda}C_{X} + d[\tfrac{d}{d\lambda}Q_{X}(g) -i_{X}\theta(g;\tfrac{d}{d\lambda}g)]\,.
\end{equation}
To simplify the notation, both the left-hand side and right-hand sides of this expression are expressed as $(d-1)$-forms --- i.e., the left-hand side is $\omega_{a_{1}\dots a_{d-1}} = \omega^{b}\epsilon_{ba_{1}\dots a_{d-1}}$ where $\epsilon_{a_{1}\dots a_{d}}$ is the positively oriented volume form of a $d$-dimensional spacetime with metric $g_{ab}$. We now define each term on the right-hand side of \eqref{eq:sympformdecomp}. The quantities $E$ are defined such that $E=0$ are the equations of motion of the theory and $C_{X}$ are the constraints. In vacuum general relativity they are given by
\begin{equation}
E^{ab}{}_{c_{1}\dots c_{d}}= -\frac{1}{16\pi G_N}G^{ab}\epsilon_{c_{1}\dots c_{d}}\,,\qquad (C_{X})_{a_{1}\dots a_{d-1}}=\frac{1}{8\pi G_N}X^{e}G_{e}{}^{f}\epsilon_{fa_{1}\dots a_{d-1}}
\end{equation}
where $G_{ab}$ is the Einstein tensor and the action of $i_{X}$ denotes contraction into the first index. The quantity $Q_{X}$ is the ``Noether charge'' 
\begin{equation}
(Q_{X})_{a_{1}\dots a_{d-2}} = -\frac{1}{16\pi G_{\textrm{N}}}\nabla_{c}X_{d}\epsilon^{cd}{}_{a_{1}\dots a_{d-2}}\,,
\end{equation}
$\theta$ is the ``pre-symplectic potential'' 
\begin{equation}
\theta_{e_{1}\dots e_{d-1}} = \frac{1}{16\pi G_{\textrm{N}}}g^{ac}g^{bd}(\nabla_{d}\tfrac{d}{d\lambda}g_{bc} - \nabla_{c}\tfrac{d}{d\lambda}g_{bd})\epsilon_{ae_{1}\dots e_{d-1}}\,
\end{equation}
and $(i_{X}\theta)_{a_{1}\dots a_{d-2}}\equiv X^{b}\theta_{ba_{1}\dots a_{d-2}}$. We emphasize that \eqref{eq:sympformdecomp} is an identity that holds for any $X^{a}$ and any $g_{ab}(\lambda)$. 

We now impose the condition that the one-parameter family \(g_{ab}(\lambda)\) satisfies Einstein's equation, so that \(E=C_{X}=0\). Integrating \eqref{eq:sympformdecomp} over \(\Sigma\) yields 
\begin{equation}
\label{eq:WSigmaQ}
W_{\Sigma}(\tfrac{d}{d\lambda}g,\mathcal{L}_{X}g)
=
\int_{S}\mathcal{Q}_{X}(g;\tfrac{d}{d\lambda}g)\,,
\end{equation}
where the boundary term is 
\begin{equation}
\label{eq:mcQX}
\mathcal{Q}_{X}(g;\tfrac{d}{d\lambda}g) \equiv \tfrac{d}{d\lambda}Q_{X}(g) -i_{X}\theta(g;\tfrac{d}{d\lambda}g)
\end{equation}
and $S$ is the boundary of $\Sigma$ which may include a conformal boundary as well as any internal boundaries such as the horizon of a black hole. It follows immediately that, for any vector field \(X^{a}\) that vanishes in a neighborhood of the boundary of \(\Sigma\),
\begin{equation}
W_{\Sigma}(\tfrac{d}{d\lambda}g,\mathcal{L}_{X}g)=0
\qquad\qquad
\textrm{(\(X\) is a small diffeomorphism)}\,.
\end{equation}
More generally, we refer to any vector field $X^{a}$ such that on the boundary we have that $\mathcal{Q}_{X}=0$ as a ``small diffeomorphism'' which defines a degeneracy direction of the symplectic form on phase space. Accordingly, two points in phase space that differ by a small diffeomorphism represent the same physical spacetime.

By contrast, if \(X^{a}\) does not vanish on the boundary, and if \(\mathcal{Q}_{X}\) admits a finite limit to \(S\), then the corresponding diffeomorphism is a ``large diffeomorphism'' and does not define a degeneracy direction of the symplectic form. An important example is provided by vector fields \(X^{a}\) that approach asymptotic symmetries in the asymptotic region. In particular, evaluating \(\mathcal{Q}_{X}\) at \(\lambda=0\) yields the variation of the Hamiltonian charge
\begin{equation}
\label{eq:HX}
\delta H_{X} = \int_{i^{0}} \mathcal{Q}_{X}(g;\delta g) = W_{\Sigma}(\delta g,\mathcal{L}_{X}g) \,.
\end{equation}
This expression can be integrated on phase space to define a conserved quantity \(H_{X}\) \cite{Iyer:1994ys}.
Of particular importance for the considerations of this paper is the case where \(X^{a}\to (\partial/\partial t)^{a}\) approaches an asymptotic time translation. Then
\begin{equation}
\label{eq:Ht}
H_{t} = M_{i^{0}}
\end{equation}
is the ADM mass at spatial infinity. Similarly if $X^{a}\to (\partial/\partial x^{i})^{a}$ is an asymptotic translation then $-H_{X}$ is the ADM momentum component $P_{i}$, if $X^{a} \to x^{i}(\partial /\partial x^{j})^{a} -  x^{j}(\partial /\partial x^{i})^{a}  $ is an asymptotic rotation then $-H_{X}$ is the angular momentum $J_{ij}=J_{[ij]}$ in the $ij$ plane and if $X^{a}\to t (\partial/\partial x^{i})^{a} + x^{i}(\partial /\partial t)^{a}$ is an asymptotic boost then $H_{X}$ is the ADM center-of-mass $C_{i}$. In total, the $\tfrac{1}{2}d(d+1)$ quantities $(M_{i^{0}},P_i,J_{ij},C_{i})$ are the ADM conserved quantities associated to an asymptotically flat spacetime $(M,g_{ab})$.

\subsubsection*{Asymptotic flatness in null directions}
\label{sec:asympflat}

In an asymptotically flat spacetime  radiation may be emitted which will, in general, carry away energy and angular momentum from the system. For example, in the merger of two black holes, a portion of the total energy is carried away by gravitational radiation, and hence the final mass differs from the initial mass. Since this radiation propagates to null infinity, the appropriate charges that describe the system at a later time are defined at null infinity. In the remainder of this section, we review the notion of asymptotic flatness in null directions. We will then present a derivation of the celebrated ``Bondi mass loss'' formula representing the energy loss due to the emission of radiation \cite{Bondi:1962px}. Our primary purpose in reviewing this derivation is that the methods employed in this section will prove useful in sec.~\ref{sec:bondi} when we consider the Bondi mass in the quantum theory. 

To impose asymptotic flatness in null directions we assume that the metric is conformal to a spacetime $(\bar{M},\bar{g}_{ab})$ with a complete future null infinity $\mathscr{I}^{+}\cong \mathbb{R}\times \mathbb{S}^{d-2}$ with null normal $n^{a}$. We can construct coordinates $(u,x^{A})$ on $\mathscr{I}^{+}$ where $n^{a}=(\partial/\partial u)^{a}$ and $x^{A}$ are arbitrary angular coordinates on the $(d-2)$-sphere cross-sections $S_{u}$ of constant $u$. The metric $q_{AB}$ on these cross-sections is the round-sphere metric.  Choosing a null vector $\ell^{a}$ transverse to $\mathscr{I}^{+}$ --- i.e., $\ell^{a}n_{a}=-1$ at null infinity --- these coordinates can be carried into the bulk of the spacetime by geodesic transport. In the physical spacetime, we use ``Bondi coordinates''\footnote{Note that this coordinate system actually differs from what is normally called “Bondi gauge”, although it has obvious similarities. See, e.g., \cite{Ishibashi:2007kb}.} $(u,r,x^{A})$ where $r$ is the affine parameter of radially outward, future-directed null geodesics. In the physical spacetime, the  metric is asymptotically flat if it admits an asymptotic expansion in powers of $1/r$ where, in any orthonormal frame, the deviation from the flat metric is 
\begin{equation}
|g_{\mu \nu} - \eta_{\mu \nu}|\sim O(1/r^{d/2-1})
\end{equation}
where the $\mu, \nu$ components denote components in any orthonormal frame. Any partial derivatives of $g_{\mu \nu} - \eta_{\mu \nu}$ with $u$ or $x^{A}$ are also $O(1/r^{d/2-1})$ whereas any partial derivatives with respect to $r$ are $O(1/r^{d/2})$.  In even dimensions, this asymptotic behavior follows from smoothness\footnote{We will need only to assume that the conformal metric $\bar{g}_{ab}$ is, at most, $C^{k}$ at $\mathscr{I}^{+}$ where $k=\textrm{max}\{\tfrac{d}{2},d-3\}$.} of the conformal metric at $\mathscr{I}^{+}$. In odd dimensions, in addition to the half-integer powers in $1/r$ one must also include an additional series with integer fall-off starting at $O(1/r^{d-3})$. We impose similar conditions at past null infinity. We refer the reader to \cite{Satishchandran:2019pyc,Hollands:2003ie} for a complete discussion of the asymptotic behavior of the metric in null directions. 

For our purposes, the radiative data is captured by the leading order behavior of the angular components of the metric. These components admit the asymptotic expansion
\begin{equation}
g_{AB}(u,r,x^{A}) = r^{2}\bigg(q_{AB} + \frac{C_{AB}}{r^{d/2-1}} + \dots \bigg)
\end{equation}
where $q_{AB}$ is the round metric on the unit sphere, $C_{AB}(u,x^{C})$ encodes the outgoing radiation field and the $\dots$ denote terms of higher-order in powers of $1/r$. The radiation is described by the trace-free part of $C_{AB}$ with respect to $q_{AB}$ which we denote as 
\begin{equation}
\sigma_{AB}(u,x^{E}) \defn -\frac{1}{2}  \bigg(q_{A}{}^{C}q_{B}{}^{D}-\frac{1}{d-2}q_{AB}q^{CD}\bigg)C_{CD}(u,x^{E})
\end{equation}
where the projection is such that $q^{AB}\sigma_{AB}=0$. The field $\sigma_{AB}$ is sometimes referred to as the asymptotic ``shear''. We additionally define the  Bondi news tensor which is given by 
\begin{equation}
N_{AB} = 2\mathcal{L}_{n}\sigma_{AB} = 2\partial_{u}\sigma_{AB}\,.
\end{equation}
As we will see, the square of the  Bondi news tensor encodes the power radiated to $\mathscr{I}^{+}$ due to gravitational waves. 

\subsubsection*{Classical Bondi mass}

We now consider the Bondi mass associated to any cut $\mathcal{C}$ of null infinity. While the Bondi mass can be independently defined as a ``charge'' on $\mathcal{C}$   associated to asymptotic time translations at $\mathscr{I}^{+}$ \cite{Wald:1999wa}, it can also be obtained by integrating \eqref{eq:WSigmaQ} on $\mathscr{I}^{+}$ where $X$ is an asymptotic time translation $t^{a}$ and $\Sigma$ is taken to be the portion of $\mathscr{I}^{+}$ to the past of the cut $\mathcal{C}$. Without loss of generality we may, by a constant rescaling of $u$ and an asymptotic supertranslation $u\to u + f(x^{A})$, choose coordinates such that $t^{a}=n^{a}$ and that $\mathcal{C}$ corresponds to a constant $u$ cross-section of $\mathscr{I}^{+}$. We will denote the subregion of null infinity to the past of a constant $u$ cut of $\mathscr{I}^{+}$ as $\mathscr{I}^{+}_{<u}$. Integrating the pullback of $\omega_{a}$ to $\mathscr{I}^{+}$ over $\mathscr{I}^{+}_{<u}$ yields \cite{Ashtekar:1987tt,Hollands:2012sf}
\begin{equation}
W_{\mathscr{I}^{+}_{<u}}(\delta_{1} g,\delta_{2}g) = -\frac{1}{16\pi G_{\textrm{N}}}\int_{\mathscr{I}^{+}_{<u}}dud\Omega~(\delta_{1}\sigma_{AB} \delta_{2}N^{AB} - \delta_{1}N_{AB} \delta_{2}\sigma^{AB})\,,
\end{equation}
where $\delta \sigma_{AB}$ and $\delta N_{AB}$ are the linearized shear and linearized news respectively, and the right-hand side converges if the radiation $N_{AB}$ decays appropriately at past infinity. To evaluate the left-hand side of \eqref{eq:WSigmaQ} with $\Sigma = \mathscr{I}^{+}_{<u}$ we evaluate the above symplectic product with $\delta_{2}g = \mathcal{L}_{X}g$ where $X$ is an asymptotic time translation. In that case, we obtain 
\begin{equation}
\label{eq:symptranslation}
W_{\mathscr{I}^{+}_{<u}}(\delta g,\mathcal{L}_{X}g) = -\frac{1}{16\pi G_{\textrm{N}}} \int_{\mathscr{I}^{+}_{<u}}du d\Omega~(\delta\sigma_{AB}\mathcal{L}_{t}N^{AB} - \delta N^{AB}\mathcal{L}_{t}\sigma_{AB}) \,.
\end{equation}
With these choices and integrating  \eqref{eq:symptranslation} by parts we obtain
\begin{equation}
\label{eq:symptranslation2}
W_{\mathscr{I}^{+}_{<u}}(\delta g,\mathcal{L}_{X}g)  = \frac{1}{16\pi G_{\textrm{N}}} \int_{\mathscr{I}^{+}_{<u}}dud\Omega ~N^{AB} \delta N_{AB} - B(u)
\end{equation}
where $B(u)$ is the boundary term 
\begin{equation}
\label{eq:boundaryterm}
B(u) = \frac{1}{16\pi G_{\textrm{N}}}\int_{\mathbb{S}^{d-2}}d\Omega~N^{AB}(u,x^{C})\delta \sigma_{AB}(u,x^{C})
\end{equation}
and we do not obtain a boundary term at past infinity since $N_{AB}\to 0$ as $u\to -\infty$. Integrating the right-hand side of \eqref{eq:WSigmaQ} and applying Eqs.~\eqref{eq:HX} and \eqref{eq:Ht} to \eqref{eq:symptranslation2} we obtain 
\begin{equation}
\label{eq:QBexp}
W_{\scri^+_{<u}}(\delta g,\L_Xg) = \delta M_{i^{0}} -\int_{\mathcal{C}(u)}\mathcal{Q}_{t}(g;\delta g)  = \frac{1}{32\pi G_{\textrm{N}}} \delta \bigg[\int_{\mathscr{I}^{+}_{<u}}dud\Omega ~ N^{2}\bigg] - B(u)
\end{equation}
where  $N^{2}=N_{AB}N^{AB}$, $\mathcal{Q}_{t}(g;\delta g)\vert_{\mathcal{C}(u)}$ is the charge $\mathcal{Q}_{t}(g;\delta g)$ evaluated at the $\mathcal{C}(u)\cong \mathbb{S}^{d-2}$ which is a constant $u$ cut of $\mathscr{I}^{+}$, and in the first term on the right-hand side we have used the fact that $N^{AB}\delta N_{AB}=\tfrac{1}{2} \delta (N^{2})$ to write this term as a total variation. While $\mathcal{Q}_{t}(g;\delta g)$ cannot be written as a total variation, it was shown in \cite{Wald:1999wa} that the quantity
\begin{equation}
\label{eq:MB}
\delta M_{B}(u) \defn \bigg[\int_{\mathcal{C}(u)}\mathcal{Q}_{t}(g;\delta g)\bigg]-B(u)
\end{equation}
is a total variation and defines the Bondi mass \cite{Wald:1999wa}.  Integrating \eqref{eq:QBexp} in phase space, using \eqref{eq:MB} as well as the assumption that the radiation decays sufficiently rapidly at early times so that \cite{Ashtekar:1979xeo,Hollands:2013cva}
\begin{equation}
\lim_{u\to -\infty}~\delta M_{\textrm{B}}(u) = \delta M_{i^{0}}
\end{equation}
then integrating \eqref{eq:QBexp} in phase space yields the ``Bondi mass-loss'' formula
\begin{equation}
\label{eq:bondimassloss}
M_{B}(u) = M_{i^{0}} - \frac{1}{32\pi G_{\textrm{N}}} \int_{\mathscr{I}^{+}_{<u}}dud\Omega ~ N^{2}\,.
\end{equation}
Importantly, the second term is negative-definite and so, $M_{i^0}-M_{B}(u)$ represents the mass lost due to the emission of radiation to the past of a constant $u$ cut of $\mathscr{I}^{+}$.

\subsubsection*{Bondi mass-loss in the presence of matter}

We obtained the Bondi mass-loss formula in vacuum general relativity. However it is straightforward to generalize the formula to include any collection of matter fields $\phi$ on $(M,g)$ where now $g$ satisfies Einstein's equation with stress energy $T_{ab}(\phi)$. We will assume, for simplicity, that the background fields vanish (i.e., $\phi(0)=0$). While there is no essential difficulty in generalizing our arguments to include background matter fields, we will neglect any such contributions to simplify the equations. The matter fields can also be endowed with a symplectic form $W^{\textrm{matt.}}_{\Sigma}(\delta_{1}\phi,\delta_{2}\phi)$ such that the full symplectic form on any solution $\Phi=(\phi,g)$ is simply 
\begin{equation}
\label{eq:sympformfull}
W_{\Sigma}(\delta_{1} \Phi,\delta_{2} \Phi)=W^{\textrm{matt.}}_{\Sigma}(\delta_{1}\phi,\delta_{2}\phi) + W_{\Sigma}(\delta_{1}g,\delta_{2}g)\,,
\end{equation}
where $\delta \Phi = (\delta \phi,\delta g)$ satisfies the decoupled, linearized equations for the matter fields on $(M,g(0))$ and $\delta g$ satisfies the vacuum linearized Einstein equation.  Repeating the analysis of the previous subsection, the symplectic product of $\delta\Phi$ and $\mathcal{L}_{X}\Phi= (\mathcal{L}_{X} g,\mathcal{L}_{X}\phi)$ is again equivalent to the integral of a boundary charge  at the boundary $S$ of $\Sigma$
\begin{equation}
\label{eq:WSigmaQ_matter}
W_{\Sigma}(\tfrac{d}{d\lambda}\Phi,\mathcal{L}_{X}\Phi)
=
\int_{S}\mathcal{Q}_{X}(\Phi;\tfrac{d}{d\lambda}\Phi)\,,
\end{equation}
where $\mathcal{Q}_{X}$ is defined similarly to \eqref{eq:mcQX} but with an additional contribution due to the presymplectic potential $i_{X}\theta(\phi,\tfrac{d}{d\lambda}\phi)$ of any matter fields (see, e.g.,  \cite{Bonga:2019bim}). One finds that the formula for the Bondi mass \eqref{eq:MB} is unchanged but now the Bondi mass-loss \eqref{eq:bondimassloss} receives an additional contribution due to the stress-energy flux of any massless fields to $\mathscr{I}^{+}$ given by 
\begin{equation}
T_{uu}^{(d-2)}(u,x^{A}) \defn \lim_{r\to \infty} r^{d-2}T_{uu}(u,r,x^{A})\,.
\end{equation}
The mass-loss formula then becomes 
\begin{equation}
\label{eq:BondimassN2Tuu}
M_{B}(u) = M_{i^{0}} - \frac{1}{32\pi G_{\textrm{N}}} \int_{\mathscr{I}^{+}_{<u}}dud\Omega ~ N^{2} - 
\int_{\mathscr{I}^{+}_{<u}}dud\Omega~T^{(d-2)}_{uu}
\end{equation}
where the contribution due to matter is negative definite for any matter fields satisfying the dominant energy condition \cite{Satishchandran:2019pyc,Bieri:2013ada}. 

Applying a similar analysis one can obtain analogous asymptotic charges on cuts of $\mathscr{I}^{+}$ associated to the full asymptotic symmetry group at null infinity (see, e.g., \cite{Wald:1999wa}). In four spacetime dimensions this group is infinite dimensional and includes the abelian group of ``supertranslations''. In $d>4$ dimensions the symmetry group can be consistently reduced to the standard Poincaré group \cite{Hollands:2016oma}. In this paper, we will be primarily focused on the Bondi mass associated to any cut $\mathcal{C}$ of $\mathscr{I}^{+}$. 

\subsection{Quantization of gravitational perturbations and quantum fields} \label{subsec:quantize}

In the previous subsection we considered the classical phase space of asymptotically flat spacetimes. In this subsection, we consider the quantization of gravitational perturbations of a background, asymptotically flat spacetime $(M,g)$. In the following, for ease of notation, we will denote perturbations $\delta g_{ab}$ which satisfy the linearized equations of motion on $(M,g_{ab})$ as the tensor field
\begin{equation}
\gamma_{ab}(x) \equiv \delta g_{ab}(x)\,,
\end{equation}
with equations of motion
\begin{equation}
\label{eq:linEE}
 -\frac{1}{2}\Box_{g}\gamma_{ab} -\frac{1}{2}\nabla_{a}\nabla_{b}\gamma^{c}{}_{c}+\frac{1}{2}\nabla_{c}\nabla_{a}\gamma^{c}{}_{b}+\frac{1}{2}\nabla_{c}\nabla_{b}\gamma^{c}{}_{a}=0 \,.
\end{equation}
We recall that two solutions $\gamma_{ab}$ and $\gamma^{\prime}_{ab}$ are considered equivalent if they differ by the gauge transformation
\begin{equation}
\label{eq:gaugetransformation}
\gamma_{ab}^{\prime} = \gamma_{ab} + 2\nabla_{(a}\xi_{b)}\,.
\end{equation}

In the quantum theory, the field $\gamma_{ab}$ is promoted to an operator satisfying the covariant commutation relations 
\begin{equation}
\label{eq:comm}
[\hat{\gamma}_{ab}(x),\hat{\gamma}_{cd}(x^{\prime})] = \ii E_{abcd}(x,x^{\prime})\hat{1}
\end{equation}
where $E_{abcd}(x,x^{\prime})$ is the advanced-minus-retarded Green's function associated with the linearized Einstein equation on any globally hyperbolic spacetime.\footnote{Since $\gamma_{ab}$ is not gauge invariant, $E_{abcd}(x,x^{\prime})$ is also not gauge invariant and therefore not uniquely defined. However, as we explain in this section, the smeared commutator function $E(f,f^{\prime})$ is unique if $f^{ab}$ and $f^{\prime ab}$ are both symmetric and divergence free \cite{Fewster:2012bj}.} The operators $\hat{\gamma}_{ab}$ act on a Hilbert space $\mathcal{H}$ that we will specify shortly. However, any notion of the operator $\hat{\gamma}_{ab}(x)$ will not be a good operator for essentially two reasons: (1) $\hat{\gamma}_{ab}(x)$ is not gauge invariant and (2) given any physical, normalizable state $\ket{\Psi}$, the action of such an operator on that state given by $\hat{\gamma}_{ab}(x)\ket{\Psi}$ will be far too singular and will, in fact, have infinite norm. The second issue essentially arises due to the universal, ultraviolet behavior of any physical quantum state. Both of these issues can be remedied by considering, instead, the smeared field operator \cite{Fewster:2012bj,Hollands:2014eia}
\begin{equation}
\hat{\gamma}(f) = \int_{M}\sqrt{-g}\,d^{d}x~\hat{\gamma}_{ab}(x)f^{ab}(x)
\end{equation}
where $f^{ab}$ is a smooth test tensor of compact support and is symmetric in its indices --- i.e., $f^{[ab]}=0$. The smearing essentially regulates the ultraviolet behavior of the operator and thereby addresses issue (2). Issue (1) is addressed by appropriately choosing the smearing tensor $f^{ab}$ so that the resulting observable is invariant under the gauge transformation \eqref{eq:gaugetransformation}. This can be achieved by choosing $f^{ab}$ to be divergence-free --- i.e., $f^{ab}$ also satisfies $\nabla_{a}f^{ab}=0$. For this class of test tensors $\gamma(f)$ is manifestly gauge invariant and satisfies the smeared versions of the distributional commutation relations \eqref{eq:comm}.

In order to do physics we not only need operators but also a Hilbert space of states on which these operators act. Given any state $\ket{\Omega}$, one can construct a Hilbert space representation $\mathcal{H}$ by a simple procedure known as the GNS construction \cite{Gelfand43,Segal47}.\footnote{Given an algebra $\mathcal{A}$ and a state $\Omega$ defined as a positive, linear map $\Omega:\mathcal{A} \to \mathbb{C}$,  one obtains a Hilbert space endowing the algebra with an inner product $\braket{a|b}\equiv \Omega(a^{\dagger}b)$ for any $a,b\in \mathcal{A}$, factoring the corresponding vector space with respect to null states and taking the Hilbert space completion.} Therefore, the relevant question is whether one can construct a ``preferred'' state $\ket{\Omega}$ on $(M,g)$. The construction of this state depends upon whether the spacetime contains a black hole or not so we consider these cases separately. 

\subsubsection*{Quantization in Spacetimes with No Black Holes}
\label{subsec:quantization}

We first consider the case where the background $(M,g)$ is a stationary spacetime. 
If the spacetime is globally stationary then it admits a globally timelike Killing field, i.e., the spacetime does not contain a black hole. In this case there exists a preferred, stationary state $\ket{\Omega}$ for the algebra of gravitational perturbations and thereby a bulk Hilbert space $\mathcal{H}$ \cite{Wald_1995}. For example, in Minkowski spacetime, this vacuum state is simply the Poincaré invariant vacuum state of the graviton. Since the spacetime is asymptotic to a flat spacetime, one can obtain a simple description of the ``out'' vacuum $\ket{\Omega_{\textrm{out}}}$ and an associated ``out'' Hilbert space $\mathcal{H}_{\textrm{out}}$ at future infinity. We now briefly describe this construction. 

In the case of asymptotically flat spacetimes without black holes, $i^{+}\cup \mathscr{I}^{+}$ is a complete Cauchy surface of $(M,g)$. By the equations of motion, the bulk graviton observables can be equivalently described by observables on any Cauchy surface $\Sigma$. Let $\hat{\gamma}(f)$ be a gauge-invariant linearized gravitational observable,
smeared with a compactly supported test tensor $f_{ab}$, and let $(Ef)_{ab}$ be
the advanced-minus-retarded solution with source $f_{ab}$. The equations of
motion imply the identity \cite{Wald_1995, Fewster:2012bj}
\begin{equation}
\label{eq:gammafsympEf}
\hat{\gamma}(f) = W_{\Sigma}(\hat{\gamma},Ef)
\end{equation}
where $\Sigma$ is a complete Cauchy surface. Eq.~\eqref{eq:gammafsympEf} relates the bulk observable to its corresponding initial data on any Cauchy surface $\Sigma$. To see how this gives an explicit map from bulk observables to operators at $\mathscr{I}^{+}$ we evaluate the right-hand side of \eqref{eq:gammafsympEf} at $\mathscr{I}^{+}$. Defining 
\begin{equation}
F_{AB}(u,x^{E}) \equiv -\frac{1}{2}\lim_{r\to \infty}r^{d/2-1} \bigg(q_{A}{}^{C}q_{B}{}^{D}-\frac{1}{d-2}q_{AB}q^{CD}\bigg)r^{-2}(Ef)_{CD}(u,r,x^{E})
\end{equation}
as the asymptotic radiation field of the linearized gravitational field $(Ef)_{ab}$ we obtain 
\begin{equation}
W_{\mathscr{I}^{+}}(\hat{\gamma},Ef) = -\frac{1}{16\pi G_{\textrm{N}}}\int_{\mathscr{I}^{+}}dud\Omega~\bigg(2\delta \hat{\sigma}_{AB}  \partial_{u}F^{AB} - \delta \hat{N}_{AB} F^{AB}\bigg)
\end{equation}
where $\delta \hat{\sigma}_{AB}$ is the asymptotic shear of $\hat{\gamma}_{ab}$ and $\delta \hat{N}_{AB} = 2\partial_{u}\delta \hat{\sigma}_{AB}$. Using the fact that $F_{AB}$ decays at asymptotically early and late times, we may integrate by parts and we obtain 
 \begin{equation}
W_{\mathscr{I}^{+}}(\hat{\gamma},Ef) = \frac{1}{8\pi G_{\textrm{N}}}\int_{\mathscr{I}^{+}} dud\Omega~\delta \hat{N}_{AB}(u,x^{A})F^{AB}(u,x^{A})\,. 
 \end{equation}
 If we define the smeared news observable at $\mathscr{I}^{+}$ as 
 \begin{equation}
\delta \hat{N}(F) \defn \int_{\mathscr{I}^{+}}dud\Omega ~\delta \hat{N}_{AB}(u,x^{A})F^{AB}(u,x^{A})\,,
 \end{equation}
 we obtain the following map between local gauge invariant observables in the bulk and radiative observables at null infinity
 \begin{equation}
 \label{eq:gammaN}
\hat{\gamma}(f) = \frac{1}{8\pi G_{\textrm{N}}}\delta \hat{N}(F)\,.
\end{equation}
Under this map, the bulk commutation relations \eqref{eq:comm} imply the following commutation relations of $\delta N_{AB}$ at $\mathscr{I}^{+}$ \cite{Ashtekar:1981hw} 
\begin{equation}
\label{eq:scricomm}
[\delta \hat{N}_{AB}(u_{1},x_{1}^{A}),\delta \hat{N}_{CD}(u_{2},x_{2}^{A})] = \,16\pi\ii G_{\mathrm N}\, Q_{ABCD}
\delta^{\prime}(u_{1}-u_{2})\delta_{\mathbb{S}^{d-2}}(x_{1}^{A},x_{2}^{A})\hat{1}
\end{equation}
where 
\begin{equation}
Q_{ABCD}=q_{A(C}q_{D)B}-\frac{1}{d-2}q_{AB}q_{CD}
\end{equation}
and $\delta^\prime$ denotes differentiation of the $\delta$-function with respect to $u_{1}$. 
At infinity, one can define a unique asymptotic state $\ket{\Omega_{\textrm{out}}}$ which is invariant under the asymptotic symmetries of $\mathscr{I}^{+}$. The state $\ket{\Omega_{\textrm{out}}}$ is essentially equivalent to the restriction of the Minkowski vacuum to null infinity. As such, this state is a Gaussian state with vanishing one-point function and two-point function given by 
\begin{equation}
\label{eq:News2pt}
\braket{\Omega_{\textrm{out}}|
\delta\hat{N}_{AB}(u_{1},x_{1}^{A})
\delta\hat{N}_{CD}(u_{2},x_{2}^{A})
|\Omega_{\textrm{out}}}
=
-\,8G_{\mathrm N}Q_{ABCD}\,
\frac{
\,
\delta_{\mathbb{S}^{d-2}}(x_{1}^{A},x_{2}^{A})
}{
(u_{1}-u_{2}-\ii 0^{+})^{2}
}\,.
\end{equation}
The GNS construction of $\ket{\Omega_{\textrm{out}}}$ with respect to the asymptotic radiative degrees of freedom yields the asymptotic Hilbert space $\mathcal{H}_{\textrm{out}}$ of graviton operators at $\mathscr{I}^{+}$.

On $\mathcal{H}_{\textrm{out}}$, the smeared field operator $\delta \hat{N}(F)$ is represented as a densely defined, unbounded operator. One can obtain a bounded operator by simply considering bounded functions of $\delta \hat{N}(F)$.\footnote{A simple example of such an operator is the ``Weyl operator'' $\hat{W}(F) = e^{\ii\delta \hat{N}(F)}\,.$} We define the full asymptotic algebra of observables $\mathcal{A}(\mathscr{I}^{+})$ as the von Neumann algebra of all bounded operators on the Hilbert space --- i.e. $\mathcal{A}(\mathscr{I}^{+}) = \mathcal{B}(\mathcal{H}_{\textrm{out}})$. One can similarly obtain subalgebras $\mathcal{A}(\mathscr{I}^{+}_{<u})\subset \mathcal{A}(\mathscr{I}^{+})$ by simply considering graviton observables to the past of a $u=\textrm{constant}$ cut of $\mathscr{I}^{+}$. These algebras trivially nest in the sense that
\begin{equation}
\mathcal{A}(\mathscr{I}^{+}_{<u_{1}}) \subset \mathcal{A}(\mathscr{I}^{+}_{<u_{2}}) \textrm{ for $u_{1}< u_{2}$}\,.
\end{equation}
Furthermore the commutant algebra
\begin{equation}
\mathcal{A}(\mathscr{I}^{+}_{<u})^{\prime}=\mathcal{A}(\mathscr{I}^{+}_{>u})
\end{equation}
is the sub-algebra of observables to the {\em future} of the $u=\textrm{constant}$ cut.

\subsubsection*{Type of the algebras}
\label{sec:type}
The algebra $\mathcal{A}(\mathscr{I}^{+})$ admits a trace which is simply the ordinary Hilbert space trace on $\mathcal{H}_{\textrm{out}}$. Therefore, given any normal state $\Psi$, one defines a density matrix $\rho_{\Psi}\in \mathcal{A}(\mathscr{I}^{+})$  such that any expectation value can be represented as 
\begin{equation}
\label{eq:trace}
\braket{a}_{\Psi} = \textrm{Tr}(a\rho_{\Psi}) \textrm{ for any $a\in \mathcal{A}(\mathscr{I}^{+})$}\,.
\end{equation}
 In the classification of von Neumann algebras introduced by Murray and von Neumann \cite{Murray:1936gtq} as well as Connes \cite{Connes:1973hg}, the algebra $\mathcal{A}(\mathscr{I}^{+})$ with this property is a Type I$_{\infty}$ algebra. 
However, this trace cannot be defined on any of the subalgebras $\mathcal{A}(\mathscr{I}^{+}_{<u})$ due to the ultraviolet entanglement of quantum fields across the cut. Moreover, these algebras do not admit {\em any} trace and so, while  all states on $\mathcal{A}(\mathscr{I}^{+}_{<u})$ are mixed, they cannot be described as density matrices  in the algebra. These algebras are Type III$_{1}$ von Neumann algebras.\footnote{We refer the reader to, e.g., \cite{Witten:2018zxz, Leutheusser:2021frk, Leutheusser:2021qhd, Sorce:2023fdx, Liu:2025krl} for further details on the properties of von Neumann algebras.} In the original classification, one also finds  ``Type II'' algebras which share some properties of both Type III and Type I algebras. In particular, similar to Type III algebras, all states are mixed in Type II algebras. However, one can define a trace (up to a state independent, multiplicative constant) and, therefore, states can be represented as density matrices on these algebras. Type II algebras have played a prominent role in recent studies of gravitational algebras (see, e.g., \cite{Chandrasekaran:2022cip, Kudler-Flam:2023qfl, Akers:2024bel, Kudler-Flam:2024psh, Chen:2024rpx, Penington:2025hrc, Herderschee:2025nsb, Klinger:2026tws, Cui:2026bcd}) and we will revisit them in sec.~\ref{subsec:schw_alg}. 

The above construction yields an algebra of ``out'' operators on a Hilbert space $\mathcal{H}_{\textrm{out}}$. We could analogously have constructed an ``in'' vacuum $\ket{\Omega_{\textrm{in}}}$ with associated ``in'' Hilbert space $\mathcal{H}_{\textrm{in}}$ which is acted upon by $\mathcal{A}(\mathscr{I}^{-})$. Since the linearized theory does not contain any ``infrared divergences'' the S-matrix $S:\mathcal{H}_{\textrm{in}}\to \mathcal{H}_{\textrm{out}}$  is well-defined and so we have that $\mathcal{H}_{\textrm{in}}\cong \mathcal{H}_{\textrm{out}}\cong \mathcal{H}$ and the bulk graviton operators are related to the boundary operators by \eqref{eq:gammaN} \cite{Ashtekar:1987tt,Prabhu:2022zcr}. 

The corresponding bulk von Neumann algebra is $\mathcal{A}(M)=\mathcal{B}(\mathcal{H})$ which is a Type I$_{\infty}$ algebra.  Subalgebras of $\mathcal{A}(M)$ obtained from restricting to subregions of the spacetime suffer from analogous ultraviolet entanglement and so are Type III$_{1}$. In particular, the ``spacelike wedge'' $\mathcal{W}_{u}\subset M$ will play a prominent role in this paper (see sec.~\ref{sec:algebra}) which consists of the set of points spacelike separated from a $u=\textrm{constant}$ cut of $\mathscr{I}^{+}$. Since the spacetime does not contain black holes, $\mathcal{W}_{u}$ is also the domain of dependence of $\mathscr{I}^{+}_{>u}\cup i^{+}$. By the above properties, the algebra $\mathcal{A}(\mathcal{W}_{u})$ is a Type III$_{1}$ algebra for any finite $u$ which satisfies 
\begin{equation}
\label{eq:spacelikewedgeTypeIII}
\mathcal{A}(\mathcal{W}_{u_{2}}) \subset \mathcal{A}(\mathcal{W}_{u_{1}}) \textrm{ for $u_{1}< u_{2}$} \quad \textrm{ and }\quad \mathcal{A}(\mathcal{W}_{u})^{\prime} = \mathcal{A}(\mathscr{I}^{+}_{<u})
\end{equation}
and so the spacelike wedge algebras nest and their commutant is the algebra of radiative observables at $\mathscr{I}^{+}$ to the past of the cut. 

\subsubsection*{Matter fields}

We now comment on the incorporation of matter fields. We will assume that all matter fields satisfy the ``time slice axiom'' in that the algebra of observables in the neighborhood of a Cauchy slice is equivalent to the algebra of its domain of dependence \cite{Brunetti:2001dx}. This property expresses determinism and ensures that the observables can be evolved to the asymptotic future and past. 

The preceding discussion carries over directly to any free field, massive or massless (e.g. scalar or the electromagnetic fields). For massive fields the asymptotic algebra is defined at future and past timelike infinity $i^{\pm}$ and the algebra of massless fields is constructed in a nearly identical manner. The absence of infrared divergences guarantees a well-defined $S$-matrix between their corresponding separable ``in'' and ``out'' Hilbert spaces and the bulk algebra $\mathcal{A}(M)$ is constructed in an identical manner. In particular, if one only considered massive fields then $i^{+}$ is a complete Cauchy surface and so, in particular, the spacelike wedge algebra $\mathcal{A}(\mathcal{W}_{u})\cong\mathcal{A}(i^{+}) \cong \mathcal{A}(M)$ since $i^{+}\subset \mathcal{W}_{u}$ for any $u$. So the algebra $\mathcal{A}(\mathcal{W}_{u})$ of just massive fields is Type I$_{\infty}$. Since, in addition to any matter fields, we will always include massless gravitons then the algebra is $\mathcal{A}(\mathcal{W}_{u}) \cong \mathcal{A}(i^{+})\;\overline{\otimes} \;\mathcal{A}(\mathscr{I}^{+}_{>u})$. The strong entanglement of these massless fields across a constant $u$ cut of $\mathscr{I}^{+}$ implies the algebra remains Type III$_{1}$ and satisfies \eqref{eq:spacelikewedgeTypeIII}.

The above arguments can be directly generalized to interacting massless fields which become sufficiently weakly interacting at large distance and late times so that the fields behave essentially as free fields at infinity. In $d>4$ dimensions, interactions generally decay sufficiently rapidly such that the S-matrix $S:\mathcal{H}_{\textrm{in}}\to \mathcal{H}_{\textrm{out}}$ is again well-defined on separable ``in'' and ``out'' Hilbert spaces \cite{Strominger:2013jfa, Prabhu:2022zcr}. This property continues to hold in four dimensions with interacting massive fields. However, for four dimensions with an interacting {\em massless} field one encounters severe infrared divergences due to the emission of an infinite number of soft, radiative quanta. The corresponding out state will not lie in the usual Fock space of radiative fields. Therefore, the standard $S$-matrix is not well-defined and the above construction is not applicable. 

Despite this, one can obtain separable ``in'' and ``out'' Hilbert spaces as well as a well-defined $S$-matrix for matter fields. The Hilbert space is known as the ``Faddeev-Kulish'' Hilbert space $\mathcal{H}^{\textrm{FK}}$ and we now briefly outline its construction (we refer the reader to, e.g., \cite{Kulish:1970ut, Kapec:2017tkm,Prabhu:2022zcr} for more details). The key point is that the emission of soft quanta is intimately tied to the enlargement of the asymptotic symmetry group to an infinite-dimensional group of large gauge transformations at infinity. These large gauge symmetries give rise to an infinite set of conserved ``large gauge charges'' at spatial infinity \cite{Strominger:2013jfa,Strom2:2014,CE,Henneaux:2018gfi,Kartik_Maxwell,Mohamed_2021}. Since we are interested in the algebra of {\em local} gauge invariant observables $\mathcal{O}$ (i.e., the electric field, the charge-current, Wilson loops) the key property is that 
\begin{equation}
[Q_{i^{0}}(\lambda),\mathcal{O}]=0
\end{equation}
for any $\mathcal{O}$ and all $\lambda(x^{A})$ where $\lambda(x^{A})$ parameterize the infinite set of large transformations. Therefore, these charges form superselection sectors and in the case of QED the  ``Faddeev-Kulish'' Hilbert space $\mathcal{H}^{\textrm{FK}}$ of definite charge $Q_{i^{0}}(\lambda)$ for all $\lambda$  was shown to be a separable Hilbert space. We assume that this property continues to hold in general gauge theories with long-range fields. Since these charges are conserved, the $S$-matrix is well-defined as a map between the space of ``in'' and ``out'' states of the same definite charge. In this paper, we will simply denote the global Hilbert space as $\mathcal{H}$. 

This completes the description of the Hilbert space of matter and linearized gravitons in any stationary spacetime. However, we note that the above construction did not use stationarity in the bulk in any essential way. The key property was simply the existence of ``in'' and ``out'' Hilbert spaces and well-defined $S$-matrix between them. As such there is no difficulty, in principle, of extending the above discussion to non-stationary asymptotically flat spacetimes though we will not need this extension for the main arguments of this paper.

\subsection{Quantization in black hole spacetimes}
\label{subsec:quantizationBH}

In the previous subsection, we have considered asymptotically flat spacetimes without black holes. The key difference is that black holes radiate: any black hole formed from gravitational collapse will emit thermal radiation and thereby evaporate \cite{Hawking:1975vcx}. However, in this paper, we will be interested in the semiclassical description of black holes as $G_{\textrm{N}}\to0$. To obtain a consistent description, we must consider this limit at fixed black hole radius $r_{\textrm{BH}}$.  In this limit, the evaporation timescale goes as $t_{\textrm{evap}}\sim r_{\textrm{BH}}^{d-1}/G_{\textrm{N}} \to \infty$ and so, although semiclassical black holes form and emit thermally, they do not evaporate. 

What quantum states accurately describe quantum fields in the vicinity of a semiclassical black hole?  After collapse, the spacetime rapidly settles down to a stationary black hole on a timescale of $t_{\textrm{settle}}\sim r_{\textrm{BH}}$. Furthermore, the quantum state rapidly approaches a stationary state. So the description of the late-time state after collapse is tantamount to studying the field on a stationary black hole background where we work in the approximation that the collapse occurred in the far distant past. 

\subsubsection*{Limitations of Hartle-Hawking and Unruh states}
\label{subsec:HHUH}
In this setting, an obvious answer to the above question is to simply choose the quantum state to be a stationary state in the full exterior of the stationary black hole. For a Schwarzschild black hole, the two quantum states that are both regular on the future horizon of the black hole and stationary in the exterior are the Hartle-Hawking state \cite{Hartle:1976tp}  and the Unruh state \cite{Unruh:1976db}.
The Hartle-Hawking state describes a black hole in perfect thermal equilibrium. In this state, the steady outgoing flux of Hawking radiation is balanced by a steady {\em incoming} flux of thermal radiation from $\mathscr{I}^{-}$. The Unruh state, on the other hand, has no incoming radiation and has only a steady outgoing flux of Hawking quanta to $\mathscr{I}^{+}$. Therefore, the Unruh state describes a stationary black hole radiating into empty space. 

The key issue that we will be concerned with in this paper is the quantum description of the Bondi mass which, by \eqref{eq:BondimassN2Tuu}, encodes the energy lost due to radiation until some retarded time $u$. For such a quantity to be well-defined, the stress-energy tensor of the quantum state must suitably decay near $\mathscr{I}^{+}$ and $i^{0}$.  As we will now explain, the stress-energy tensors in {\em both} states fail to appropriately decay at large distance and/or early and late times. 

In the Hartle-Hawking state, the asymptotic behavior at infinity is that of a thermal bath in Minkowski spacetime and so the stress-energy approaches a constant at null infinity
\begin{equation}
\lim_{r\to \infty}\braket{\hat{T}_{uu}(u,r,x^{A})}_{\Omega_{\textrm{HH}}} =C
\end{equation}
where the constant $C$ is theory dependent.  The total power at the retarded time $u$  
\begin{equation}
L_{\textrm{HH}}(u) = \lim_{r\to \infty}r^{d-2}\int_{\mathbb{S}^{d-2}}d\Omega~\braket{\hat{T}_{uu}(u,r,x^{A})}_{\Omega_{\textrm{HH}}} = \infty 
\end{equation}
diverges in the limit as $r\to \infty$. In the Unruh state, there is no incoming thermal radiation and the stress-energy actually decays as $1/r^{d-2}$
\begin{equation}
\label{eq:Unruhflux}
\lim_{r\to \infty}r^{d-2}\braket{\hat{T}_{uu}(u,r,x^{A})}_{\Omega_{\textrm{UH}}} =\frac{L_{\textrm{UH}}(u)}{\textrm{Vol}(\mathbb{S}^{d-2})}
\end{equation}
where $\textrm{Vol}(\mathbb{S}^{d-2})$ is the area of the $d-2$ sphere and $L_{\textrm{UH}}\geq 0$ is the total expected power radiated at retarded time $u$ which is {\em finite} in the Unruh state.\footnote{Since $\braket{T_{ab}\xi^{b}}$ is conserved where $\xi^{a}$ is the horizon Killing field, this outward flux is compensated by an equal and opposite negative energy flux into the horizon $\mathcal{H}^{+}$ of the black hole.} The precise value of $L_{\textrm{UH}}$ depends on the species of particles under consideration \cite{Page:1976df}. However, since the Unruh state is stationary, $L_{\textrm{UH}}$ is actually {\em constant} and so the total expected energy radiated to null infinity from $u=-\infty$ to any finite retarded time $u$ 
\begin{equation}
\lim_{r\to \infty}r^{d-2}\int_{-\infty}^{u}du^{\prime}\int_{\mathbb{S}^{d-2}}d\Omega~\braket{\hat{T}_{uu}(u^{\prime},r,x^{A})}_{\Omega_{\textrm{UH}}} = \int_{-\infty}^{u} du'~L_{\textrm{UH}}
\end{equation}
is divergent. The expected radiated power due to gravitons is also constant and the total radiated energy also diverges. While the initial ADM mass of the system is finite, the expected Bondi mass $\braket{\hat{M}_{\textrm{B}}(u)}$ will diverge for all $u$. We will return to this point at the end of sec.~\ref{subsec:RegularizedBondiMass}. 

\subsubsection*{A Hilbert space of states that decay at infinity and are regular on the horizon}
\label{subsec:Hilbreg}

The actual quantum state in the semiclassical collapse spacetime approaches the Minkowski vacuum at spatial infinity. The error we made above is in approximating {\em both} the spacetime and the quantum state to be stationary. A better approximation would be to take the spacetime to be stationary but require that (1) the quantum state appropriately decays at infinity and (2) is regular on the horizon. The precise notion of regularity we require is that the state must be a ``Hadamard'' state (see, e.g., \cite{Wald_1995} for more details). The decay at infinity ensures that the integral \eqref{eq:EnergyUH} converges for any finite $u$.  A large class of states for gravitons and matter fields satisfying (1) and (2) in any stationary, asymptotically flat black hole spacetime was constructed in \cite{Chen:2024rpx}. We now briefly review the construction of this Hilbert space, which we will refer to henceforth as the vacuum-at-infinity Hilbert space.

Consider a (two-sided) stationary, asymptotically flat black hole spacetime $(M,g)$ and let $\Sigma$ be any spacelike Cauchy surface. The goal will be to construct a Hilbert space $\mathcal{H}_{\Sigma}$ on $\Sigma$ satisfying (1) and (2). We first note that if the spacetime were {\em globally} stationary then, for linear fields,\footnote{The uniqueness of the vacuum state in stationary spacetimes was proven in \cite{Ashtekar:1975zn}, and \cite{Sahlmann:2000fh} proved that the vacuum is also Hadamard.} there exists a unique vacuum state $\ket{\Omega_{\Sigma}}$ which is Hadamard everywhere on $\Sigma$ and approaches the Minkowski vacuum at infinity \cite{Wald_1995}. Using the arguments of the previous subsection this result can be directly extended to interacting theories which approach free fields at infinity. A black hole spacetime, however, admits no global timelike Killing field. In
Schwarzschild, the horizon Killing field is timelike throughout the exterior but becomes null on the horizon. Its associated ground state is the ``Boulware state'' which is singular on both the future and past horizons, and hence fails to be Hadamard \cite{Boulware:1974dm}.  The basic strategy for building $\mathcal{H}_{\Sigma}$ is to use a version of a ``deformation argument'' of Fulling, Narcowich and Wald \cite{Fulling:1981cf}. 

\begin{figure}
    \centering
    \includegraphics[width=1\linewidth]{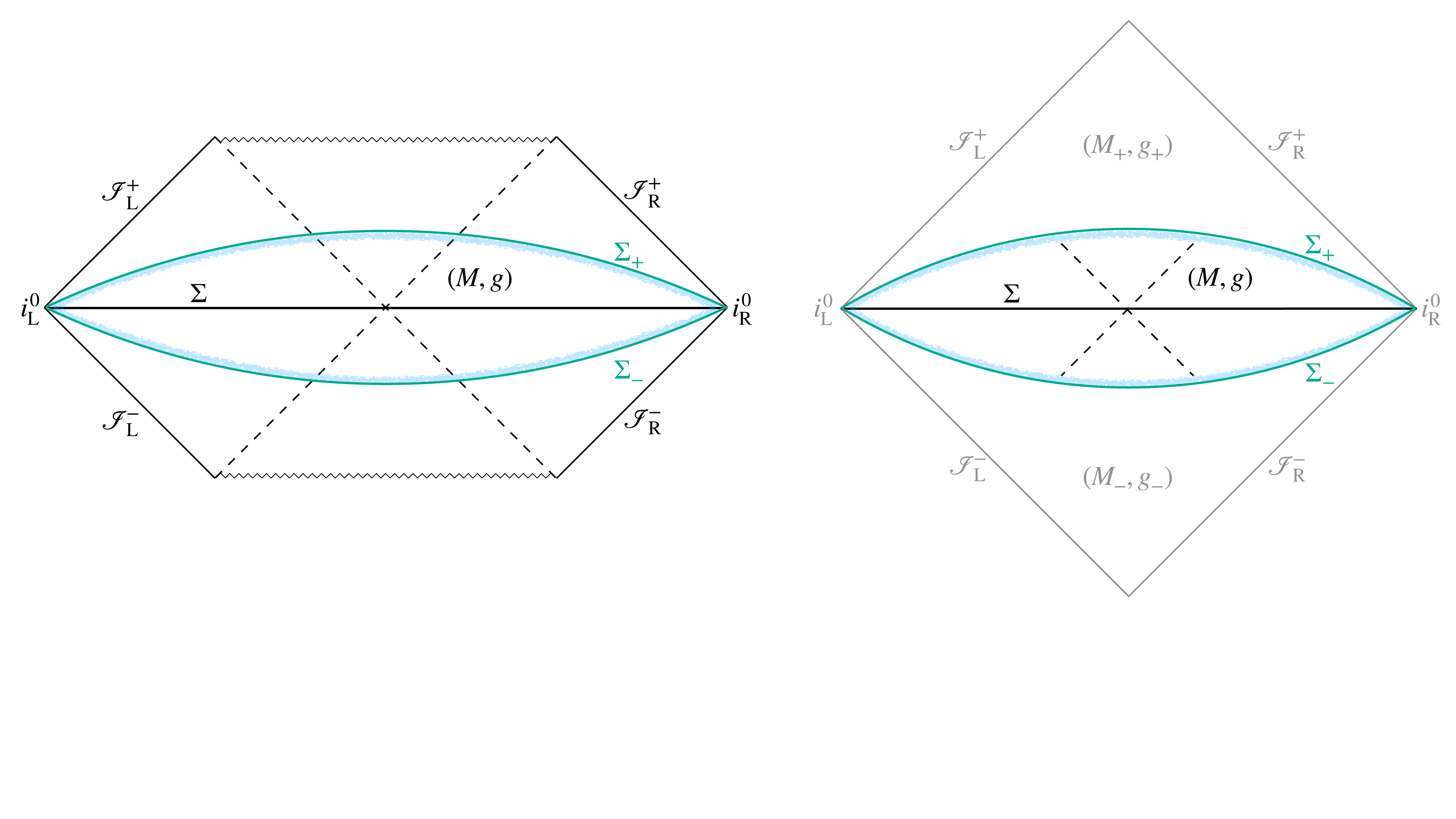}
    \vspace{-3cm}
    \caption{Deformation of a two-sided Schwarzschild spacetime $(M,g)$ away from a Cauchy slice $\Sigma$ \cite{Fulling:1981cf, Chen:2024rpx}. The spacetimes to the future of a Cauchy slice $\Sigma_+$ and to the past of a Cauchy slice $\Sigma_-$ are smoothly deformed to the stationary spacetimes $(M_+,g_+)$ and $(M_-,g_-)$, respectively, which are isometric to $(M,g)$ outside a neighborhood of the black hole. The light blue shaded region corresponds to a metric that interpolates between the black hole spacetime and the stationary spacetimes.}
    \label{fig:bh}
\end{figure}

Let $\Sigma_{+}$ be a spacelike Cauchy surface which is strictly to the future of $\Sigma$ and let $\Sigma_{-}$ be a spacelike Cauchy surface that is strictly to the past of $\Sigma$ (see Fig.~\ref{fig:bh}). The spacetime to the future of $\Sigma_{+}$ can be smoothly deformed to a stationary spacetime $(M_{+},g_{+})$ which is isometric to $(M,g)$ outside a neighborhood of the black hole but has a globally timelike Killing vector (and hence no horizon). We apply a similar construction to obtain a stationary spacetime $(M_{-},g_{-})$ to the past of $\Sigma_{-}$.  
This is done in such a way that the deformed spacetime remains globally hyperbolic. The new spacetime is now an asymptotically flat spacetime with no black holes and so admits canonical ``in'' and ``out'' Hilbert spaces $\mathcal{H}_{\Sigma_{+}}$ and $\mathcal{H}_{\Sigma_{-}}$ built from the associated ``in'' and ``out'' stationary vacua. By the arguments of the previous section, there exists a well-defined $S$-matrix and so we obtain $\mathcal{H}_{\Sigma}$ by evolving backwards and forwards from $\mathcal{H}_{\Sigma_{+}}$ and $\mathcal{H}_{\Sigma_{-}}$ respectively. The resulting Hilbert space $\mathcal{H}_{\Sigma}$ contains a dense set of Hadamard states which decay to the Minkowski vacuum at infinity.\footnote{\label{foot:defindep}By the following argument \cite{Witten:2021jzq}, we expect that the Hilbert space $\mathcal{H}$ should not depend on the choices of deformation. Since we require that the deformed spacetime be isometric to Schwarzschild outside a neighborhood of the black hole, the deformation on $\Sigma_{+}$ is compact. As a result, time evolution on the deformed spacetime gives a canonical unitary identification between $\mathcal{H}_{\Sigma_{-}}$ and $\mathcal{H}_{\Sigma_{+}}$. However, $\mathcal{H}_{\Sigma_{-}}$ is manifestly independent of the deformation to the future of $\Sigma$, while $\mathcal{H}_{\Sigma_{+}}$ is manifestly independent of the past deformation. It follows that both can be identified with a canonical $\H_{\Sigma}$ that depends only on the original undeformed Schwarzschild spacetime. A similar argument applies to stationary but not static spacetime where we require that the deformed spacetime be isometric to Kerr outside a neighborhood of the ergoregion.}

We obtain the global vacuum-at-infinity Hilbert space $\mathcal{H}$ by simply evolving $\mathcal{H}_{\Sigma}$ off of $\Sigma$ in the black hole spacetime $(M,g)$.  By a result of Fredenhagen and Haag, any state $\ket{\Phi}\in \mathcal{H}$ satisfying (1) and (2) above should approach the Unruh state at late times and so accurately describes the late-time behavior of an isolated black hole formed from collapse \cite{Fredenhagen:1989kr}. The radiated power $L_{\Phi}(u)$ vanishes as $u\to -\infty$ and approaches $L_{\textrm{UH}}$ as $u\to \infty$ for any such $\Phi\in \mathcal{H}$. Furthermore, for any finite $u$, the total radiated energy is finite 
\begin{equation}
\label{eq:EnergyUH}
\lim_{r\to \infty}r^{d-2}\int_{-\infty}^{u}du^{\prime}\int_{\mathbb{S}^{d-2}}d\Omega~\braket{\hat{T}_{uu}(u^{\prime},r,x^{A})}_{\Phi} = \int_{-\infty}^{u} du'~L_{\Phi}(u')<\infty
\end{equation}
for a dense set of states $\Phi\in\mathcal{H}$. If we denote $\mathcal{R}$ as the right black hole exterior, we note that all states in $\mathcal{H}$ are mixed when restricted to $\mathcal{A}(\mathcal{R})$ due to entanglement with the black hole interior. As such the algebra $\mathcal{A}(\mathcal{R})$ is a Type III$_{1}$ algebra \cite{Fredenhagen:1984dc}. Similarly, any local subalgebra of $\mathcal{A}(\mathcal{R})$ --- such as the spacelike wedge algebra $\mathcal{A}(\mathcal{W}_{u})$ --- is also Type III$_{1}$.

\section{The Bondi Mass} \label{sec:bondi}
In this section we consider the notion of the Bondi mass as an operator in quantum gravity. In sec.~\ref{subsec:BMisnotanop}, we show that, in contrast to the ADM mass at $i^{0}$, the Bondi mass is {\em not} a well-defined operator at null infinity \cite{Bousso:2017xyo} in both the perturbative and nonperturbative theory. Nevertheless, in sec.~\ref{subsec:RegularizedBondiMass} we construct a ``regularized'' Bondi mass operator which plays a similar role to the Bondi mass but is a well-defined operator on the Hilbert space.

\subsection{The ADM mass at $i^{0}$ vs. the Bondi mass at $\mathscr{I}^{+}$}
\label{subsec:BMisnotanop}

The fact that the Bondi mass at $\mathscr{I}^{+}$ is not a good operator was previously argued in \cite{Bousso:2017xyo} where it was shown that vacuum fluctuations of $\hat{M}_{\textrm{B}}(u)$ diverge in the limit to $\mathscr{I}^{+}$. In this section we explain how these conclusions naturally arise within the context of von Neumann algebras at null infinity. Furthermore, from these arguments we will also see that one cannot obtain a good operator by simply smearing $\hat{M}_{\textrm{B}}$ over $\mathscr{I}^{+}$. In other words, the Bondi mass cannot be defined as an operator at $\mathscr{I}^{+}$. To prove this claim, we first consider the simplest scenario of linearized gravitons and matter fields on a fixed, background, stationary spacetime $(M,g)$. At the end of this subsection, we explain that these conclusions generalize to non-stationary spacetimes in the full non-perturbative theory.

\subsubsection*{The ADM mass }

We first review the status of the ADM mass which {\em is} a well-defined quantum operator that generates asymptotic time translations. Consider a stationary spacetime without black holes (e.g., flat spacetime or the spacetime of a stationary star). Let $\bar{M}_{i^{0}}$ be the background ADM mass which, in the absence of any black holes, is $O(1)$ in powers of $G_{\textrm{N}}$. The background ADM mass is simply a multiple of the identity and so does not generate time translations on the Hilbert space $\mathcal{H}$ of gravitons and matter. To obtain a non-trivial operator we will have to consider the ADM mass at higher perturbative orders. At next order $\delta M_{i^{0}}$ corresponds to $O(\sqrt{G_{\textrm{N}}})$ perturbations of the spacetime. However, evaluating  \eqref{eq:HX} with $X^{a}=t^{a}$ the timelike Killing vector field we obtain 
\begin{equation}
\delta M_{i^{0}} = W_{\Sigma}(\gamma,\mathcal{L}_{t}g) + W^{\textrm{matt.}}_{\Sigma}(\delta \phi ,\mathcal{L}_{t}\phi)=0
\end{equation}
which vanishes since the background fields satisfy $\mathcal{L}_{t}g_{ab}=0$ and $\phi$ vanishes on the background. To obtain a non-trivial operator one must go to second order in perturbation theory.  Expanding about $\lambda=0$, $M_{i^{0}}(\lambda)=M_{i^{0}}(0)+\lambda \delta M_{i^{0}} + \lambda^{2}\delta^{2} M_{i^{0}}+\dots$, the second-order coefficient satisfies \cite{Hollands:2012sf}
\begin{equation}
\label{eq:Mi0}
\delta^{2}M_{i^{0}} =\frac{1}{2} W^{\textrm{GR}}_{\Sigma} (\gamma,\mathcal{L}_{t}\gamma)  + \frac{1}{2}W^{\textrm{matt.}}_{\Sigma}(\delta \phi ,\mathcal{L}_{t}\delta\phi)\,.
\end{equation}
where\footnote{We note that our convention for the second-order variation of the mass differs from that of Hollands and Wald in \cite{Hollands:2012sf} who define $\delta^{2}M_{i^{0}} \equiv \tfrac{d^{2}}{d\lambda^{2}}M_{i^{0}}(\lambda)\big\vert_{\lambda=0}$. The reason for the factor of $1/2$ in our definition is to ensure that $\delta^{2}M_{i^{0}}$ is the Taylor coefficient of the expansion of the full ADM mass $M_{i^{0}}(\lambda)$ about $\lambda=0$.} $\delta^{2}M_{i^{0}} \equiv \frac{1}{2}\tfrac{d^{2}}{d\lambda^{2}}M_{i^{0}}(\lambda)\big\vert_{\lambda=0}$. 
Both sides of this expression are well-defined in the semiclassical limit.

The right-hand side of \eqref{eq:Mi0} is the classical Hamiltonian 
\begin{equation}
H= \mathcal{E}_{\Sigma}(\gamma)  + \int_{\Sigma}\sqrt{h}d^{d-1}x~n^{a}t^{b}T_{ab}[\delta \phi]\,,
\end{equation}
where the first term is the so-called ``canonical energy'' of gravitational perturbations
\begin{equation}
\mathcal{E}_{\Sigma}(\gamma) = \frac{1}{2}W^{\textrm{GR}}_{\Sigma} (\gamma,\mathcal{L}_{t}\gamma)\,.
\end{equation}
The second term is equivalent to the Hamiltonian of the matter fields\footnote{The integral is equivalent to $\tfrac{1}{2}W^{\textrm{matt.}}_{\Sigma}(\delta \phi ,\mathcal{L}_{t}\delta\phi)$ up to a boundary term which vanishes at infinity. We will later be considering the case where $\Sigma$ is a partial Cauchy surface; however, these boundary terms will not be important for our purposes.} which can be expressed in terms of an integral of the stress-energy of the matter perturbation  over a Cauchy slice.

In the quantum theory, for any spacetime without a black hole, $\delta^{2}\hat{M}_{i^{0}}$ is the order one perturbation of the ADM mass,
\begin{equation}
\delta^{2}\hat{M}_{i^{0}} = \lim_{G_{\textrm{N}}\to 0} [\hat{M}_{i^{0}} - \bar{M}_{i^{0}}] \qquad\quad \textrm{ (no black holes)}\,.
\end{equation}
The Hamiltonian is the sum of two well-defined operators on $\mathcal{H}$ that generate time translations of the gravitons and quantum fields, respectively,
\begin{equation}
\hat{H} = \mathcal{E}_{\Sigma}(\hat{\gamma}) + \hat{H}_{\textrm{QFT}}
\end{equation}
 for 
 \begin{equation}
 \label{eq:HQFT}
\hat{H}_{\textrm{QFT}} = \int_{\Sigma}\sqrt{h}d^{d-1}x~n^{a}t^{b}\hat{T}_{ab}[\delta \hat{\phi}]\,.
 \end{equation}
Via the constraints, the sum of these operators is equal to the (second-order) ADM mass at spatial infinity 
\begin{equation}
\label{eq:ADMHam}
\delta^{2}\hat{M}_{i^{0}} = \mathcal{E}_{\Sigma}(\hat{\gamma}) + \hat{H}_{\textrm{QFT}} \,.
\end{equation}
Then $\delta^{2}\hat{M}_{i^{0}}$ is equivalent to the Hamiltonian on $\mathcal{H}$. 

The key difference when the background spacetime contains a black hole is that the background ADM mass $\bar{M}_{i^{0}}\sim O(1/G_{\textrm{N}})$ is divergent in the semiclassical limit. We can obtain a finite quantity by considering higher orders in $G_{\textrm{N}}$. As an example, we consider the case where the background is a Schwarzschild black hole. At next order, $\delta M_{i^{0}}$ is generally non-vanishing. Indeed, letting $X^{a}=t^{a}$ in \eqref{eq:HX} and choosing $\Sigma$ to be a Cauchy surface from $i^{0}$ to the bifurcation surface yields the first law \cite{Bardeen:1973gs,Iyer:1994ys}
\begin{equation}
\delta M_{i^{0}} = \frac{\kappa}{8\pi G_{\textrm{N}}}\delta A
\end{equation}
where $\kappa$ is the surface gravity of the black hole. The perturbed ADM mass $\delta M_{i^{0}}$ corresponds to linearized, stationary perturbations which change the mass of the black hole. In the quantum theory, a non-vanishing $\delta M_{i^{0}}$ corresponds to stationary fluctuations of the ADM mass  at $O(1/\sqrt{G_{\textrm{N}}})$ in addition to the Hilbert space $\mathcal{H}$ of dynamical (time-dependent) fluctuations of the spacetime. Since $\delta M_{i^{0}}$ is independent of the graviton perturbations we can freely choose the quantum state of this mode. To obtain a well-defined semiclassical limit, we choose a ``microcanonical ensemble'' of quantum states such that $\delta \hat{M}_{i^{0}}=0$ --- i.e., the fluctuations of the full ADM mass $M_{i^{0}}$ are $O(1)$ \cite{Chandrasekaran:2022eqq}.  The $O(1)$ fluctuations of the ADM mass are thereby encoded in 
\begin{equation}
\delta^{2}\hat{M}_{i^{0}} = \lim_{G_{\textrm{N}}\to 0}[\hat{M}_{i^{0}} - \bar{M}_{i^{0}}] \qquad\quad \textrm{ (black holes)}
\end{equation}
which has a well-defined semiclassical limit. 

Varying \eqref{eq:WSigmaQ_matter} again yields a relationship between the second-order ADM mass and the Hamiltonian of quantum fields and gravitons. For a two-sided, Schwarzschild black hole we obtain
\begin{equation}
\label{eq:ADMHamLR}
\delta^{2}\hat{M}_{i^{0}}^{\textrm{R}} - \delta^{2}\hat{M}_{i^{0}}^{\textrm{L}} = \mathcal{E}_{\Sigma}(\hat{\gamma}) + \hat{H}_{\textrm{QFT}} 
\end{equation}
where $\Sigma$ is a complete Cauchy surface for the (two-sided) black spacetime and so receives contributions from the second-order ADM mass of the left and right wedges which we denote with a superscript L and R respectively. The extension of the algebra to include the ADM mass and its impact on the algebraic structure was considered in \cite{Witten:2021unn, Chandrasekaran:2022eqq, Kudler-Flam:2023qfl}. We will revisit this point in sec.~\ref{subsec:schw_alg}. The above discussion can be straightforwardly generalized to any Kerr-Newman black hole where $t^{a}$ is replaced with the horizon Killing field and the left-hand side of \eqref{eq:ADMHamLR} receives additional contributions due to the second-order total charge $\delta^{2}Q^{\textrm{L/R}}_{i^{0}}$ and total angular momentum $\delta^{2}J^{\textrm{L/R}}_{i^{0}}$.

\subsubsection*{The Bondi mass }
\label{subsec:Bondimassatscri}

We now explain that, in contrast to the ADM mass which is a good operator even in the semiclassical limit, the semiclassical Bondi mass is {\em not} a well-defined operator at $\mathscr{I}^{+}$. Taking two variations of \eqref{eq:BondimassN2Tuu}, the proposed semiclassical Bondi mass operator satisfies
\begin{equation}
\label{eq:MB2}
\delta^{2}\hat{M}_{\textrm{B}}(u) = \delta^{2}\hat{M}_{i^{0}} - \frac{1}{32\pi G_{\textrm{N}}} \int_{\mathscr{I}^{+}_{<u}}dud\Omega ~ :\delta \hat{N}^{AB}\delta \hat{N}_{AB}: - 
\int_{\mathscr{I}^{+}_{<u}}dud\Omega~\hat{T}^{(2)}_{uu}\,.
\end{equation}
Here the first term generates time translations of gravitons and matter fields and the second term is equivalent to the Hamiltonian of the graviton fields restricted to $\mathscr{I}^{+}$ except that the integrals have been cut off at some retarded time $u$.  The third term is the Hamiltonian of the matter fields but integrated over $\mathscr{I}^{+}_{<u}$. 

Therefore, for an operator $\mathcal{O}$ on $\mathscr{I}^{+}$ to the past of $u$, the Bondi mass commutes with $\mathcal{O}(u^{\prime})$ for $u^{\prime}<u$ --- i.e., the action of the ADM translates the operator forward in retarded time and the action of the second and third terms translates the operator back. However, due to the ultralocality of quantum fields at $\mathscr{I}^{+}$ (see, e.g., \eqref{eq:scricomm} and \cite{Sewell:1982zz,Wall:2011hj,Prabhu:2022zcr}), any operator $\mathcal{O}(u^{\prime})$ with $u^{\prime}>u$ commutes with all operators to the past of $u$ and so it commutes with the second and third terms in \eqref{eq:MB2}.  Therefore, the Bondi mass generates time translations on $\mathscr{I}^{+}$ on any $\mathcal{O}(u^{\prime})$ for $u^{\prime}>u$. In summary 
\begin{equation}
\label{eq:MBOcomm}
[\delta^{2}\hat{M}_{\textrm{B}}(u), \mathcal{O}(u^{\prime}) ]=
\begin{cases}
-\ii\mathcal{L}_{t}\mathcal{O}(u^{\prime}) \quad\quad &\textrm{for $u^{\prime}>u$}\\
0& \textrm{for $u^{\prime}<u$}
\end{cases}
\end{equation}
for any $\mathcal{O}\in \mathcal{A}(\mathscr{I}^{+})$ with the requisite support. The corresponding action of the Bondi mass on $\mathcal{H}$ is highly singular and so cannot be defined as a quantum operator. In terms of the quantization of fields at $\mathscr{I}^{+}$ described in the previous section, this singularity arises due to the $\delta-$function singularity on $\mathbb{S}^{d-2}$ in the vacuum $2$-point function \eqref{eq:News2pt}. Indeed, expanding in spherical harmonics and applying \eqref{eq:News2pt} and \eqref{eq:comm} one formally obtains
\begin{equation}
\label{eq:fluxscrilessu}
\braket{\Psi|\mathcal{E}^{2}_{\mathscr{I}^{+}_{<u}}|\Psi} = \frac{d(d-3)}{16\pi^{2}}\lim_{L  \to \infty}N_{L}\bigg(\lim_{\epsilon \to 0^{+}}\int_{-\infty}^{u}du_{1}\int_{-\infty}^{u}du_{2} \frac{1}{(u_{1}-u_{2}-i\epsilon)^{4}}\bigg) + \dots 
\end{equation}
for any $\Psi\in \mathcal{H}$ where the ``$\dots$'' denotes terms which are finite, $L$ is an angular cutoff of spherical harmonics and $N_{L}=\sum_{\ell = 0}^{L}g_{\ell}$ is the sum of the degeneracy $g_{\ell}$ of degree $\ell$ spherical harmonics on $\mathbb{S}^{d-2}$. Both factors are divergent. The second term diverges due to the sharp cutoff in the retarded time $u$ of the integrals. The first term diverges since $N_{L}\sim L^{d-2}$ for large $L$. The fluctuations of the stress-energy flux \eqref{eq:MB2} similarly diverge for any finite $u$. Consequently, while $\delta^{2}M_{B}$ can be defined as a sesquilinear form (i.e., the matrix elements of  $\delta^{2}M_{B}$ are finite), its fluctuations diverge
\begin{equation}
\braket{\Psi|\delta^{2}M_{\textrm{B}}(u)^{2}|\Psi} = \infty
\end{equation}
for all $\Psi\in \mathcal{H}$ and so it is not a well-defined operator on $\mathcal{H}$. Furthermore, smearing $\delta^{2}M_{\textrm{B}}(f)$ in retarded time with some test function $f(u)$ does not yield a well-defined operator since the angular singularity still diverges.\footnote{A similar analysis shows that the full Bondi mass aspect $m_{B}(u,x^{A})$ is similarly ill-defined even if one considered the smeared object $m_{B}(f)$ for any $f(u,x^{A})$ on $\mathscr{I}^{+}$ \cite{Prabhu:2022zcr}. The Bondi angular momentum also suffers from similar divergences.} The divergence is entirely due to the fact that we have restricted ourselves to null infinity and are considering the Bondi mass at a finite duration of retarded time. Indeed, for spacetimes without black holes (and hence without late-time Hawking radiation), we may instead consider the limit $u\to \infty$ in \eqref{eq:fluxscrilessu} first at any fixed angular cut-off $L$, then the retarded time integrals in the first term of \eqref{eq:fluxscrilessu} vanish and then one may safely take $L\to \infty$. Correspondingly, the total energy $\mathcal{E}_{\mathscr{I}^{+}}$ flux through all of $\mathscr{I}^{+}$ is a good operator that generates time translations of fields on $\mathscr{I}^{+}$.\footnote{We expect that this convergence is well-defined as a limit of sesquilinear forms on $\mathscr{I}^{+}$ though we will not attempt to make this precise. In the following section, we will provide a definition of a ``regularized'' Bondi mass where this limit converges in the strong operator topology in any spacetime where the total energy flux through $\mathscr{I}^{+}$ is a good operator.} 

Finally, we comment on the generalization of these arguments to full, non-perturbative quantum gravity. Since black holes evaporate at finite $G_{\textrm{N}}$ we need not distinguish between spacetimes with or without black holes. We assume that the ADM mass $\hat{M}_{i^{0}}$ continues to be a well-defined operator in full quantum gravity. However, we note that if the Bondi mass is not a good operator perturbatively, one does not expect it to be a good operator at finite $G_{\textrm{N}}$. This is because any massless fields propagating to infinity decay and so are ``weak field'' at $\mathscr{I}^{+}$. As such, the quantization algebra $\mathcal{A}(\mathscr{I}^{+})$ of perturbative gravitons and matter at $\mathscr{I}^{+}$ is actually equivalent to the algebra of observables at $\mathscr{I}^{+}$ in {\em full} quantum gravity. In particular, one can define the full, non-perturbative news operator $\hat{N}_{AB}$ and vacuum state $\ket{\Omega_{\textrm{out}}}$  by  the analogs of \eqref{eq:scricomm} and \eqref{eq:News2pt} where $\delta \hat{N}_{AB}$ is simply replaced by $\hat{N}_{AB}$ \cite{Ashtekar:1981hw}. One can attempt to define a Bondi mass operator $\hat{M}_{\textrm{B}}(u)$ in the full non-perturbative theory satisfying the analog of \eqref{eq:MB2} with the replacement $\delta \hat{N}_{AB}\to  \hat{N}_{AB}$ and $\hat{T}^{(d-2)}_{uu}$ is the flux of matter stress-energy at finite $G_{\textrm{N}}$. Since the GNS Hilbert space is equivalent to $\mathcal{H}_{\textrm{out}}$ at $\mathscr{I}^{+}$, it follows from the above arguments that $\braket{\Psi|M^{2}_{\textrm{B}}|\Psi}=\infty$. Therefore, if the ADM mass is a good operator in quantum gravity, then the Bondi mass cannot be.

\subsection{A regularized semiclassical Bondi mass}
\label{subsec:RegularizedBondiMass}

While the Bondi mass at $\mathscr{I}^{+}$ is not a well-defined operator, we will explain in this section that one can obtain a ``regularized'' Bondi mass operator in the bulk of spacetime, which plays an analogous role to the Bondi mass at $\mathscr{I}^{+}$ but is a {\em good} operator on $\mathcal{H}$. We again, for simplicity, restrict to vacuum gravity and we will simply state the straightforward generalization of the resulting formulae to include matter fields. 

\begin{figure}
    \centering
    \includegraphics[width=0.7\linewidth]{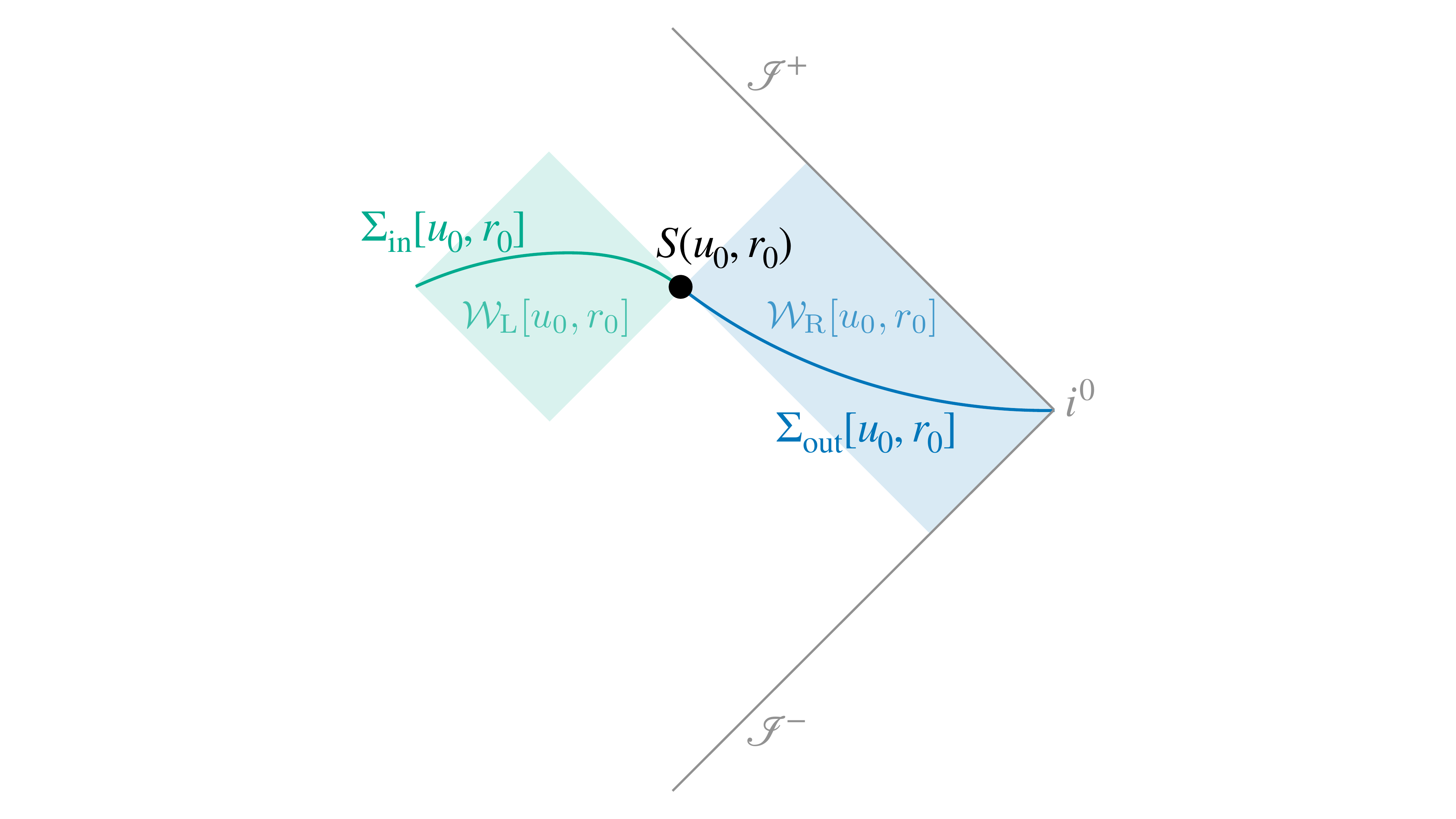}
    \vspace{-.4cm}
    \caption{A spacelike surface $\Sigma_{\textrm{out}}[u_{0},r_{0}]$ whose boundary is $i^{0}$ and $S(u_{0},r_{0})$. The spacelike surface $\Sigma_{\textrm{in}}[u_{0},r_{0}]$ is a smooth extension of $\Sigma_{\textrm{out}}[u_{0},r_{0}]$ such that $\Sigma_{\textrm{in}}[u_{0},r_{0}]\cup \Sigma_{\textrm{out}}[u_{0},r_{0}]$ is a smooth, spacelike Cauchy surface for the full spacetime.}
    \label{fig:cauchy}
\end{figure}

The Bondi mass at $\mathscr{I}^{+}$ is uniquely defined as the asymptotic charge associated to asymptotic time translations at null infinity \cite{Ashtekar:1981bq,Wald:1999wa}. In this section, our goal is to obtain an analogous quantity on any large but finite sphere $S(u_{0},r_{0})$ of radius $r_{0}$ at retarded time $u_{0}$ in the Bondi-like coordinates defined in  sec.~\ref{sec:asympflat}. In contrast to the charges defined at infinity, there exist in the literature many inequivalent definitions of the classical mass on a finite sphere (see, e.g., \cite{Hawking1968,Penrose1982,Bartnik1989,DouganMason1991,BrownYork1993,LiuYau2003,LiuYau2006,WangYau2009PRL,Ashtekar:2024xwl} and references therein). However it is unclear which, if any, of these objects admit a well-defined quantum description. We will not attempt to single out a unique notion of Bondi mass on $S$. Instead, we construct a local, second-order semiclassical observable $\hat{M}_{B}^{(2)}(u_{0},r_{0})$ that captures certain aspects of the classical Bondi mass but also is a well-defined operator on $\mathcal{H}$. We note that the superscript $(2)$ indicates that, in a similar manner to the second order Bondi mass at $\mathscr{I}^{+}$, our regularized, semiclassical Bondi mass will be a function of the second-order metric. Additionally, we will show that the regularized semiclassical Bondi mass generates time translations of bulk observables in a manner that naturally generalizes \eqref{eq:MBOcomm}.

We first construct the classical, second-order Bondi mass on $S(u_{0},r_{0})$. To obtain this, integrate \eqref{eq:HX} with $X=t^{a}$ over a spacelike surface $\Sigma_{\textrm{out}}[u_{0},r_{0}]$ which ranges from spatial infinity to 
$S(u_{0},r_{0})$ (see Fig.~\ref{fig:cauchy}). This integral is equal to the difference of two boundary terms 
\begin{equation}
\label{eq:regMB1}
\delta M_{i^{0}}-M^{(1)}_{\textrm{B}}(u_{0},r_{0})   = -\int_{\Sigma_{\textrm{out}}[u_{0},r_{0}]} \sqrt{h}d^{d-1}x ~n_{a}\omega^{a}(\gamma,\mathcal{L}_{t}g) = 0
\end{equation}
where the right-hand side vanishes since the background $g_{ab}$ is stationary and the second term on the left-hand side is the charge $\mathcal{Q}_{t}$ evaluated at the sphere $S(u_{0},r_{0})$
\begin{equation}
M^{(1)}_{\textrm{B}}(u_{0},r_{0}) \equiv \int_{S(u_{0},r_{0})}\mathcal{Q}_{t}(g(\lambda);\tfrac{d}{d\lambda}g)\bigg\vert_{\lambda=0}\,.
\end{equation}
We will refer to this quantity as the first-order semiclassical Bondi mass which depends locally on the background metric $g_{ab}$ and the perturbed metric $\gamma_{ab}$. We have that 
\begin{equation}
M^{(1)}_{B}(u_{0},r_{0}) = \delta M_{i^{0}} = 0
\end{equation}
since, as discussed in the previous subsection, we assume that the perturbed ADM mass vanishes in all stationary spacetimes. 

To obtain a non-trivial relation we consider an additional variation of \eqref{eq:regMB1}. We define the second variation of the regularized Bondi mass as  
\begin{equation}
M_{B}^{(2)} \equiv \frac{1}{2}\int_{S(u_{0},r_{0})} \frac{d}{d\lambda} \mathcal{Q}_{t}(g(\lambda);\tfrac{d}{d\lambda}g)\bigg\vert_{\lambda=0}
\end{equation}
which now also depends on the second-order metric $\delta^{2}g_{ab}$ and satisfies 
\begin{equation}
 \delta^{2} M_{i^{0}} -M^{(2)}_{\textrm{B}}(u_{0},r_{0})  = \mathcal{E}_{\Sigma_{\textrm{out}}[u_{0},r_{0}]}(\gamma)
\end{equation}
where the right-hand side is the canonical energy for the metric perturbations integrated over $\Sigma_{\textrm{out}}[u_{0},r_{0}]$ 
\begin{equation}
\mathcal{E}_{\Sigma_{\textrm{out}}[u_{0},r_{0}]}(\gamma) = -\frac{1}{2}\int_{\Sigma_{\textrm{out}}[u_{0},r_{0}]} \sqrt{h}d^{d-1}x ~n_{a}\omega^{a}(\gamma,\mathcal{L}_{t}\gamma)\,.
\end{equation}
Following sec.~\ref{subsec:firstorderclass}, this formula can be straightforwardly generalized to include perturbative matter fields which yields
\begin{equation}
M^{(2)}_{\textrm{B}}(u_{0},r_{0})  =  \delta^{2} M_{i^{0}}-\mathcal{E}_{\Sigma_{\textrm{out}}[u_{0},r_{0}]}(\gamma) - \int_{\Sigma_{\textrm{out}}[u_{0},r_{0}]}\sqrt{h}d^{d-1}x~n^{a}t^{b}T_{ab}[\delta \phi]\,,
\end{equation}
where the second and third terms of the right-hand side are simply the Hamiltonian generators of the gravitons and quantum fields, where the integrals are truncated at the sphere $S(u_{0},r_{0})$. 

While $M_{B}^{(2)}(u_{0},r_{0})$ is locally constructed from  $g_{ab}$,  $\gamma_{ab}$ and $\delta^{2}g_{ab}$ on $S(u_{0},r_{0})$ it, by itself, cannot generally be expressed as the second variation of a local quantity on $S(u_{0},r_{0})$. This is analogous to the situation at  $\mathscr{I}^{+}$ where, to obtain the Bondi mass $\delta M_{\textrm{B}}(u_{0})$ in \eqref{eq:MB} as a total variation,  we had to subtract $B(u_{0})$ which was a local quantity constructed in terms of the asymptotic perturbed metric on the cut. Building on the work of \cite{Harlow:2019yfa}, this procedure was generalized in \cite{Chandrasekaran:2021vyu} to ``quasi-local'' charges defined on a timelike worldtube\footnote{In this case, $\Gamma$ can be taken to be the worldtube obtained by translating $S(u_{0},r_{0})$ with respect to the timelike Killing field.}  $\Gamma$ containing $S(u_{0},r_{0})$. It was shown that one can construct an analogous, locally constructed term $B^{(2)}(u_{0},r_{0})$ on $S(u_{0},r_{0})$  to obtain an integrable, second-order charge $\delta^{2}M_{\textrm{B}}(u_{0},r_{0})$. While we could equivalently work with such a quantity, this approach would introduce a considerable amount of additional technical machinery without changing any of the main arguments or conclusions of the paper. For our purposes it will suffice to work with $M_{B}^{(2)}(u_{0},r_{0})$. Indeed, as emphasized above, we will argue that the quantity $M_{B}^{(2)}(u_{0},r_{0})$ itself can be made into a well-defined, local operator on the semiclassical Hilbert space $\mathcal{H}$.

In the quantum theory, the proposed second-order, semiclassical Bondi mass is the local operator $\hat{M}_{\textrm{B}}^{(2)}$ on $S(u_{0},r_{0})$ which satisfies 
\begin{equation}
\label{eq:regMB2}
\hat{M}^{(2)}_{\textrm{B}}(u_{0},r_{0})  =  \delta^{2} \hat{M}_{i^{0}}-\mathcal{E}_{\Sigma_{\textrm{out}}[u_{0},r_{0}]}(\hat{\gamma}) - \int_{\Sigma_{\textrm{out}}[u_{0},r_{0}]}\sqrt{h}d^{d-1}x~n^{a}t^{b}\hat{T}_{ab}[\delta \hat{\phi}]\,. 
\end{equation}

Before proving that $\hat{M}^{(2)}_{\textrm{B}}$ can be smeared into a well-defined operator, we note that it directly follows from \eqref{eq:regMB2} that  $\hat{M}^{(2)}_{\textrm{B}}$ generates time translations of bulk observables. Since this is a key property of the regularized Bondi mass for the arguments of sec.~\ref{sec:algebra}, we spell this out more precisely. The sphere $S(u_{0},r_{0})$ divides the spacetime into two parts as depicted in Fig.~\ref{fig:cauchy}. We denote bulk region spacelike separated from $S(u_{0},r_{0})$ and of smaller radius $r<r_{0}$ as $\mathcal{W}_{\textrm{L}}[u_{0},r_{0}]$. Similarly, we denote the region spacelike separated from $S(u_{0},r_{0})$ of larger radius $r>r_{0}$ as $\mathcal{W}_{\textrm{R}}[u_{0},r_{0}]$. 

Any physical graviton or quantum field operator $\mathcal{O}_{\textrm{L}}$ with support in $\mathcal{W}_{\textrm{L}}[u_{0},r_{0}]$ commutes with all observables in $\mathcal{W}_{\textrm{R}}[u_{0},r_{0}]$ and therefore commutes with the second and third terms on the right-hand side of \eqref{eq:regMB2}. For spacetimes without black holes this property simply follows from causality. However, as we explain in sec.~\ref{subsec:schw_alg}, for spacetimes with black holes the quantum field operators in $\mathcal{R}$ must be suitably ``dressed'' to the left or right boundary. In language of \cite{Chandrasekaran:2022eqq}, we ``dress'' the QFT and graviton operators to the right boundary so that right ADM mass generates time translations of $\mathcal{O}\in \mathcal{A}(\mathcal{R})$ and all quantum field observables commute with the left ADM mass (see sec.~\ref{subsec:i0bh} for more details). In black hole spacetimes, the action of the Bondi mass we describe below will pertain only to observables in $\mathcal{R}$ that are dressed to the right boundary. In this setting, the subregion $\mathcal{W}_{\textrm{L}}[u_{0},r_{0}]$ will should be interpreted as bulk region in $\mathcal{R}$ that is spacelike separated from $S(u_{0},r_{0})$ and of smaller radius. 

With this distinction in mind, the action of the Bondi mass is given by
\begin{equation}
[\hat{M}^{(2)}_{\textrm{B}}(u_{0},r_{0}),\mathcal{O}_{\textrm{L}}] = [\delta^{2}\hat{M}_{i^{0}},\mathcal{O}_{\textrm{L}}]=-\ii\mathcal{L}_{t}\mathcal{O}_{\textrm{L}}\,.
\end{equation}
We now consider any QFT or graviton observable $\mathcal{O}_{\textrm{R}}$ with support in $\mathcal{W}_{\textrm{R}}[u_{0},r_{0}]$. For any such observable, the commutation with the second and third terms of the right-hand side of \eqref{eq:regMB2} are equivalent to the commutation with (minus) the full Hamiltonian. This cancels the action of ADM mass and so we have that $[\hat{M}^{(2)}_{\textrm{B}},\mathcal{O}_{\textrm{R}}]=0$. In summary, for observables spacelike separated from $S(u_{0},r_{0})$, we have that 
\begin{equation}
\label{eq:MBOcomm2}
[\hat{M}^{(2)}_{\textrm{B}}(u_{0},r_{0} ), \mathcal{O}(u^{\prime}) ]=
\begin{cases}
-\ii\mathcal{L}_{t}\mathcal{O}(u^{\prime})  \qquad &\textrm{supp}(\mathcal{O})\in \mathcal{W}_{\textrm{L}}[u_{0},r_{0}]\\
0\quad \quad \quad &\textrm{supp}(\mathcal{O})\in \mathcal{W}_{\textrm{R}}[u_{0},r_{0}]\,.
\end{cases}
\end{equation}
$\hat{M}^{(2)}_{\textrm{B}}(u_{0},r_{0} )$ will not, in general, have a simple, local action on observables to the future or past of $S(u_{0},r_{0})$. The existence of a local operator satisfying \eqref{eq:MBOcomm2} is the key property of $\hat{M}^{(2)}_{\textrm{B}}(u_{0},r_{0} )$ that will be important for the arguments of the following sections.

We now argue that $\hat{M}^{(2)}_{\textrm{B}}$ can be made into a well-defined operator on $\mathcal{H}$. As written, the operator $\hat{M}^{(2)}_{\textrm{B}}(u_{0},r_{0})$ defined at a sharp moment of retarded time cannot be a good operator, as can be directly seen from \eqref{eq:regMB2}. While $\delta^{2}\hat{M}_{i^{0}}$ is a good operator, the integrals of the Hamiltonian generators of the gravitons and quantum fields over a partial Cauchy surface are not well-defined operators and their fluctuations will be UV divergent. 

\subsubsection*{Spacetimes with no black holes}

To see how to obtain a well-defined operator we first consider the case where the spacetime does not contain any black holes. In this setting we can, for now, avoid any potential ``infrared issues'' due to the fact that $\Sigma_{\textrm{out}}[u_{0},r_{0}]$ is non-compact by writing the stress-energy integral as 
\begin{equation}
\label{eq:inout}
\int_{\Sigma_{\textrm{out}}[u_{0},r_{0}]}\sqrt{h}d^{d-1}x~n^{a}t^{b}\hat{T}_{ab}[\delta \hat{\phi}] = \hat{H}_{\textrm{QFT}} - \int_{\Sigma_{\textrm{in}}[u_{0},r_{0}]}\sqrt{h}d^{d-1}x~n^{a}t^{b}\hat{T}_{ab}[\delta \hat{\phi}]
\end{equation}
where $\hat{H}_{\textrm{QFT}}$ is given by \eqref{eq:HQFT} where $\Sigma = \Sigma_{\textrm{out}}[u_{0},r_{0}]\cup \Sigma_{\textrm{in}}[u_{0},r_{0}]$ is a complete Cauchy surface which contains the sphere $S(u_{0},r_{0})$.  $\Sigma_{\textrm{out}}[u_{0},r_{0}]$ is the non-compact portion of the Cauchy surface starting at $S(u_{0},r_{0})$ and ending at spatial infinity, and $\Sigma_{\textrm{in}}[u_{0},r_{0}]$ is the compact portion of the Cauchy surface ranging from the origin $r=0$ to the sphere $S(u_{0},r_{0})$. In this decomposition, $\hat{H}_{\textrm{QFT}}$ is a well-defined operator and so the ultraviolet fluctuations entirely arise from the second term on the right-hand side. Nevertheless, we can cure these divergences by smearing the operator in time. Given a real smearing function $f(u)$ normalized so that $\int_{\mathbb{R}} du f(u)=1$, we consider  
\begin{equation}
\label{eq:Ein}
\hat{E}^{\textrm{QFT}}_{\textrm{in}}[f;u_{0},r_{0}] \equiv \int_{\mathbb{R}}du^{\prime} f(u^{\prime}-u_{0})  \int_{\Sigma_{\textrm{in}}[u^{\prime},r_{0}]}\sqrt{h}d^{d-1}x~n^{a}t^{b}\hat{T}_{ab}[\delta \hat{\phi}]\,.
\end{equation}
It follows from the general results of Borchers \cite{Borchers1964FieldOA} as well as Witten and Strohmaier \cite{Strohmaier:2023opz} that smearing any local field in a timelike direction yields an operator that is well-defined on any Hadamard state. As such, $\hat{E}^{\textrm{QFT}}_{\textrm{in}}[f;u_{0},r_{0}]$ is a densely defined operator on $\mathcal{H}$. Indeed, in four dimensional flat spacetime, the fluctuations of this operator were computed in \cite{Bousso:2017xyo} for the choice of smearing function $f(u)=\tfrac{\delta u}{\pi(u^{2}+\delta u^{2})}$ and found to be of the order\footnote{A useful heuristic for the scaling found in \cite{Bousso:2017xyo} is to evolve the sphere at fixed radius $r_{0}$ for an interval $\delta u$ and smear the energy operator over the resulting timelike cylinder. The smearing suppresses Rindler frequencies ($\omega\delta u\gg1$), giving an effective frequency scale ($\omega_{\mathrm{eff}}\sim\delta u^{-1}$). The leading vacuum fluctuations are localized in a shell of thickness (O($\delta u$)) around the sphere, containing ($N\sim(r_{0}/\delta u)^{2}$) approximately independent correlation cells. Adding their energy fluctuations in quadrature gives $\textrm{Std}(M_{B}^{(2)})\sim\sqrt{N}$ with $\omega_{\mathrm{eff}}\sim r_{0}/\delta u^{2}$, and hence $\textrm{Var}(M_{B}^{(2)})\sim r_{0}^{2}/\delta u^{4}$.}
\begin{equation}
\braket{\hat{E}^{\textrm{QFT}}_{\textrm{in}}[f;u_{0},r_{0}]^{2}} \sim \frac{r_{0}^{2}}{\delta u^{4}} \quad \quad\quad \quad \textrm{ ($d=4$)}\,. 
\end{equation}
It follows that the non-compact integral 
\begin{align}
\label{eq:Eout}
\hat{E}^{\textrm{QFT}}_{\textrm{out}}[f;u_{0},r_{0}]& \equiv \int_{\mathbb{R}}du^{\prime} f(u^{\prime}-u_{0})  \int_{\Sigma_{\textrm{out}}[u^{\prime},r_{0}]}\sqrt{h}d^{d-1}x~n^{a}t^{b}\hat{T}_{ab}[\delta \hat{\phi}]   \nonumber \\
&= \hat{H}_{\textrm{QFT}} - \hat{E}^{\textrm{QFT}}_{\textrm{in}}[f;u_{0},r_{0}]
\end{align}
is also a well-defined operator on $\mathcal{H}$. 

For the gravitational contribution we similarly consider the splitting 
\begin{equation}
\label{eq:Egravsplit}
\mathcal{E}_{\Sigma}(\hat{\gamma})=\mathcal{E}_{\Sigma_{\textrm{in}}[u_{0},r_{0}]} (\hat{\gamma})+\mathcal{E}_{\Sigma_{\textrm{out}}[u_{0},r_{0}]} (\hat{\gamma})\,.
\end{equation}
We recall that $\mathcal{E}_{\Sigma}(\hat{\gamma})$ is a densely defined operator on $\mathcal{H}$. However, in addition to the UV divergences of the individual terms on the right-hand side of \eqref{eq:Egravsplit}, neither term is a priori invariant under small, linearized diffeomorphisms that are non-vanishing on $S(u_{0},r_{0})$.  In particular, applying a linearized diffeomorphism that vanishes at infinity to $\mathcal{E}_{\Sigma_{\textrm{out}}[u_{0},r_{0}]}$ we find that this quantity changes by 
\begin{equation}
\label{eq:Egauge}
\mathcal{E}_{\Sigma_{\textrm{out}}[u_{0},r_{0}]}(\hat{\gamma} + \mathcal{L}_{\xi}g) = \mathcal{E}_{\Sigma_{\textrm{out}}[u_{0},r_{0}]}(\hat{\gamma})+\frac{1}{2}\int_{S(u_{0},r_{0})} d\Omega~\big[- \mathcal{Q}_{\xi}(g,\mathcal{L}_{t}\hat{\gamma})+ \mathcal{Q}_{[t,\xi]}(g,\hat{\gamma})+\mathcal{Q}_{[t,\xi]}(g,\mathcal{L}_{\xi}g)\big]
\end{equation}
an operator localized on $S(u_{0},r_{0})$ that depends only on the background metric and $\hat{\gamma}_{ab}$. The lack of linearized gauge invariance arises from the fact that we did not specify $S(u_{0},r_{0})$ in a gauge invariant way. A trivial way to make this quantity gauge invariant is to completely fix the gauge and choose $S(u_{0},r_{0})$ in that gauge. Then the splitting is trivially gauge invariant when performed in that gauge. Since the distinction between different choices of gauge corresponds to operators localized on $S(u_{0},r_{0})$, properties such as \eqref{eq:MBOcomm2} will be unaffected by this choice. A more covariant approach would be to invariantly define the sphere $S(u_{0},r_{0})$ at large $r_{0}$. 
For example, $S(u_{0},r_{0})$ could be defined by the intersection of a future lightcone $\mathcal{N}_{u_{0}}$ anchored at $\mathscr{I}^{+}$ at $u=u_{0}$ and a past light cone $\mathcal{N}_{v_{0}}$ anchored at $\mathscr{I}^{-}$ at advanced time $v=v_{0}$ so that their intersection is at the areal radius $r=r_{0}$. 
In either case, it is clear that the relevant quantity can be defined in a manner invariant under small, linearized diffeomorphisms. As an abuse of notation, we will continue to denote this quantity as $\mathcal{E}_{\Sigma_{\textrm{out}}[u_{0},r_{0}]}$ with the dressing left implicit. A similar discussion applies to $\mathcal{E}_{\Sigma_{\textrm{in}}[u_{0},r_{0}]}$.

With this understanding we can now apply the same treatment to the graviton energy on $\Sigma_{\textrm{out}}[u_{0},r_{0}]$ 
\begin{equation}
\hat{E}^{\textrm{GR}}_{\textrm{out}}[f;u_{0},r_{0}] \equiv \int_{\mathbb{R}}du^{\prime} f(u^{\prime}-u_{0}) \mathcal{E}_{\Sigma_{\textrm{out}}[u',r_{0}]} (\hat{\gamma})
\end{equation}
which yields a densely defined operator $\hat{E}^{\textrm{GR}}_{\textrm{out}}[f;u_{0},r_{0}]$ on $\mathcal{H}$. Therefore, the smeared Bondi mass 
\begin{equation}
\label{eq:MB2reg}
\hat{M}_{B}^{(2)}(f;u_{0},r_{0}) \equiv \int_{\mathbb{R}}du^{\prime} f(u^{\prime}-u_{0})  \hat{M}^{(2)}_{B}(u',r_{0})
\end{equation}
is also a well-defined operator on $\mathcal{H}$ which satisfies
\begin{equation}
\label{eq:MBreg}
\hat{M}^{(2)}_{\textrm{B}}(f;u_{0},r_{0}) = \delta^{2} \hat{M}_{i^{0}}-\hat{E}^{\textrm{GR}}_{\textrm{out}}[f;u_{0},r_{0}] - \hat{E}^{\textrm{QFT}}_{\textrm{out}}[f;u_{0},r_{0}]
\end{equation}
where we have used the fact that the ADM mass is conserved and $\int_{\mathbb{R}} du f(u)=1$ so that the first term on the right-hand side is independent of the smearing function $f$. 

We note that we have not attempted to rigorously establish the domains on which the smeared Bondi mass operator is defined. While the above arguments strongly indicate that $M_{\textrm{B}}^{(2)}$ can be defined as a densely defined, self-adjoint operator, we will not attempt a rigorous proof of this here. This would involve, in particular, proving that the operators $\hat{H}$, $\hat{E}^{\textrm{GR}}_{\textrm{in}}$ and $\hat{E}^{\textrm{QFT}}_{\textrm{in}}$ admit a self-adjoint extension on a common, dense domain. While we believe this can be established, we will simply assume that these properties are satisfied on some dense domain in the remainder of this paper.

\subsubsection*{Spacetimes with Black Holes}

\begin{figure}
    \centering
    \includegraphics[width=.9\linewidth]{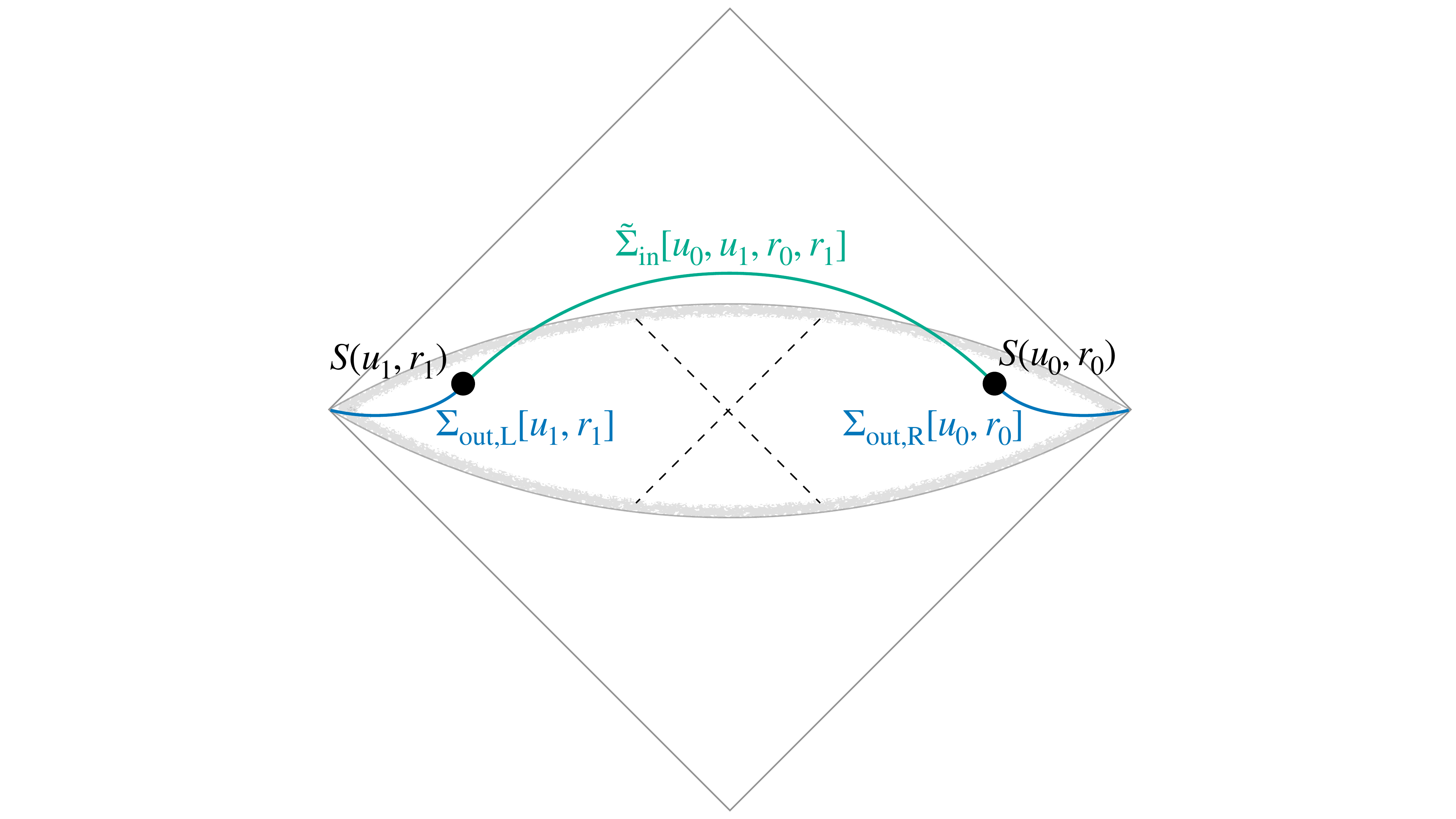}
    \vspace{-.1cm}
    \caption{The deformed Cauchy surface $\tilde{\Sigma}=\Sigma_{\textrm{out,L}}[u_{1},r_{1}]\cup \tilde{\Sigma}_{\textrm{in}}[u_{0},u_1,r_{0},r_1]\cup \Sigma_{\textrm{out,R}}[u_{0},r_{0}]$, where $\tilde{\Sigma}_{\textrm{in}}[u_{0},u_1,r_{0},r_1]$ is a compact, partial Cauchy surface that avoids the interior of the black hole region in the undeformed part of the spacetime, and hence lies entirely in the stationary part of the spacetime.}
    \label{fig:bondi}
\end{figure}

We now consider the construction of a second-order, semiclassical Bondi mass in a stationary (two-sided) black hole spacetime. As in the previous subsection, the key issue is whether the following non-compact integral
\begin{equation}
\label{eq:noncompact}
    \hat{E}^{\textrm{QFT}}_{\textrm{out,R}}[f;u_{0},r_{0}] \equiv \int_{\mathbb{R}}du^{\prime} f(u^{\prime}-u_{0})  \int_{\Sigma_{\textrm{out,R}}[u^{\prime},r_{0}]}\sqrt{h}d^{d-1}x~n^{a}t^{b}\hat{T}_{ab}[\delta \hat{\phi}] 
\end{equation}
is a well-defined operator on the Hilbert space $\mathcal{H}$ of quantum fields and gravitons on a stationary black hole spacetime, where now $\Sigma_{\textrm{out,R}}[u_0,r_0]$ is a partial Cauchy surface which ranges from $S(u_0,r_0)$ to spatial infinity in the ``right'' exterior $\mathcal{R}$ of the black hole, shown in Fig.~\ref{fig:bondi}. The major difference is that the black hole spacetime has two non-compact ends and so we cannot rewrite the integral as the total Hamiltonian minus an integral over a compact, partial Cauchy surface as we did in the previous subsection. 

Nevertheless, it is intuitively clear that, at least for sufficiently large spheres $S(u_{0},r_{0})$ with $r_{0}\gg r_{\textrm{BH}}$,~\eqref{eq:noncompact} should be no more ill-defined than in a spacetime without a black hole. In this case, the spacetime in the exterior of $S(u_{0},r_{0})$ is essentially a flat spacetime. Furthermore, by construction, the states in $\mathcal{H}$ are well approximated at large distances by states in a flat spacetime. Since we concluded in the previous section that~\eqref{eq:Eout} is well-defined in a flat spacetime, the analogous integral~\eqref{eq:noncompact} should also be well-defined in this setting. 

To show that this is indeed the case we again utilize the ``deformation argument'' presented in sec.~\ref{subsec:Hilbreg}. We recall that, to construct the vacuum-at-infinity Hilbert space, we deformed the spacetime to the future and past of the spacelike Cauchy surfaces $\Sigma_{+}$ and $\Sigma_{-}$ to the globally stationary spacetimes $(M_{+},g_{+})$ and $(M_{-},g_{-})$ respectively. In particular, the deformed spacetime $(\tilde{M},\tilde{g})$ does not contain a black hole and is globally stationary except for the ``black hole'' region in the undeformed part of the spacetime. We now consider two surfaces $\Sigma_{\textrm{out,R}}[u_{0},r_{0}]$ and $\Sigma_{\textrm{out,L}}[u_{1},r_{1}]$ anchored on the spheres $S(u_{0},r_{0})$ and $S(u_{1},r_{1})$, respectively, which both lie entirely in the undeformed part of the spacetime with $r_{0}\gg r_{\textrm{BH}}$ and $r_{1}\gg r_{\textrm{BH}}$. We then choose a smooth Cauchy surface $\tilde{\Sigma}=\Sigma_{\textrm{out,L}}[u_{1},r_{1}]\cup \tilde{\Sigma}_{\textrm{in}}[u_{0},u_1,r_{0},r_1]\cup \Sigma_{\textrm{out,R}}[u_{0},r_{0}]$ of $\tilde{M}$ where $\tilde{\Sigma}_{\textrm{in}}[u_{0},u_1,r_{0},r_1]$ is a smooth, spacelike extension of $\Sigma_{\textrm{out,L}}[u_{1},r_{1}]\cup \Sigma_{\textrm{out,R}}[u_{0},r_{0}]$ such that $\tilde{\Sigma}_{\textrm{in}}[u_{0},u_1,r_{0},r_1]$ lies entirely in the stationary part of the spacetime --- i.e., we choose $\tilde{\Sigma}_{\textrm{in}}[u_{0},u_1,r_{0},r_1]$ so that it ``bends'' into $M_{+}$ or $M_{-}$ and thereby avoids the black hole region of the undeformed part of the spacetime as depicted in Fig.~\ref{fig:bondi}. Therefore, $\tilde{\Sigma}_{\textrm{in}}[u_{0},u_1,r_{0},r_1]$ is a compact, partial Cauchy surface. By the arguments of the previous section, it follows that
\begin{equation} \label{eq:energy-sum}
    \hat{E}^{\textrm{QFT}}_{\textrm{out,R}}[f;u_{0},r_{0}] + \hat{E}^{\textrm{QFT}}_{\textrm{out,L}}[f;u_{1},r_{1}] = \hat{\tilde{H}}_{\textrm{QFT}} - \hat{\tilde{E}}^{\textrm{QFT}}_{\textrm{in}}[f;u_{0},u_1,r_{0},r_1]
\end{equation}
is a densely defined operator on $\mathcal{H}$, where 
\begin{equation}
\hat{\tilde{E}}_{\textrm{in}}^{\textrm{QFT}}
[f;u_{0},u_{1},r_{0},r_{1}]
\equiv
\int_{\mathbb{R}} d\tau\, f(\tau)
\int_{\tilde{\Sigma}_{\textrm{in}}[u_{0}+\tau,u_{1}+\tau,r_{0},r_{1}]}
\sqrt{h}\,d^{d-1}x\,
n^{a}t^{b}\hat{T}_{ab}[\delta\hat{\phi}]
\end{equation}
and $\hat{\tilde{H}}_{\textrm{QFT}}$ is the Hamiltonian on $\tilde{\Sigma}$. 

Showing that the left and right smeared energies are separately densely defined operators on $\mathcal{H}$ requires further input. Let $\A(\mathcal{W}_{\rm R}[u_0,r_0])$ and $\A(\mathcal{W}_{\rm L}[u_1,r_1])$ be the algebras generated by the matter fields to the right of $S(u_0,r_0)$ (for $r>r_0$) and to the left of $S(u_1,r_1)$ (for $r>r_1$), respectively. In appendix~\ref{app:split} we show that the algebras $\A(\mathcal{W}_{\rm R}[u_0,r_0])$ and $\A(\mathcal{W}_{\rm L}[u_1,r_1])$ satisfy a split inclusion, so we have
\begin{align}
\A(\mathcal{W}_{\rm R}[u_0,r_0]) \subseteq \mathcal{N} \subseteq \A(\mathcal{W}_{\rm L}[u_1,r_1])'
\end{align}
for some Type I factor $\mathcal{N}$. 

Now, if \eqref{eq:energy-sum} is a densely defined self-adjoint operator, we can exponentiate it to generate a family of unitary operators. Thanks to the (so-far formal) splitting $\hat{E}^{\textrm{QFT}}_{\textrm{out,R}}+\hat{E}^{\textrm{QFT}}_{\textrm{out,L}}$, these operators generate automorphisms of $\mathcal{N}$. Since $\mathcal{N}$ is Type I, this automorphism must be inner and hence can be written as a product of unitaries on $\mathcal{N}$ and $\mathcal{N}'$. We identify $\hat{E}^{\textrm{QFT}}_{\textrm{out,R}}$ and $\hat{E}^{\textrm{QFT}}_{\textrm{out,L}}$ with the logarithms of these respective unitary operators. 

Consequently, $\hat{E}^{\textrm{QFT}}_{\textrm{out,R}}[f;u_{0},r_{0}]$ is a densely defined operator on $\mathcal{H}$. Furthermore, it is clear from the above construction that its action on $\mathcal{H}$ is independent of the choice of deformation of the spacetime. Finally, while we have assumed that $r_{0}\gg r_{\textrm{BH}}$, it is clear that $\hat{E}^{\textrm{QFT}}_{\textrm{out,R}}[f;u_{0}',r_{0}']$ is a densely defined operator for any $u_{0}'$ and $r_{0}'>r_{BH}$ since the difference $\hat{E}^{\textrm{QFT}}_{\textrm{out,R}}[f;u_{0}',r_{0}'] - \hat{E}^{\textrm{QFT}}_{\textrm{out,R}}[f;u_{0},r_{0}]$ is a smeared stress-energy integral over a compact, spacelike surface. A similar procedure applies to the energy flux of gravitons $\hat{E}^{\textrm{GR}}_{\textrm{out,R}}[f;u_{0},r_{0}]$, which can be shown to be similarly well-defined in a black hole spacetime for all $r_{0}>r_{BH}$. Therefore, the regularized Bondi mass $\hat{M}^{(2)}_{\textrm{B}}(f;u_{0},r_{0})$ is also a well-defined operator on the Hilbert space of a black hole spacetime.

\section{The Algebra of Observables Associated to a Cut of $\scri^+$}\label{sec:algebra}
In this section we consider the von Neumann algebra of observables associated to any cut $u=u_{0}$ of $\mathscr{I}^{+}$ in semiclassical quantum gravity. We begin, in sec.~\ref{subsec:i0}, by first considering the analogous construction of the algebra at spatial infinity and illustrate how this algebra gives rise to the familiar bulk algebras studied previously \cite{Kudler-Flam:2023qfl,Chen:2024rpx,Klinger:2026tws}. We then, in sec.~\ref{subsec:construct}, use a similar approach to define and construct the  algebra of a cut of $\mathscr{I}^{+}$. We analyze the properties of this algebra in stationary spacetimes with no black holes in sec.~\ref{subsec:algnoblackholes} and, in sec.~\ref{subsec:schw_alg}, we consider the properties of the algebra in black hole spacetimes. 

\subsection{The algebra at spatial infinity} \label{subsec:i0}

Prior to considering the algebra of a cut $u=u_{0}$ of $\mathscr{I}^{+}$, we first consider the analogous construction of an algebra associated to spatial infinity $i^0$. As we will see, the construction of the algebra at spatial infinity is a slightly simpler setting which avoids some of the complications that will arise when we consider the construction of algebras at finite cuts of $\mathscr{I}^{+}$. This is essentially due to the fact that the ADM mass is a well-defined operator at $i^{0}$ whereas the Bondi mass, as we have emphasized, needs regularization.  

Since $i^0$ is an idealized boundary point, local matter and graviton observables cannot be assigned support at $i^0$ itself and must, in any case, be appropriately smeared. We therefore consider, instead, the algebra of observables in an asymptotic {\em neighborhood} of $i^0$  and define the algebra at spatial infinity by appropriately taking a ``limit'' in which this neighborhood shrinks to $i^0$. To parameterize these neighborhoods and consider this limit, we define coordinates in a neighborhood of spatial infinity. Since the Bondi coordinates introduced in sec.~\ref{sec:asympflat} are adapted to future null infinity, it will be convenient to introduce a ``double null'' coordinate chart $(u,v,x^{A})$ which covers spatial infinity. Here $u=\textrm{const.}$ labels outgoing null hypersurfaces which, as in the Bondi-like coordinates, intersect $\mathscr{I}^{+}$ at cuts of constant retarded time $u$. We define an analogous foliation of ingoing $v=\textrm{const.}$ null hypersurfaces which intersect $\mathscr{I}^{-}$ at cuts of constant advanced time $v$. These hypersurfaces intersect at a codimension-$2$ surface $S(u_{0},v_{0})$ which, in a stationary spacetime are round spheres with radius  $r _{0}\approx\tfrac{1}{2}(v_{0}-u_{0})$ near spatial infinity. Choosing a sphere
$S(u_0,v_0)$ at an arbitrary fixed retarded time $u_0$ and advanced time $v_0$, we define the neighborhood of spatial infinity as the region $\mathcal{W}_{\rm R}$ which is spacelike separated from $S(u_0,v_0)$ and of larger radius (this region is equivalent to the wedge regions we defined previously --- see Fig.~\ref{fig:cauchy}). In our double null coordinates we denote this neighborhood as $\mathcal{W}_{\rm R}[u_0,v_0]$ which is given by 
\begin{equation}
    \mathcal{W}_{\rm R}[u_0, v_0] := \{(u,v,x^{A}) : v_0 < v < \infty \text{ and }  -\infty < u < u_0\}\,.
\end{equation}
The regions $\mathcal{W}_{\rm R}(u_0,v_0)$ contain $i^{0}$ and in the limit as $u_{0}\to -\infty$ and $v_{0}\to +\infty$ we have that $\mathcal{W}_{\rm R}(u_0,v_0)\to i^{0}$.

The semiclassical algebra of observables associated to the neighborhood $\mathcal{W}_{\rm R}[u_0,v_0]$ is constructed from all observables that can be measured in this asymptotic region. In addition to the matter and graviton observables $\delta \hat{\phi}(f)$ and $\hat{\gamma}(f)$ smeared in $\mathcal{W}_{\rm R}[u_0,v_0]$, this includes the second-order ADM mass $\delta^2\hat M_{i^0}$ at spatial infinity, which is a well-defined operator on the Hilbert space $\mathcal H$. To obtain a complete algebra of observables, we consider the von Neumann algebra generated by these operators. Formally this amounts to including all bounded functions of the generators and closing under weak operator limits.\footnote{More precisely, these are bounded operators whose matrix elements arise as limits of matrix elements of operators in the $*$-algebra. See, e.g., \cite{Witten:2018zxz} for further details.} By the von Neumann double commutant theorem \cite{MR641217}, this algebra can equivalently be characterized as the double commutant of these observables, i.e.
 \begin{equation}
\label{eq:i0_double_commutant}
    \mathcal{A}_{u_0,v_0}(i^0) \equiv \left(\left\{\delta \hat{\phi}(f), \hat{\gamma}(f), \delta^2\hat{M}_{i^0}~\forall~ \operatorname{supp}(f)\subset \mathcal{W}_{\rm R}[u_0, v_0]\right\}\right)''\,.
\end{equation}
 We note that we did not explicitly include the regularized Bondi mass observables at finite radius in the generators of the algebra. This is because, by \eqref{eq:regMB2}, the regularized Bondi mass differs from the ADM mass by QFT and graviton operators affiliated with the algebra. Therefore, after taking the double commutant, the algebra $ \mathcal{A}_{u_0,v_0}(i^0)$ includes all Bondi mass observables in $\mathcal{W}_{\rm R}[u_0, v_0]$. Finally, the Hilbert space on which the double commutant in \eqref{eq:i0_double_commutant} is taken depends on whether the spacetime contains a black hole. Since the construction is otherwise identical in the two cases, we suppress this distinction until we turn explicitly to black-hole spacetimes.

This defines a family of algebras parametrized by $u_{0}$ and $v_{0}$. Importantly, since $ \mathcal{W}_{\rm R}[u_2, v_2] \subset \mathcal{W}_{\rm R}[u_1, v_1]$ for $u_{2} < u_{1}$ and $v_{2} > v_{1}$ this family of  algebras manifestly  nests: 
\begin{equation}
      \mathcal{A}_{u_2,v_2}(i^0)\subset \mathcal{A}_{u_1,v_1}(i^0)\,.
\end{equation}
It follows that we can define the limiting algebra $\mathcal{A}(i^0)$ associated to $i^{0}$ as the intersection 
\begin{equation} \label{eq:intersecti0}
    \A(i^0) \equiv \bigcap_{u_{0}, v_{0}} \mathcal{A}_{u_{0},v_{0}}(i_{0})
\end{equation}
for all  $u_{0}$ and $v_{0}$. A standard result says that this intersection is itself always a von Neumann algebra, and we have seen that it is naturally associated with $i^0$. 

In quantum field theory {\em without} gravity (i.e., without including the second-order ADM mass $\delta^2\hat{M}_{i^0}$ in the definition of $\mathcal{A}_{u_0,v_0}(i^0)$), the algebra $\A({i^0})$ obtained from their intersection is trivial: it contains only multiples of the identity $\hat{1}$. We will show that the situation is drastically different if we include quantum gravitational effects. In fact, $\mathcal{A}_{u_0,v_0}(i^0)$ is actually {\em independent} of both times $u_0$ and $v_0$. The intersection defining $\mathcal{A}(i^0)$ is therefore trivial in the opposite sense: every algebra in the intersection is identical, and hence already equal to $\mathcal{A}(i^0)$. To illustrate this we will, in the following, separately consider the algebra for spacetimes with and without black holes. 

\subsubsection*{Spacetimes with no black holes}
We first consider the algebra  $\mathcal{A}_{u_0,v_0}(i^0)$ for some fixed $u_{0}$ and $v_{0}$. The key point is that, in addition to the quantum field observables in the asymptotic region  $\mathcal{W}_{\rm R}[u_0, v_0]$ we also have access to the ADM mass $\delta^{2}\hat{M}_{i^{0}}$. Since $\delta^{2}\hat{M}_{i^{0}}$ is a self-adjoint operator, the unitary $e^{\ii\delta^{2}\hat{M}_{i^{0}} t}$ is an element of $\mathcal{A}_{u_0,v_0}(i^0)$. If $a$ is any QFT or graviton observable in $\mathcal{A}_{u_0,v_0}(i^0)$, it follows that 
\begin{equation}
a_{t}\equiv e^{\ii\delta^{2}\hat M_{i^{0}}t}ae^{-\ii\delta^{2}M_{i^{0}}t} 
\end{equation}
is also an element of $\mathcal{A}_{u_0,v_0}(i^0)$. By evolving all QFT and graviton operators in $\mathcal{A}_{u_0,v_0}(i^0)$ backwards and forwards in time for all $t\in \mathbb{R}$ we therefore obtain all operators in the exterior $r>r_{0}$ of a worldtube of radius $r_{0}=\tfrac{1}{2}(v_{0}-u_{0})$. The next step is to apply the timelike tube theorem which states that the double commutant of the algebra of operators near a worldline includes all operators in its surrounding ``timelike envelope'' (i.e., the set of points
that can be reached via a timelike curve that starts and ends on the worldline) \cite{Borchers:1961,Strohmaier:2023opz}. In this case, the region in the exterior of a worldtube of radius $r_{0}$ contains a worldline which extends from $i^{-}$ to $i^{+}$ and so its timelike envelope is all of the operators in the full spacetime. Hence the algebra $\mathcal{A}_{u_0,v_0}(i^0)$ is actually equivalent to the full bulk algebra $\mathcal{A}(M)\cong \mathcal{B}(\mathcal{H})$ and so is independent of $u_{0},v_{0}$. Therefore, taking the intersection leaves the algebra unchanged, and
\begin{equation}
    \mathcal{A}(i^0) = \bigcap_{u_{0},v_0} \mathcal{A}_{u_{0},v_{0}}(i^{0}) = \mathcal{A}(M) \cong \mathcal{B}(\mathcal{H})
\end{equation}
and so $\mathcal{A}(i^0)$ is a Type I factor \cite{Marolf:2008mf}.

\subsubsection*{Spacetimes with black holes} \label{subsec:i0bh}

We now consider the algebra associated to spatial infinity in spacetimes that contain a black hole. The analysis bears many similarities to the analysis of spacetimes without black holes so we will focus on the key differences that arise in black hole spacetimes. We focus, for simplicity, on the algebra in a Schwarzschild spacetime however our conclusions can be straightforwardly generalized to any stationary black hole spacetime \cite{Kudler-Flam:2023qfl,Chen:2024rpx}. 

The first difference is the Hilbert space on which the algebras are defined. In a spacetime without a black hole, the ADM mass $\delta^{2}\hat{M}_{i^{0}}$ is equivalent to the Hamiltonian $\hat{H}$ of the quantum field by Eq.~\eqref{eq:ADMHam} and so is not an independent degree of freedom of the quantum gravitational field. However, as emphasized in \cite{Chandrasekaran:2022eqq}, this is not the case if the spacetime contains a black hole. In this case, $\hat{H}$ is given by the difference of the left and right ADM masses (see Eq.~\eqref{eq:ADMHamLR}). To obtain the algebra of physical observables in the right black hole exterior $\mathcal{R}$ we seek a representation on which (1) $\delta^{2}\hat{M}^{\textrm{R}}_{i^{0}}$ generates time translations of all bulk observables in $\mathcal{R}$ and (2) commutes with the left ADM mass $\delta^{2}\hat{M}^{\textrm{L}}_{i^{0}}$ in the left black hole exterior $\mathcal{L}$. Following \cite{Chandrasekaran:2022eqq}, such a representation can be obtained by enlarging the Hilbert space to 
\begin{equation}
\mathcal{H}\otimes L^{2}(\mathbb{R})_{\textrm{L}}
\end{equation}
where $\mathcal{H}$ is the Hilbert space of quantum fields and gravitons and $L^{2}(\mathbb{R})_{\textrm{L}}$ corresponds to the Hilbert space of fluctuations of the left ADM mass which simply acts as a multiplication operator. Therefore, by construction, all quantum field observables in $\mathcal{R}$ satisfy (2) and, furthermore, the right ADM mass 
\begin{equation}
\delta^{2}\hat{M}^{\textrm{R}}_{i^{0}} \equiv \hat{H} + \delta^{2}\hat{M}^{\textrm{L}}_{i^{0}}
\end{equation}
generates time translations of all bulk observables in $\mathcal{R}$ since it is equivalent to $\hat{H}$ up to a term that lies in the commutant. We will refer to this quantization as ``dressing'' the QFT and graviton operators to the right boundary. 

\begin{figure}
    \centering
    \includegraphics[width=1\linewidth]{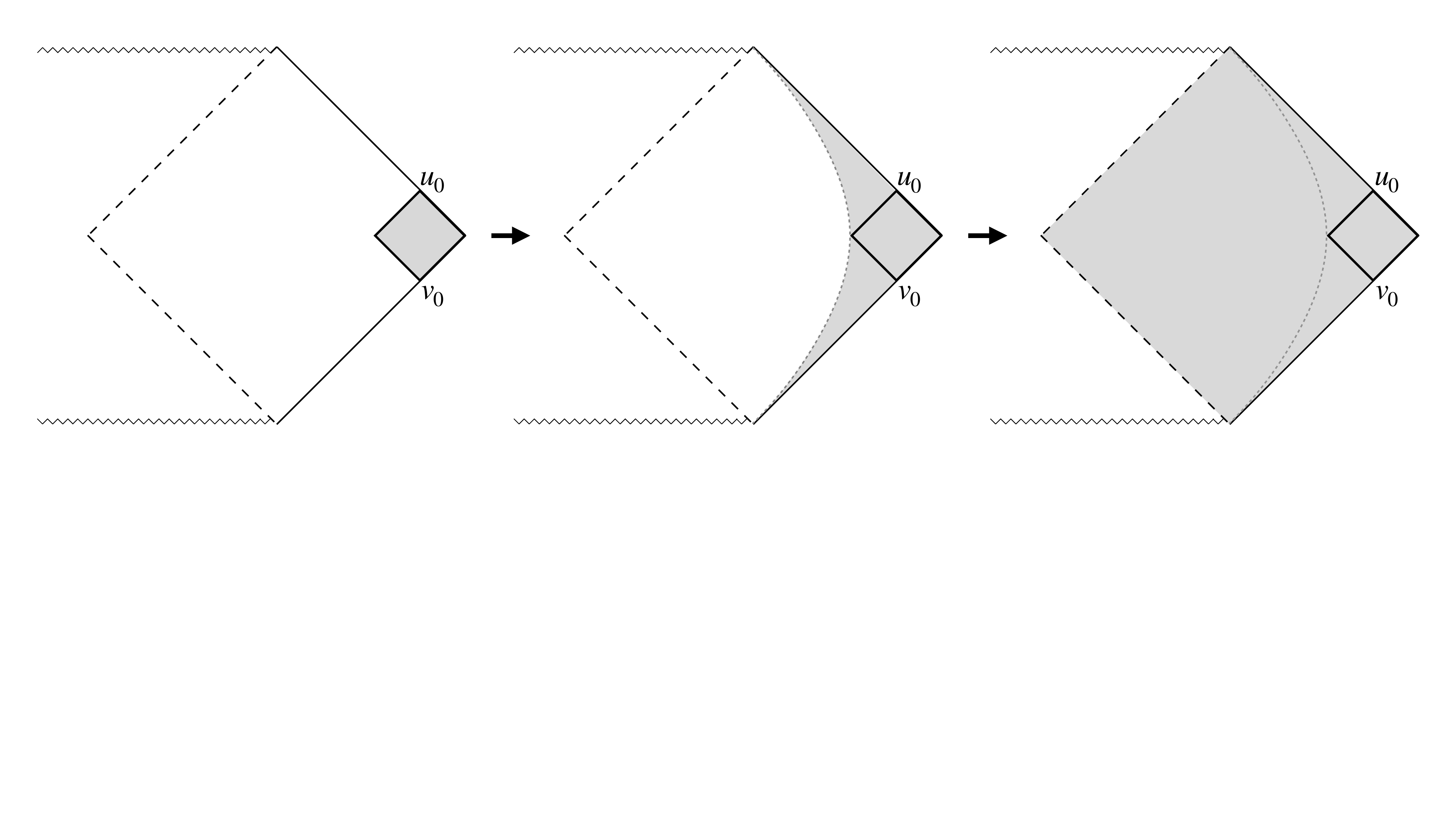}
    \vspace{-4.6cm}
    \caption{The algebra of observables associated to a subregion near spatial infinity in a Schwarzschild background. 1) We start from a subregion anchored to $i^0$. 2) We evolve with the ADM mass. 3) We apply the timelike tube theorem.}
    \label{fig:spatial}
\end{figure}

To distinguish them from the ordinary QFT algebras acting on $\mathcal{H}$, we will denote algebras on the larger Hilbert space $\mathcal{H}\otimes L^{2}(\mathbb{R})_{\textrm{L}}$ as $\tilde{\mathcal{A}}$.
With the representation of the left and right ADM mass now defined, the regularized Bondi mass operators also have a well-defined action on this larger Hilbert space. The algebras $\tilde{\mathcal{A}}_{u_0,v_0}(i^0)$ are defined similarly to the case without black holes, by the right-hand side of \eqref{eq:i0_double_commutant} where $\delta^{2}\hat{M}_{i^{0}}$ is replaced by $\delta^{2}\hat{M}^{\textrm{R}}_{i^{0}}$ and the region $\mathcal{W}_{\textrm{R}}\subset \mathcal{R}$.\footnote{In Kerr the algebra would also include the ADM angular momentum \cite{Kudler-Flam:2023qfl}.} Finally, the algebra $\tilde{\mathcal{A}}(i^0)$ is analogously defined as the intersection of algebras defined in successively smaller neighborhoods of $i^{0}$,
\begin{equation}
\label{eq:intersectionBH_i0}
    \tilde{\mathcal{A}}(i^0) = \bigcap_{u_{0},v_{0}} \tilde{\mathcal{A}}_{u_0,v_0}(i^0)
\end{equation}
on the Hilbert space $\mathcal{H}\otimes L^{2}(\mathbb{R})_{\textrm{L}}$. The key point is that each algebra in a neighborhood of spatial infinity includes the right ADM mass and so we can once again use $\delta^{2}\hat{M}_{i^{0}}^{\textrm{R}}$ to evolve all quantum field theory operators to the past and future.  The timelike tube theorem then implies that $\tilde{\mathcal{A}}_{u_0,v_0}(i^0)$ contains every QFT and graviton observable in the right exterior. Since it also contains $\delta^{2}\hat{M}_{i^{0}}^{\textrm{R}}$ we have that $\tilde{\mathcal{A}}(\mathcal{R}) \subseteq \tilde{\mathcal{A}}_{u_0,v_0}(i^0)$. Conversely, it is clear by its definition that $\tilde{\mathcal{A}}_{u_0,v_0}(i^0) \subset \tilde{\mathcal{A}}(\mathcal{R})$ since they both consist of QFT and graviton operators dressed to the right boundary together with the right ADM mass. Therefore we have that $\tilde{\mathcal{A}}_{u_0,v_0}(i^0) = \tilde{\mathcal{A}}(\mathcal{R})$ and so independent of $u_0$ and $v_{0}$. It follows that 
\begin{equation}
\tilde{\mathcal{A}}(i^{0}) = \tilde{\mathcal{A}}(\mathcal{R})\,,
\end{equation}
so the algebra $\tilde{\mathcal{A}}(i^{0})$ is equivalent to the full bulk algebra in the right exterior of the black hole, as shown in Fig.~\ref{fig:spatial}. This algebra has the form of a crossed product $\mathcal{A}\rtimes\mathbb{R}_{H}$, where $\mathbb{R}_{H}$ is the translation group generated by the Hamiltonian $\hat H$ \cite{Witten:2021unn}.

The analogous semiclassical algebra for the exterior $\mathcal{R}$ of an AdS-Schwarzschild black hole was first analyzed in \cite{Chandrasekaran:2022eqq}. In these spacetimes, the Hartle-Hawking state is normalizable and is thermal with respect to $\hat H$. Therefore, flow generated by the Hamiltonian is also the modular flow of the Hartle-Hawking state in $\mathcal{R}$. It was recognized in \cite{Chandrasekaran:2022eqq} that the AdS-Schwarzschild algebra $\tilde{\mathcal{A}}(\mathcal{R})$ is a {\em modular crossed-product} and so, by a theorem of Takesaki, $\tilde{\mathcal{A}}(\mathcal{R})$ is a Type II$_{\infty}$ factor \cite{takesaki1973duality}. 

While the Hartle-Hawking state is not normalizable for asymptotically flat black holes, it was noted in \cite{Kudler-Flam:2023qfl} that the Unruh state is well-defined on these spacetimes. Within the Hilbert space associated to the Unruh vacuum, it was argued in \cite{Kudler-Flam:2023qfl} that one can generalize the conclusions of \cite{Chandrasekaran:2022eqq} and show that the semiclassical algebra in the exterior of any stationary black hole is Type II$_{\infty}$. However, as emphasized in sec.~\ref{subsec:quantizationBH}, while the ADM mass can be defined on this Hilbert space, we cannot simultaneously define a regularized Bondi mass. This was immaterial to the above discussion since we really only needed the ADM mass to obtain a non-trivial algebra $\tilde{\mathcal{A}}(i^{0})$. However, as we will see in the following subsections, to obtain a non-trivial algebra associated to a cut of $\mathscr{I}^{+}$ one must consider the vacuum-at-infinity Hilbert space $\mathcal{H}$ of quantum fields, on which the ADM mass {\em and} the regularized Bondi mass are both well-defined. 

It was argued by two of us in \cite{Chen:2024rpx} that the algebra $\tilde{\mathcal{A}}(\mathcal{R})$ on $\mathcal{H}\otimes L^{2}(\mathbb{R})$ is a Type II$_{\infty}$ factor. The key idea is that while the Hartle-Hawking state is non-normalizable, it can be realized as a faithful, normal, and semifinite weight on the algebra. We provide a proof of this statement in appendix~\ref{app:fns}. Since the theorem of Takesaki actually applies to this larger class of weights, it follows that $\tilde{\mathcal{A}}(\mathcal{R})$ is also a modular crossed product with respect to modular flow of the Hartle-Hawking weight. Finally, we note that the commutant of any Type II$_{\infty}$ is also a Type II$_{\infty}$. The commutant of $\tilde{\mathcal{A}}(\mathcal{R})$ consists of all QFT observables in $\mathcal{L}$ dressed to the left boundary together with the left ADM mass. We denote this algebra as 
\begin{equation}
    \tilde{\mathcal{A}}(\mathcal{L}) = \tilde{\mathcal{A}}(\mathcal{R})^{\prime}
\end{equation}
which is then also a Type II$_{\infty}$ factor.

\subsection{The algebra of a cut of null infinity}
\label{subsec:construct}

We now repeat the construction introduced in sec.~\ref{subsec:i0} for a finite cut $u=u_0$ of $\scri^+$. The principal new ingredient is that the ADM mass used at spatial infinity must now be replaced by a family of regularized Bondi masses. However, since the QFT operators cannot be defined at a cut and the Bondi mass cannot be defined at infinity we consider, as before, an asymptotic neighborhood of the cut. We define this neighborhood using Bondi-like coordinates of sec.~\ref{sec:asympflat}. We first thicken the cut to a retarded time interval $(u_{0},u_{0}+\delta u)$ of width $\delta u>0$ and then extend this interval into the bulk along past-directed null geodesics to a radius $r=r_{\textrm{min}}$. In these coordinates we have that\footnote{In a similar manner to the Bondi mass (see discussion below Eq.~\eqref{eq:Egauge}), the region $\r[u_0, \delta u, \rmin]$ can also be defined in an invariant way by, for example, considering the intersection of a pair of past-directed null surfaces anchored at $u=u_{0}$ and $u=u_{0}+\delta u$ with a future-directed null surface from $\mathscr{I}^{-}$ at an advanced time $v=v_{\textrm{min}}$ so that $r_{\textrm{min}}\approx \tfrac{1}{2}(v_{\textrm{min}}-u_{0})$. While important, this distinction will not affect our analyses in any material way since we will show that the holographic algebras \eqref{eq:intersect} are actually independent of the choice of asymptotic neighborhood $\mathcal{R}$.}
\begin{equation} \label{eq:strip}
    \r[u_0, \delta u, \rmin] := \{(u,r,x^{A}) : \rmin < r < \infty \text{ and } u_0 < u < u_0+\delta u\}\,.
\end{equation}

Similarly to the algebra in a neighborhood of spatial infinity, the algebra of observables associated to this neighborhood consists of all QFT and graviton observables in $\r[u_0, \delta u, \rmin]$ together with the family of Bondi masses whose defining spheres and smearing intervals also lie in $\r[u_0, \delta u, \rmin]$. The von Neumann algebra associated to this neighborhood is
\begin{equation}
\label{eq:double_commutant}
    \mathcal{A}_{\delta u,r_{\textrm{min}}}(u_{0}) \equiv \left(\left\{\delta \hat{\phi}(f), \hat{\gamma}(f), \hat{M}_{\textrm{B}}^{(2)}(f;u,r)~\forall~ (u,r)\in\r[u_0, \delta u, \rmin]\right\}\right)''\,,
\end{equation}
where $\operatorname{supp}(f)\subset \r[u_0, \delta u, \rmin]$ for all $f$. We note that, as explained in the previous subsection, the Hilbert space on which we take the double commutant will depend on whether the spacetime contains a black hole or not. If the spacetime contains a black hole then all operators are dressed to the right boundary and we will denote the algebras as $\tilde{\mathcal{A}}_{\delta u,r_{\textrm{min}}}(u_{0})$ as described in sec.~\ref{subsec:i0bh}. Since the following construction applies equally to $\tilde{\mathcal{A}}_{\delta u,r_{\textrm{min}}}(u_{0})$ as it does for $\mathcal{A}_{\delta u,r_{\textrm{min}}}(u_{0})$, we will suppress this distinction for now. 

In either case, the algebras trivially nest in the sense that $\A_{\delta u_1, r_{\textrm{min,1}}} (u_0)\subset \A_{\delta u_2, r_{\textrm{min,2}}} (u_0)\,$ for $\delta u_{1}< \delta u_{2}$ and $r_{\textrm{min,1}} > r_{\textrm{min,2}}$ and so, following the logic of the previous subsection, we may define the algebra $\mathcal{A}(u_{0})$ associated to just the cut $u=u_{0}$ as an intersection 
\begin{equation} \label{eq:intersect}
    \A({u_0}) = \bigcap_{\rmin,\, \delta u} \mathcal{A}_{\delta u,r_{\textrm{min}}}(u_{0})
\end{equation}
for all $\delta u>0$ and $r_{\textrm{min}}$. In an analogous manner to the algebra at spatial infinity, we will show that this algebra is also non-trivial. Indeed, as before, adjoining the family of regularized Bondi masses will imply that $\mathcal{A}_{\delta u,r_{\textrm{min}}}(u_{0})$ is actually independent of $\delta u$ and $r_{\textrm{min}}$ and we again have that every algebra in the intersection is already equal to $\A({u_0})$.

Prior to considering the full intersection algebra \eqref{eq:intersect} we will first  consider some general properties of $\mathcal{A}_{\delta u,r_{\textrm{min}}}(u)$ for all $u$ and any fixed $\delta u$ and $r_{\textrm{min}}$. These properties will play a key role in subsequent arguments in the remainder of this paper. The first property is the behavior of these algebras under the action of ADM mass. Let $U(t)$ be the unitary group generated by $\delta^{2}\hat{M}_{i^{0}}$,
\begin{equation}
U(t) = e^{-\ii t \delta^{2}\hat{M}_{i^{0}}} \,,
\end{equation}
with action on the algebra given by 
\begin{equation}
\label{eq:timetranslate}
U^{\dagger}(t)\mathcal{A}_{\delta u,r_{\textrm{min}}}(u_{0}) U(t)= \mathcal{A}_{\delta u,r_{\textrm{min}}}(u_{0}+t)\,.
\end{equation}
This property trivially follows from the fact that the ADM mass generates time translations of any bounded function of the smeared operators $\delta \hat{\phi}$, $\hat{\gamma}$ and $\hat M_{\textrm{B}}^{(2)}$. Taking the double commutant yields the desired relation. 
This is simply the familiar statement in ordinary QFT that the Hamiltonian generates time translations and \eqref{eq:timetranslate} simply reflects the fact that this property is unchanged when we include quantum gravitational effects. 

The second property is that the algebras satisfy a non-trivial nesting relation. In particular we will show that, if $u_{1}>u_{0}$ then the algebra $\mathcal{A}_{\delta u,r_{\textrm{min}}}(u_{1})$ is a strict {\em subalgebra} of $\mathcal{A}_{\delta u,r_{\textrm{min}}}(u_{0})$ ,
\begin{equation}
\label{eq:inclusion}
\mathcal{A}_{\delta u,r_{\textrm{min}}}(u_{1}) \subsetneq \mathcal{A}_{\delta u,r_{\textrm{min}}}(u_{0}) \quad \quad \textrm{ for any $u_{1}> u_{0}$}
\end{equation}
and fixed $\delta u$ and $r_{\textrm{min}}$. Therefore, due to the inclusion of the semiclassical Bondi mass, the family of algebras $\mathcal{A}_{\delta u,r_{\textrm{min}}}(u)$  associated to different cuts are not independent but form a nested family.

To prove \eqref{eq:inclusion}, let $a_{1}$ be any bounded function of the smeared operators  $\delta \hat{\phi}$ and $\hat{\gamma}$. We choose a sphere $S(u^{\prime},r^{\prime})$ in the defining asymptotic region for $\mathcal{A}_{\delta u,r_{\textrm{min}}}(u_{0})$ such that it is spacelike separated and at a larger radius than the support of $a_{1}$ --- i.e., $a_{1}$ lies in the region $\mathcal{W}_{\textrm{L}}[u^{\prime},r^{\prime}]$ in Fig.~\ref{fig:cauchy}. In particular, this implies that  $a_{1}$ is also spacelike separated from the unitary
\begin{equation}
    U_{\textrm{B}}(t)=e^{-\ii t \hat M_{\textrm{B}}^{(2)}(f;u^{\prime},r^{\prime})} \in \mathcal{A}_{\delta u,r_{\textrm{min}}}(u_{0})
\end{equation}
defined with respect to the sphere $S(u^{\prime},r^{\prime})$. Conjugating $a_{1}$ with $U_{\textrm{B}}(-t)$ for $t>0$ we obtain 
\begin{equation}
a_{1}(t)=U_{\textrm{B}}^\dag(-t)a_{1}U_{\textrm{B}}(-t)\,,
\end{equation}
where $a_1(0)=a_1$. By \eqref{eq:MBOcomm2}, conjugation with $U_{\textrm{B}}(t)$ is equivalent to backwards time translation of $a_{1}$ for all $t$ as long as $a_{1}(t)$ remains spacelike separated from $S(u^{\prime},r^{\prime})$. Let $t_{\ast}$ denote the time at which $a_{1}(t)$ first becomes null separated  from $\hat M_{\textrm{B}}^{(2)}(f;u^{\prime},r^{\prime})$. We therefore have that 
\begin{equation}
\label{eq:a1backwards}
a_{1}(t)=U_{\textrm{B}}^{\dagger}(-t)a_{1}U_{\textrm{B}}(-t) \quad \quad \textrm{ $t\in [0,t_{\ast})$}\,.
\end{equation} 
If $t_{0}<t_{\ast}$ denotes a time such that $a_{1}(t_{0})$ is supported in $\mathcal{R}[u_{0},\delta u,r_{\textrm{min}}]$ then by \eqref{eq:timetranslate}, $a_{1}(t_{0})\in \mathcal{A}_{\delta u,r_{\textrm{min}}}(u_{0})$.
It follows that 
\begin{equation}
a_{1} = U_{\textrm{B}}(-t_{0})a_{1}(t_{0})U_{\textrm{B}}^{\dagger}(-t_{0}) \in \mathcal{A}_{\delta u,r_{\textrm{min}}}(u_{0})\,.
\end{equation}
We now prove that bounded functions of the $\hat{M}_{\textrm{B}}^{(2)}(f;u_1,r_1)$ on a sphere $S(u_{1},r_{1})$ at the later retarded time $u_{1}>u_{0}$ are also in the algebra. We note that the difference between any two unsmeared Bondi masses can be formally written as 
\begin{equation}
\hat{M}_{\textrm{B}}^{(2)}(u_1,r_1) - \hat{M}_{\textrm{B}}^{(2)}(u_0,r_0) = -\mathcal{E}_{\Sigma_{12}}(\hat{\gamma}) - \int_{\Sigma_{12}}\sqrt{h}d^{d-1}x~n^{a}t^{b}T_{ab}[\delta \hat{\phi}]
\end{equation}
where we choose $r_{0}$ such that there exists a spacelike surface $\Sigma_{12}$ which interpolates between the later sphere $S(u_{1},r_{1})$ and the earlier sphere $S(u_{0},r_{0})$. Smearing both sides in time yields an expression that is well-defined as a densely defined operator. Therefore, if $\hat{M}_{\textrm{B}}^{(2)}(f;u_0,r_0)$ is affiliated with $\mathcal{A}_{\delta u,r_{\textrm{min}}}(u_{0})$ then its difference with any regularized Bondi mass at a later retarded time is simply given by quantum field and graviton operators in the region $u>u_{0}$. Since we just argued that all such operators are affiliated with the algebra $\mathcal{A}_{\delta u,r_{\textrm{min}}}(u_{0})$ then so is any regularized Bondi mass to the future of $u=u_{1}$. 

Since the above argument applies for any bounded generator of $\mathcal{A}_{\delta u,r_{\textrm{min}}}(u_{1})$, taking the double commutant yields \eqref{eq:inclusion}. The reverse argument does not work (and indeed we shall see that the inclusion is strict): any Bondi mass at retarded times $u>u_{1}$ does not act as a time-translation of observables supported at $u<u_{1}$. From a physical perspective, the inclusion \eqref{eq:inclusion} reflects the fact that while both of the algebras $\mathcal{A}_{\delta u, r_{\textrm{min}}}(u_{0})$ and $\mathcal{A}_{\delta u, r_{\textrm{min}}}(u_{1})$ have access to all observables to their future, they do not have access to the degrees of freedom that have radiated to $\mathscr{I}^{+}$ to the past --- i.e. $\mathcal{A}_{\delta u, r_{\textrm{min}}}(u_{1})$ cannot reconstruct the radiation observables in $\mathcal{A}_{\delta u, r_{\textrm{min}}}(u_{0})$ over the interval $(u_{0},u_{1})$.

\subsection{Spacetimes with no black holes}
\label{subsec:algnoblackholes}
We now use the properties of the preceding subsection to prove that $\mathcal{A}(u_{0})$ on $\mathcal{H}$ is a non-trivial algebra of observables in semiclassical quantum gravity. We first consider the algebra for spacetime without black holes. While our analysis will apply quite generally to any stationary spacetime without a black hole (e.g., the spacetime of a star,...), for definiteness we will consider the case of Minkowski spacetime. 

\begin{figure}
    \centering
    \includegraphics[width=\linewidth]{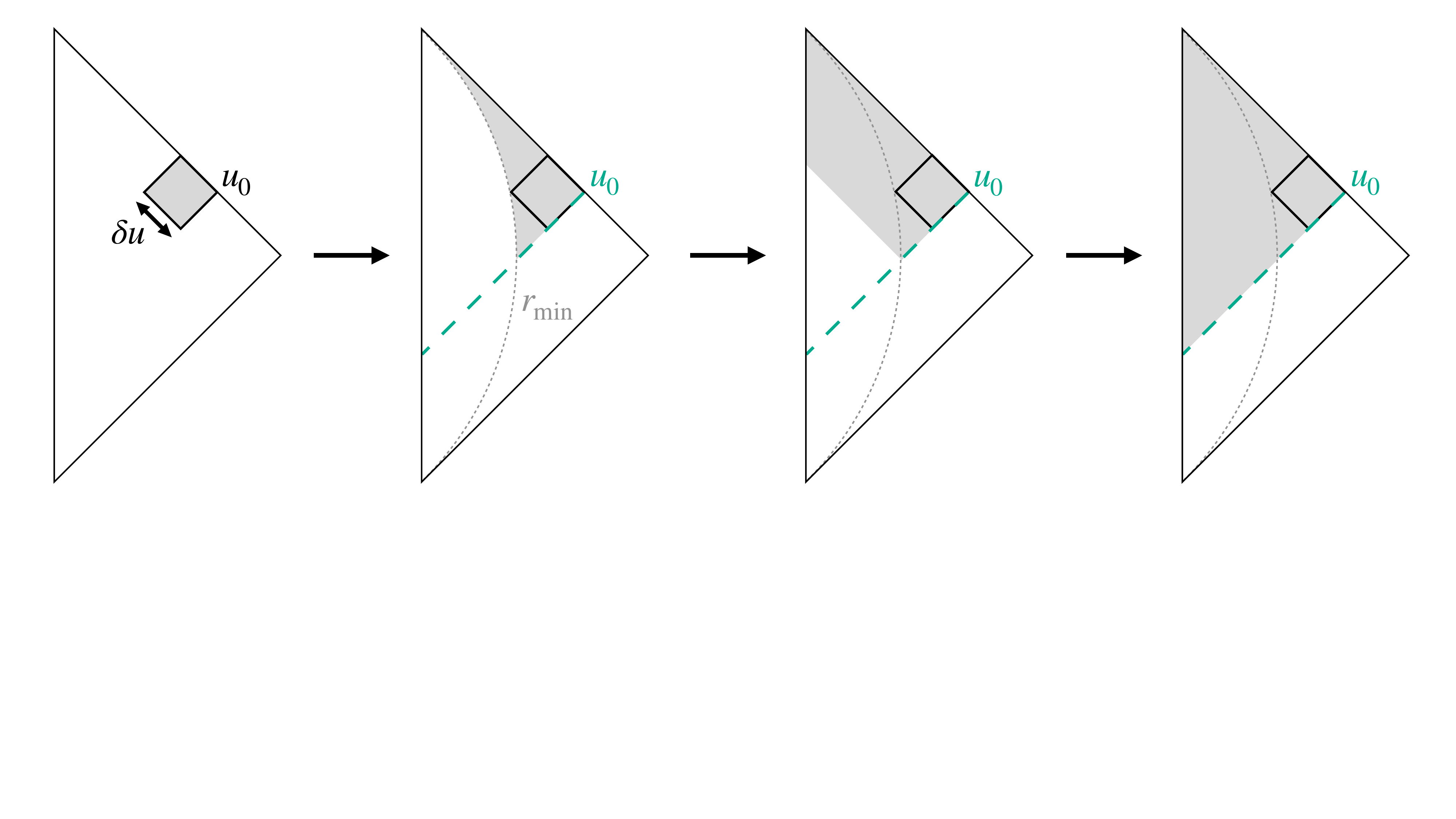}
    \vspace{-4cm}
    \caption{The algebra of observables associated to a subregion near asymptotic infinity in Minkowski. 1) We start from a subregion anchored to $\scri^+$ at $u_0$. 2) We evolve with the Bondi mass. 3) We apply the timelike tube theorem. 4) We apply the time slice axiom.}
    \label{fig:mink4}
\end{figure}

Starting with the algebra $\mathcal{A}_{\delta u,r_{\textrm{min}}}(u_{0})$ we note that the arguments of the previous subsection already imply that this algebra is independent of $\delta u$. Indeed, using \eqref{eq:a1backwards} we see that the algebra already contains all observables in the region $r<r_{\textrm{min}}$ and $u>u_{0}$. This step is depicted in the second panel of Fig.~\ref{fig:mink4}. Next we apply the timelike tube theorem which then implies that the algebra contains the observables in its timelike envelope as depicted in the third panel of Fig.~\ref{fig:mink4}.  We note that the resulting algebra has support on a Cauchy slice for the spacetime region $\mathcal{W}_{u_{0}}$ spacelike separated from the $u=u_0$ cut. The time slice axiom therefore shows that the algebra in fact includes all of $\mathcal{W}_{u_{0}}$ (as illustrated by the fourth panel of Fig.~\ref{fig:mink4}). 

It turns out that we are now done and there are no further operators hidden in $\mathcal{A}_{\delta u,r_{\textrm{min}}}(u_{0})$. In other words, for any $\delta u>0$ and any $r_{\textrm{min}}$, we have
\begin{equation}
\mathcal{A}_{\delta u,r_{\textrm{min}}}(u_{0}) = \mathcal{A}(\mathcal{W}_{u_{0}})\,,
\end{equation}
the von Neumann algebra of all observables in the spacelike wedge. To see this, note that in Minkowski space, any regularized Bondi mass $\hat M_{\rm{B}}^{(2)}(f;u^{\prime},r^{\prime})$ can be written as an integral over quantum fields in $\mathcal{W}_{u_{0}}$. It follows that all the operators used to originally define $\mathcal{A}_{\delta u,r_{\textrm{min}}}(u_{0})$ are contained in $\mathcal{A}(\mathcal{W}_{u_{0}})$; combined with the converse argument above, we see that the two algebras are equal. 
It directly follows that 
\begin{equation}
\label{eq:entwedgeflat}
    \mathcal{A}(u_{0}) = \bigcap_{\rmin,\, \delta u} \mathcal{A}_{\delta u,r_{\textrm{min}}}(u_{0}) = \mathcal{A}(\mathcal{W}_{u_{0}})\,,
\end{equation}
and, from the arguments in sec.~\ref{sec:type}, we then see that $\A(u_{0})$ is a Type III$_{1}$ factor.\footnote{It would be interesting to check if an analogous Type III$_1$ algebraic structure exists in AdS with radiative degrees of freedom, leaky boundary conditions, or when coupled to an external bath (as described in e.g. \cite{Penington:2019npb, Almheiri:2019hni, Compere:2019bua, Ciambelli:2024kre}).} More precisely, the algebra factorizes into radiation operators at $\scri^+$ and massive field operators at $i^+$ \cite{Prabhu:2022zcr}
\begin{equation}
\label{eq:Au0tensor}
    \mathcal{A}(u_{0})\cong \mathcal{A}(\mathscr{I}^{+}_{u>u_{0}})\;\overline{\otimes} \;\mathcal{A}(i^{+})\,,
\end{equation}
where $\mathcal{A}(i^{+})\cong \mathcal{B}(\mathcal{H}_{i^{+}})$ is a Type I$_{\infty}$ factor. The commutant of $\mathcal{A}(u_{0})$ must commute with all Bondi mass operators in the spacelike wedge $\mathcal{W}_{u_{0}}$, and therefore is simply given by massless fields on $\mathscr{I}^{+}_{<u_{0}}$,
\begin{equation}
\mathcal{A}(u_{0})^{\prime} = \mathcal{A}(\mathscr{I}^{+}_{<u_{0}})\,.
\end{equation}
Both $\mathcal{A}(\mathscr{I}_{>u_{0}}^{+})$ and its commutant $\mathcal{A}(\mathscr{I}_{<u_{0}}^{+})$ are Type III$_{1}$ factors due to the entanglement with radiative degrees of freedom across the cut $u=u_{0}$, so we see that the Type III behavior of the holographic algebra $\mathcal{A}(u_{0})$ is entirely due to its failure to reconstruct observables to the past of the $u=u_{0}$ cut of $\mathscr{I}^{+}$. It also follows from \eqref{eq:entwedgeflat} that the holographic algebras nest
\begin{equation}
\label{eq:nestnoBH}
\mathcal{A}(u_{1})\subset \mathcal{A}(u_{0})
\end{equation}
for any $u_{1}>u_{0}$, as we argued on general grounds in \eqref{eq:inclusion}. 

Finally, we comment on the relationship between the holographic algebras $\mathcal{A}(u_{0})$ associated to a $u=u_{0}$ cut of $\mathscr{I}^{+}$ and the algebra $\mathcal{A}(i^{0})$ constructed in sec.~\ref{subsec:i0}. While the algebras $\mathcal{A}(u)$ for all $u\in \mathbb{R}$ are Type III$_{1}$, it is intuitively clear that these algebras should ``limit'' to the Type I algebra at spatial infinity as $u\to -\infty$. To make this precise we note that  the inclusion structure of the algebra given by \eqref{eq:nestnoBH} and the fact that the holographic algebras are appropriately continuous\footnote{It follows from \eqref{eq:entwedgeflat} that the holographic algebras satisfy $\mathcal{A}(u_{0}) = \big(\bigcup_{\epsilon>0}\mathcal{A}(u_{0}+\epsilon)\big)''$. Roughly speaking this condition simply states that there are no degrees of freedom localized on the cut $u=u_{0}$.} implies that the algebra $\mathcal{A}(u_{0})$ is equivalent to the (von Neumann) union of all algebras $\mathcal{A}(u)$ for $u>u_{0}$
\begin{equation}
\label{eq:Au0}
\mathcal{A}(u_{0}) = \bigg(\bigcup_{u>u_{0}}\mathcal{A}(u)\bigg)''
\end{equation}
where, since the union of von Neumann algebras need not be a von Neumann algebra, we complete the algebra by taking the double commutant. Therefore, a natural definition of the limiting algebra as $u_{0}\to -\infty$ is simply  
\begin{equation}
\label{eq:Ainf}
\mathcal{A}(-\infty) \equiv \bigg(\bigcup_{u}\mathcal{A}(u)\bigg)''\,.
\end{equation}
Its commutant is equivalent to 
\begin{equation}
\label{eq:Ainfprime}
\mathcal{A}(-\infty)'=  \bigcap_{u}\mathcal{A}(u)'=\bigcap_{u}\mathcal{A}(\mathscr{I}_{<u})\,,
\end{equation}
the intersection of all radiation operators over the intervals $(-\infty,u)$ of $\mathscr{I}^{+}$ as $u\to -\infty$. This algebra is trivial --- i.e., it consists only of multiples of the identity $\hat{1}$. Its commutant is $\mathcal{B}(\mathcal{H})$, the algebra of all bounded operators on the Hilbert space. Therefore, we have that 
\begin{equation}
\mathcal{A}(-\infty) = \mathcal{B}(\mathcal{H})=\mathcal{A}(i^{0})\,,
\end{equation}
the algebra in the limit as $u\to -\infty$ is equivalent to the algebra at spatial infinity.

\subsection{Spacetimes with black holes} \label{subsec:schw_alg}
We now consider the algebra $\mathcal{A}(u_{0})$ in spacetimes that contain a stationary black hole. As when considering algebras at spatial infinity, we will focus primarily on the differences from Minkowski and on Schwarzschild black holes. We comment briefly on the generalization of the construction to, e.g., Reissner-Nordström or Kerr at the end of this subsection.

\begin{figure}
    \centering
    \includegraphics[width=1\linewidth]{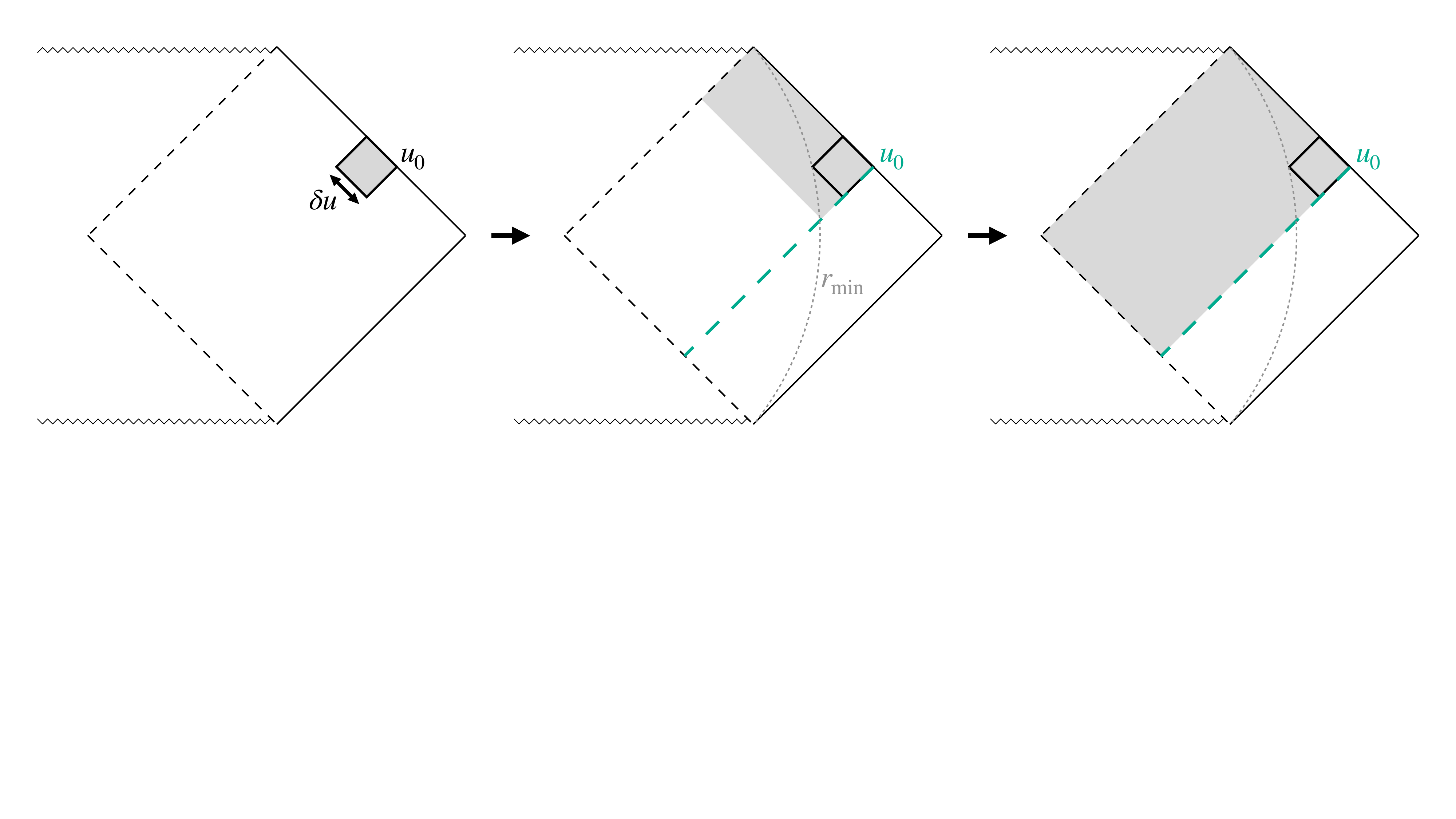}
    \vspace{-4.6cm}
    \caption{The algebra of observables associated to a subregion near asymptotic infinity in a Schwarzschild background. 1) We start from a subregion anchored to $\scri^+$ at $u_0$. 2) We evolve with the Bondi mass and apply the timelike tube theorem. 3) We evolve with the Bondi mass a second time.}
    \label{fig:schw4}
\end{figure}

The construction closely parallels the black hole algebra at spatial infinity discussed in sec.~\ref{subsec:i0bh}. We use the same enlarged Hilbert space $\mathcal H\otimes L^2(\mathbb R)_{\mathrm L}$, where $\H$ is the vacuum-at-infinity Hilbert space introduced in sec.~\ref{subsec:Hilbreg}, and the same right-boundary dressing of the QFT and graviton observables. The role played by the right ADM mass at spatial infinity is now played by the family of regularized Bondi masses whose defining spheres and smearing intervals lie in the neighborhood of the cut. Thus, for every $\delta u>0$ and $r_{\min}>r_{\mathrm{BH}}$, we obtain a neighborhood algebra $\tilde{\mathcal A}_{\delta u,r_{\min}}(u_0)$ on $\mathcal H\otimes L^2(\mathbb R)_{\mathrm L}$. Following sec.~\ref{subsec:construct}, we therefore define the algebra $\tilde{\mathcal{A}}(u_{0})$ associated to a finite cut $u=u_0$ as the intersection of these neighborhood algebras,
\begin{equation}
\label{eq:intersectionBH}
    \tilde{\mathcal{A}}(u_{0}) = \bigcap_{\rmin > r_{\textrm{BH}},\, \delta u} \tilde{\mathcal{A}}_{\delta u,r_{\textrm{min}}}(u_{0})
\end{equation}
which acts on $\mathcal{H}\otimes L^{2}(\mathbb{R})_{\textrm{L}}$. We first show that $\tilde{\mathcal{A}}(u_{0})$ is independent of $\delta u$ and $r_{\textrm{min}}$. Starting with the algebra $\tilde{\mathcal{A}}_{\delta u,r_{\textrm{min}}}(u_{0})$ we first evolve the operators to the future with the collection of Bondi mass operators in the algebra and then apply the timelike tube theorem. This procedure is identical to the first two steps of sec.~\ref{subsec:algnoblackholes} as shown in the second panel of Fig.~\ref{fig:schw4}. This step yields all operators in a timelike envelope that (at sufficiently late times) extends all the way down to the black hole horizon. 

In Minkowski space, after this step we were essentially done: the domain of dependence of the timelike envelope was the full wedge $\mathcal{A}(\mathcal{W}_{u_{0}})$ spacelike to the cut $u_0$, and the algebra $\mathcal{A}(u_{0})$ was just the QFT algebra $\mathcal{A}(\mathcal{W}_{u_{0}})$. For a black hole, we have one more step. Consider an arbitrary operator $a_0$ in the right exterior spacelike to $u_0$. Because it is spacelike to $u_0$, we can evolve it forwards in time using a regularized Bondi mass $\hat M_{\textrm{B}}^{(2)}(f;u^{\prime},r^{\prime})$ for some sufficiently large time $t$ that
\begin{align}
    a_{1} = e^{\ii t \hat M_{\rm{B}}^{(2)}(f;u^{\prime},r^{\prime})} a_{0}e^{-\ii t \hat M_{\textrm{B}}^{(2)}(f;u^{\prime},r^{\prime})}
\end{align}
is contained in the timelike envelope. However, it follows immediately that
\begin{align}
    a_{0} = e^{-\ii t \hat M_{\rm{B}}^{(2)}(f;u^{\prime},r^{\prime})} a_{1}e^{\ii t \hat M_{\textrm{B}}^{(2)}(f;u^{\prime},r^{\prime})}
\end{align}
where the right-hand side is now a product of operators that we already showed were contained in $\tilde{\mathcal{A}}_{\delta u, r_{\textrm{min}}}(u_{0})$. It follows that $\tilde{\mathcal{A}}_{\delta u, r_{\textrm{min}}}(u_{0})$ contains all operators in the intersection $\mathcal{R}\cap \mathcal{W}_{u_{0}}$ of the right wedge with the spacelike wedge associated to the cut $u=u_{0}$. In particular, this means that $\tilde{\mathcal{A}}_{\delta u, r_{\textrm{min}}}(u_{0})$ is independent of both $\delta u$ and $r_{\textrm{min}}$ and hence is also equal to the intersection algebra $\tilde{\mathcal{A}}(u_{0})$.

We now show that $\tilde{\mathcal{A}}_{\delta u, r_{\textrm{min}}}(u_{0})$ contains no further QFT or graviton operators. Let $a$ be any QFT or graviton operator contained in $\tilde{\mathcal{A}}_{\delta u,r_{\textrm{min}}}(u_{0})$. Since the defining generators of the algebra are dressed to the right boundary, they commute with all QFT observables in $\mathcal{A}(\mathcal{L})$ and so $a\in \mathcal{A}(\mathcal{R})$. Moreover, we again have that the radiation operators to the past of the cut $u=u_{0}$ commute with both the QFT and Bondi mass observables in $\tilde{\mathcal{A}}_{\delta u,r_{\textrm{min}}}(u_{0})$
\begin{equation}
    \mathcal{A}(\mathscr{I}^{+}_{<u_{0}}) \subset \tilde{\mathcal{A}}_{\delta u,r_{\textrm{min}}}(u_{0})^{\prime}\,.
\end{equation}
Finally we note that, by relative Haag duality across the cut, we also have $\mathcal{A}(\mathcal{R})\cap \mathcal{A}(\mathscr{I}^{+}_{<u_{0}})^{\prime}=\mathcal{A}(\mathcal{R}\cap \mathcal{W}_{u_{0}})$. We therefore conclude that, indeed, $a\in \mathcal{A}(\mathcal{R}\cap \mathcal{W}_{u_{0}})$. Since the preceding timelike tube argument showed that every QFT and graviton observable in $\mathcal{R}\cap \mathcal{W}_{u_{0}}$ lies in $\tilde{\mathcal{A}}_{\delta u,r_{\textrm{min}}}(u_{0})$ we obtain 
\begin{equation}
    \tilde{\mathcal{A}}_{\delta u,r_{\textrm{min}}}(u_{0}) \cap \mathcal{A}(\mathcal{R}) = \mathcal{A}(\mathcal{R}\cap \mathcal{W}_{u_{0}})\,.
\end{equation}

We now include the regularized Bondi mass generators. We consider the algebra $\tilde{\mathcal{A}}(\mathcal{R}\cap \mathcal{W}_{u_{0}})$, which includes all Bondi mass operators in $\mathcal{R}\cap \mathcal{W}_{u_{0}}$,
\begin{equation}
\label{eq:AWcapR}
    \tilde{\mathcal{A}}(\mathcal{R}\cap \mathcal{W}_{u_{0}}) \equiv \left(\left\{\delta \hat{\phi}(f), \hat{\gamma}(f), \hat{M}_{\textrm{B}}^{(2)}(f;u,r)~\forall~ (u,r)\in\mathcal{R}\cap \mathcal{W}_{u_{0}}\right\}\right)''\,,
\end{equation}
where $\operatorname{supp}(f)\subset \mathcal{R}\cap\mathcal{W}_{u_{0}}$. Since every regularized Bondi mass operator in $\tilde{\mathcal{A}}(\mathcal{R}\cap \mathcal{W}_{u_{0}})$ differs from a regularized Bondi mass operator in $\tilde{\mathcal{A}}_{\delta u,r_{\textrm{min}}}(u_{0})$ by QFT and graviton operators in $\mathcal{A}(\mathcal{R}\cap \mathcal{W}_{u_{0}})$, it follows that 
$\tilde{\mathcal{A}}(\mathcal{R}\cap \mathcal{W}_{u_{0}})\subset \tilde{\mathcal{A}}_{\delta u,r_{\textrm{min}}}(u_{0})$. 

By construction, we already have the reverse inclusion $ \tilde{\mathcal{A}}_{\delta u,r_{\textrm{min}}}(u_{0})\subset \tilde{\mathcal{A}}(\mathcal{R}\cap \mathcal{W}_{u_{0}})$, so we find $\tilde{\mathcal{A}}_{\delta u,r_{\textrm{min}}}(u_{0}) = \tilde{\mathcal{A}}(\mathcal{R}\cap \mathcal{W}_{u_{0}})\,.$ 
Therefore, taking the intersection of these algebras for all $\delta u$ and $r_{\textrm{min}}$, we obtain
\begin{equation} \label{eq:Au0BH}
    \tilde{\mathcal{A}}(u_{0}) = \tilde{\mathcal{A}}(\mathcal{R}\cap \mathcal{W}_{u_{0}})\,.
\end{equation}

Finally we note that the above arguments straightforwardly generalize to any (subextremal) static black hole spacetime. For example, an identical procedure can be applied to a Reissner-Nordström black hole. For stationary but not static black hole spacetimes such as Kerr, there is a further complication due to the fact that the asymptotically timelike Killing field $t^{a}$ becomes spacelike in the ergoregion near the horizon and so one cannot use the Bondi mass to evolve the near-horizon algebra into the timelike tube. To obtain these observables, we must additionally include the asymptotic angular momentum associated to the axial Killing field $\psi^{a}$ in the algebra. While the angular momentum on a cut of $\mathscr{I}^{+}$ suffers from the same divergences as the Bondi mass we found in sec.~\ref{subsec:Bondimassatscri}, one can construct a densely defined ``regularized Bondi angular momentum'' in the bulk of the spacetime by simply repeating the construction of sec.~\ref{subsec:RegularizedBondiMass} with $t^{a}$ replaced by $\psi^{a}$. The orbits of the Killing field $t^{a} + \Omega_{\textrm{H}}\psi^{a}$, where $\Omega_{\textrm{H}}$ is the angular velocity of the horizon, are null on the horizon and timelike in the ergoregion near the horizon. Therefore, we can repeat the above argument for operators in the ergoregion by, instead, evolving with a linear combination of the regularized Bondi mass and the regularized Bondi angular momentum. Hence, for any stationary black hole, the algebra $\tilde{\mathcal{A}}(u_{0})$ is given by \eqref{eq:Au0BH}.

\subsubsection*{Properties and type of the algebra}
\label{subsubsec:proptypBH}
We now use the structure of its commutant to show that $\tilde\A(u_0)$ is a Type III$_1$ factor. As before, the commutant contains the algebra of radiation observables $\mathcal{A}(\mathscr{I}^{+}_{u<u_{0}}) \subset \tilde{A}(\mathcal{R})$ on $\mathcal{H}$ to the past of the $u=u_{0}$ cut, which is a Type III$_{1}$ factor. We note that this is the only subalgebra of $\tilde{\mathcal{A}}(\mathcal{R})$ that lies in the commutant of $\tilde{\mathcal{A}}(u_{0})$. Therefore the commutant satisfies
\begin{equation}
\label{eq:commutantRninL}
    \tilde{\mathcal{A}}(u_{0})' \cap \tilde{\mathcal{A}}(\mathcal{R}) = \mathcal{A}(\mathscr{I}^{+}_{u<u_{0}})\,.
\end{equation}
Additionally, if we denote the algebra $\tilde{\mathcal{A}}(\mathcal{L})$ as the observables in the left wedge $\mathcal{L}$ that commute with the right ADM mass $\delta^{2}\hat{M}_{i^{0}}^{\textrm{R}}$, then this algebra is also in the commutant of  $\tilde{\mathcal{A}}(u_{0})$. The full commutant can therefore be written as
\begin{equation}
\label{eq:commutant2}
    \tilde{\mathcal{A}}(u_{0})^{\prime} = \tilde{\mathcal{A}}(\mathcal{L}) \vee \mathcal{A}(\mathscr{I}^{+}_{u<u_{0}}) \equiv (\tilde{\mathcal{A}}(\mathcal{L}) \cup\mathcal{A}(\mathscr{I}^{+}_{u<u_{0}}))'' \,.
\end{equation}
The algebra $\tilde{\mathcal{A}}(\mathcal{L})$ was constructed in \cite{Chandrasekaran:2022eqq, Kudler-Flam:2023qfl,Chen:2024rpx}, and it was shown that this algebra is a Type II$_{\infty}$ factor. We will comment on this algebra further in the following subsection. 

It follows from the proof in appendix~\ref{app:split} that $\tilde\A(\L)$ and $\A(\scri^{+}_{u<u_{0}})$ satisfy a split inclusion. This allows us to write the commutant of $\tilde\A(u_0)$ as
\begin{equation}
    \tilde\A(u_0)' = \tilde\A(\L)\vee\A(\scri^{+}_{u<u_{0}}) \cong \tilde\A(\L)\;\overline{\otimes}\;\A(\scri^{+}_{u<u_{0}})\,.
\end{equation}
The tensor product of a Type III$_1$ factor with any factor, and in particular with a Type II$_\infty$ factor, results in a Type III$_1$ factor \cite{Connes:1973hg, Connes1977}. Then $\tilde\A(u_0)'$ is a Type III$_1$ factor, and we hence conclude that $\tilde\A(u_0)$ is also a Type III$_1$ factor.

As before, the Type III nature of $\tilde{\mathcal{A}}(u_{0})$ has a simple, physical origin. The algebra is strongly entangled with radiative degrees of freedom to the past of $u=u_{0}$ and the Hilbert space therefore fails to factorize across the cut. This is a characteristic feature of Type III algebras. The above argument illustrates that the additional entanglement with the left wedge does not change this property. 

We now consider the ``limit'' of the semiclassical algebras $\tilde{\mathcal{A}}(u_{0})$ at asymptotically early times and their relationship to the algebra $\tilde{\mathcal{A}}(i^{0})$ of sec.~\ref{subsec:i0}. By the properties established at the beginning of this section, the holographic algebras again satisfy 
\begin{equation}
\tilde{\mathcal{A}}(u_{1})\subset \tilde{\mathcal{A}}(u_{0})
\end{equation}
for any $u_{1}>u_{0}$. Using this nesting property, we define the limiting algebra at past infinity in a similar manner to \eqref{eq:Ainf}
\begin{equation}
\tilde{\mathcal{A}}(-\infty) \equiv \bigg(\bigcup_{u}\tilde{\mathcal{A}}(u)\bigg)''\,.
\end{equation} 
Using \eqref{eq:commutant2}, the commutant is given by
\begin{equation}
\label{eq:Adressinfprime}
\tilde{\mathcal{A}}(-\infty)'=\bigcap_{u}\big[\tilde{\mathcal{A}}(\mathcal{L})\;\overline{\otimes} \; \mathcal{A}(\mathscr{I}_{<u})\big]=\tilde{\mathcal{A}}(\mathcal{L})\;\overline{\otimes} \;\bigg[ \bigcap_{u}\mathcal{A}(\mathscr{I}_{<u})\bigg]\,.
\end{equation}
As before, the term in the brackets is a trivial algebra and so we find that the limiting algebra at asymptotically early times is indeed equivalent to the algebra at spatial infinity
\begin{equation}
\label{eq:Adressinfprime2}
\tilde{\mathcal{A}}(-\infty)'=\tilde{\mathcal{A}}(\mathcal{L})\implies  \tilde{\mathcal{A}}(-\infty)= \tilde{\mathcal{A}}(i^{0})\,.
\end{equation}

\subsubsection*{Entropy and Type II algebras}
We conclude this section by considering the entropy of the holographic algebras. The algebra $\tilde{\mathcal{A}}(i^{0})=\tilde{\mathcal{A}}(\mathcal{R})$ of spatial infinity is Type II$_{\infty}$. For such algebras, one can construct a trace which is defined up to a state-independent multiplicative constant. Given a choice of trace and a state $\hat{\Psi}\in \mathcal{H}\otimes L^{2}(\mathbb{R})$, one can define a density matrix $\rho_{\hat{\Psi}}$ by 
\begin{equation}
\braket{a}_{\hat{\Psi}} = \textrm{tr}(a\rho_{\hat{\Psi}})
\end{equation}
where we note that $\textrm{tr}(\cdot)$ is {\em not} the Hilbert space trace but a ``renormalized'' trace that is faithful, normal, and semifinite on $\tilde{\mathcal{A}}(\mathcal{R})$. 
For ``semiclassical states'' $\ket{\hat{\Psi}} = \ket{\psi}\otimes \ket{f}$ where $\ket{\psi}\in \mathcal{H}$ and $f\in L^{2}(\mathbb{R})$ is a slowly varying function of $\delta^{2}\hat{M}_{i^{0}}^{\textrm{L}}$ --- i.e., it is sharply peaked in the conjugate ``time'' variable --- the von Neumann entropy of the density matrix $\rho^{\textrm{R}}_{\hat{\Psi}}$ was shown to be equivalent to the generalized entropy \cite{Chandrasekaran:2022eqq, Kudler-Flam:2023qfl, Chen:2024rpx, Klinger:2026tws}
\begin{equation}
\label{eq:vonneumann}
S_{\textrm{vN}}(\rho^{\textrm{R}}_{\hat{\Psi}}) = - \textrm{tr}(\rho^{\textrm{R}}_{\hat{\Psi}}\log(\rho^{\textrm{R}}_{\hat{\Psi}})) \approx S^{(\mathcal{R})}_{\textrm{gen}}[\hat{\Psi}] + C + \dots 
\end{equation}
where the first term is the generalized entropy 
\begin{equation}
\label{eq:Sgen}
 S^{(\mathcal{R})}_{\textrm{gen}}[\hat{\Psi}]\equiv \frac{\braket{A_{\mathrm{BH}}}_{\hat{\Psi}}}{4G_{\textrm{N}}}+S^{(\mathcal{R})}_{\textrm{vN}}(\rho_{\psi})\,,
\end{equation}
$C$ is a state independent, divergent constant, and the $\dots$ indicate terms that depend on the wavefunction $f$.\footnote{A formula for the density matrix for state $\ket{\hat{\Psi}} = \ket{\psi}\otimes \ket{f}$ for any $f$ was provided in \cite{Jensen:2023yxy} and the precise control over the neglected terms in \eqref{eq:vonneumann} was obtained in \cite{Kudler-Flam:2023hkl}.} As emphasized in \cite{Susskind:1994sm}, the generalized entropy is well-defined even if the individual terms in \eqref{eq:Sgen} are ill-defined in semiclassical quantum gravity. 

We now consider the entropy of states on the algebras $\tilde{\mathcal{A}}(u_{0})$. Since the algebra is Type III$_{1}$ the von Neumann entropy of any quantum state is divergent. However, as we will now explain, the problematic divergence only involves vacuum entanglement near the cut $u=u_0$ at null infinity. The modes near the horizon behave like in a Type II$_\infty$ algebra, with a divergent but renormalizable entropy that equals the generalized entropy for semiclassical states. 

The easiest way to see this is to recall that the commutant algebra has the form 
\begin{equation} \label{eq:commutantII}
    \tilde{\mathcal{A}}(u_{0})' =\tilde{\mathcal{A}}(\mathcal{L}) \;\overline{\otimes}\; \mathcal{A}(\mathscr{I}^{+}_{u<u_{0}})\,,
\end{equation}
where $\tilde{\mathcal{A}}(\mathcal{L})$ consists of left-exterior operators dressed to the left boundary, while $\mathcal{A}(\mathscr{I}^{+}_{u<u_{0}})$ consists of early-time radiation operators at null infinity. The latter algebra is Type III$_1$ while the former is Type II$_\infty$. As a result, the Type III$_1$ entanglement of $\tilde{\mathcal{A}}(u_{0})$ can only come from entanglement with $\mathcal{A}(\mathscr{I}^{+}_{u<u_{0}})$, i.e. from vacuum entanglement of radiation modes near $u=u_0$. Moreover, just like for $\tilde{\mathcal{A}}(\mathcal{R})$, the renormalized entropy of $\tilde{\mathcal{A}}(\mathcal{L})$ is equal, for semiclassical states, to the generalized entropy $S^{(\mathcal{L})}_{\textrm{gen}}$ of the left exterior. 

We consider this renormalized entropy on $\tilde{\mathcal{A}}(\mathcal{L})$ and  regularize the entropy of $\mathcal{A}(\mathscr{I}^{+}_{u<u_{0}})$ in the conventional manner for QFT algebras. While we will give a more precise description of these ``cutoff algebras'' shortly, the key point is that, in any regularization scheme,  we obtain a semiclassical entropy on $\tilde{\mathcal{A}}(\mathcal{L}) \otimes \mathcal{A}(\mathscr{I}^{+}_{u<u_{0}})$ that is equal to the ``generalized entropy''
\begin{equation}\label{eq:sgenLcupu<u0}
 S^{(\mathcal{L} \cup \mathscr{I}^{+}_{u<u_{0}})}_{\textrm{gen}}[\hat{\Psi}]\equiv \frac{\braket{A_{\mathrm{BH}}}_{\hat{\Psi}}}{4G_{\textrm{N}}}+S^{(\mathcal{L} \cup \mathscr{I}^{+}_{u<u_{0}})}_{\textrm{vN}}(\rho_{\psi})\,.
\end{equation}
Unlike a true generalized entropy, this is not expected to be UV-finite, because there is no area term to cancel the divergences from the cut at $u=u_0$. However, it does include the Bekenstein-Hawking entropy associated with the black hole horizon.

To obtain an entropy associated to $\mathcal{R}\cap \mathcal{W}_{u_{0}}$, we first consider the case where $\hat{\Psi}$ is a pure state. Since the entropy of a pure state on a von Neumann algebra is equal to the entropy of the same state on its commutant then, if we appropriately ``regularize'' the contribution from the entanglement at the cut $u=u_{0}$, the entropy of semiclassical states on $\tilde{\mathcal{A}}(u_{0})$ should also be equal to \eqref{eq:sgenLcupu<u0}, or equivalently to 
\begin{equation}\label{eq:sgenRcapu>u0}
 S^{(\mathcal{R}\cap \mathcal{W}_{u_{0}})}_{\textrm{gen}}[\hat{\Psi}]\equiv \frac{\braket{A_{\mathrm{BH}}}_{\hat{\Psi}}}{4G_{\textrm{N}}}+S^{(\mathcal{R}\cap \mathcal{W}_{u_{0}})}_{\textrm{vN}}(\rho_{\psi})\,.
\end{equation}
By purifying the state using an arbitrary reference system, one can confirm that \eqref{eq:sgenRcapu>u0} continues to hold even for mixed states on $\mathcal{H} \otimes L^2(\mathbb{R})$.

We can make the meaning of \eqref{eq:sgenRcapu>u0} more precise by regulating the $ \mathcal{A}(\mathscr{I}^{+}_{u<u_{0}})$ using the split property. We start by considering a larger algebra $\mathcal{A}(\mathcal{W}_{\textrm{R}}[u_{1},r_{1}])\supset \mathcal{A}(\mathscr{I}^{+}_{u<u_{0}})$ where $u_{1}>u_{0}$ and $r_{1}\gg r_{\textrm{BH}}$, and we recall that $\mathcal{W}_{\textrm{R}}[u_{1},r_{1}]$ is the region of larger radius and spacelike separated from the sphere $S(u_{1},r_{1})$ (see Fig.~\ref{fig:cauchy}). Since these algebras are nested with a spacelike collar, the results of appendix \ref{app:split} imply that there exists a Type I factor $\mathcal{N}_{u_{1},r_{1}}$ such that 
\begin{equation}
    \A(\scri^+_{u<u_0}) \subset \N_{u_1,r_1} \subset \A(\mathcal W_{\rm R}[u_1,r_1])\,.
\end{equation}
The algebra $\mathcal{N}_{u_{1},r_{1}}$ regulates the Type III entanglement across the cut while still approximating\footnote{The algebra $\mathcal{N}_{u_{1},r_{1}}$ approximates $\A(\scri^+_{u<u_0})$ in the following sense: If $S_{i}(u_{i},r_{i})$ are a sequence of spheres that approach the $u=u_{0}$ cut of $\mathscr{I}^{+}$ then we can obtain a nested family of Type I factors $\mathcal{N}_{u_{i},r_{i}}\supset  \A(\scri^+_{u<u_0})$ which ``limit'' to $\A(\scri^{+}_{u<u_0})$. More precisely, the intersection over this family of algebras yields  $\bigcap_{i}\mathcal{N}_{u_{i},r_{i}} =  \A(\scri^{+}_{u<u_0})$.} the radiation modes on $\mathscr{I}^{+}_{<u_{0}}$ in the limit as $u_{1}$ approaches $u_{0}$ for arbitrarily large $r_{1}$. The algebra 
\begin{equation}
\tilde{\mathcal{A}}(\mathcal{L})\:\overline{\otimes} \:\mathcal{N}_{u_{1},r_{1}}
\end{equation}
is Type II$_{\infty}$ and so \eqref{eq:sgenLcupu<u0} describes the renormalized entropy on this algebra. The commutant algebra $\left(\tilde\A(\L)\;\overline\otimes\; \N_{u_1,r_1}\right)'$ is also Type II$_\infty$ and is a regularization of $\tilde{\mathcal{A}}(u_{0})$. It is this regularized algebra that has a renormalized entropy given by \eqref{eq:sgenRcapu>u0}.

\subsection{(Ultra)locality at null infinity} \label{subsec:partial}
We conclude this section with a very brief discussion of the local spatial structure at null infinity and its consequences. Thus far in this paper, we have focused on the algebra associated with a full cut of null infinity. We found that this algebra is non-trivial and equivalent to a certain bulk algebra deep in the interior of the spacetime. Although we focus on cuts at a fixed $u=u_0$, all the results of this paper straightforwardly generalize to cuts $\mathcal{C}_{u(x^{A})}$ where the retarded time varies across the generators of $\mathscr{I}^{+}$. The nesting property becomes 
\begin{equation}
\A({u_1 (x^{A})}) \subseteq \A({u_0 (x^{A})})
\end{equation}
if $u_1(x^{A}) \geq u_0(x^{A})$ for all $x^A$.

\begin{figure}
    \centering
    \includegraphics[width=.7\linewidth]{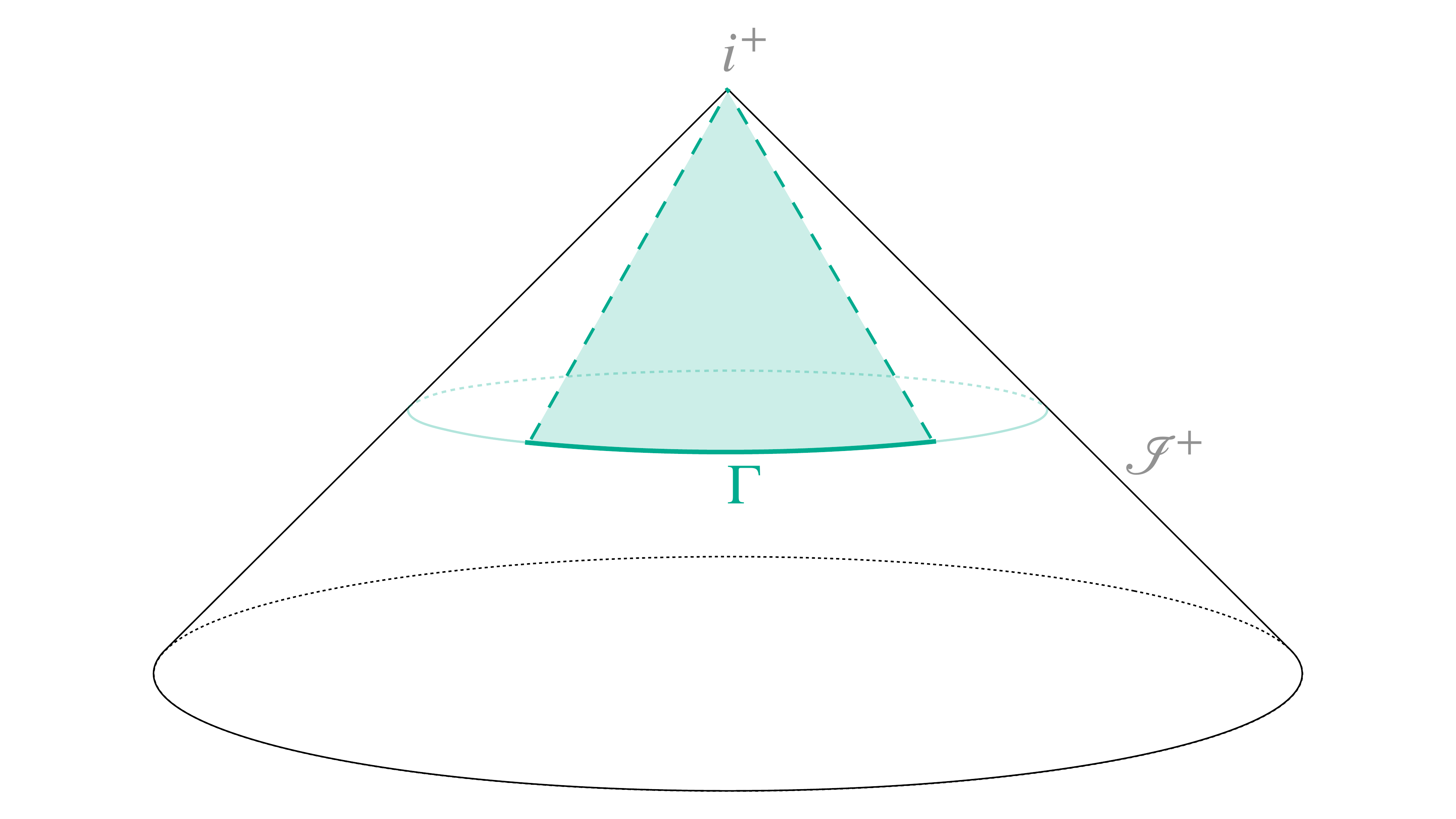}
    \vspace{-.1cm}
    \caption{A partial cut $\Gamma \subset \mathcal{C}_{u_0}$ of $\mathscr I^+$. Because the classical time-translation generators are ultralocal, the algebra localized on $\Gamma$ reconstructs only the future tail $\mathscr I^+_{\Gamma,>u_0}$, while its bulk entanglement wedge degenerates to the boundary without covering any open bulk region.}
    \label{fig:partial}
\end{figure}

It is also natural to ask what algebra and bulk region are associated with only a portion $\Gamma \subset \mathcal{C}_{u_{0}}$ of a cut of $\mathscr{I}^{+}$ --- i.e., a partial cut which contains only the generators of $\mathscr{I}^{+}$ within some angular window as depicted in Fig.~\ref{fig:partial}. 
Classically, the time-translation generators are ultralocal. For any smooth function $f(x^{A}-\bar{x}^{A})$ centered about a generator at $\bar{x}^{A}$, we can define a ``supertranslation charge'' $M_{\textrm{B}}(f;u_{0},\bar{x}^{A})$ which is localized in an arbitrarily small angular neighborhood of that generator and generates translations within only this neighborhood \cite{Wald:1999wa}. We expect that the supertranslation charges $M(f;u_{0},\bar{x}^{A})$ can be regularized analogously to the full Bondi mass $M_{\textrm{B}}(u_{0})$ to obtain a densely defined quantum operator localized within an arbitrary bulk neighborhood $\mathcal{R}_{\Gamma}$ of $\Gamma$.\footnote{The supertranslation charges at $\mathscr{I}^{+}$ can similarly be related to supertranslation charges at $i^{0}$ which are also conserved \cite{Strom1:2014,Strom2:2014,Prabhu:2019fsp,Magdy:2023diu}. In black hole spacetimes, the charges at $i^{0}$ must be quantized in the manner described in \cite{Klinger:2026tws}.} It follows that the algebra associated to the partial cut $\Gamma$ reconstructs the entire ``future tail'' $\mathscr{I}^{+}_{\Gamma,>u_{0}}$. This reflects the Carrollian character of time evolution at null infinity. 

There is, however, no corresponding open, bulk region. The future light cone of any bulk point intersects the generators of $\mathscr{I}^{+}$ outside of any proper angular subregion $\Gamma$. Hence, a local bulk operator is causally visible in the complementary portion of null infinity. Since the theory at $\mathscr{I}^{+}$ is ultralocal, the operators on this portion of null infinity commute with all operators that can be measured in $\Gamma$ alone. The entanglement wedge of $\Gamma$ therefore degenerates to its future tail on $\mathscr{I}^{+}$, with no nontrivial bulk interior. This differs sharply from AdS/CFT, where a proper boundary subregion has a nontrivial bulk entanglement wedge bounded by a quantum extremal surface \cite{Ryu:2006bv, Hubeny:2007xt, Engelhardt:2014gca}.

\section*{Acknowledgements}
We would like to thank Jan Boruch, Jonah Kudler-Flam, Ted Jacobson, Ana-Maria Raclariu, Steve Shenker, and Edward Witten for valuable discussions. This work was supported in part by the Leinweber Institutes for Theoretical Physics at UC Berkeley and Stanford, the Princeton Gravity Initiative at Princeton University, the Department of Energy through DE-SC0019380 and DE-FOA0002563, by AFOSR award FA9550-22-1-0098, and by Sloan and Packard Fellowships. 

\appendix 
\section{The Split Property in Black Hole Spacetimes} \label{app:split}
In this appendix we will prove the split property in a Schwarzschild spacetime for the von Neumann algebras associated with two regions $\mathcal{W}_{1}$ and $\mathcal{W}_{2}$ which have disjoint support separated by a finite, spacelike distance. The split property is the statement that the (von Neumann) union of their corresponding algebras is isomorphic to a spatial tensor product 
\begin{equation}
\label{eq:split1}
\mathcal{A}(\mathcal{W}_{1}) \vee \mathcal{A}(\mathcal{W}_{2}) \cong \mathcal{A}(\mathcal{W}_{1})\;\overline{\otimes}\;\mathcal{A}(\mathcal{W}_{2}) \,.
\end{equation}
Equivalently, there exists a Type I factor $\mathcal{N}$ such that 
\begin{equation}
\label{eq:split2}
\mathcal{A}(\mathcal{W}_{1}) \subset \mathcal{N} \subset \mathcal{A}(\mathcal{W}_{2})^{\prime}\,.
\end{equation}
Roughly speaking, the split property states that there is a ``finite'' amount of entanglement between regions $\mathcal{W}_{1}$ and $\mathcal{W}_{2}$. For bounded regions with strictly positive spacelike separation, the split property is satisfied for a large class of quantum field theories \cite{Buchholz:1989zz,Verch:1993fs,Fewster:2016mzz}. In essence, the spacelike separation provides a UV cutoff and the fact that the regions are bounded provides an IR cutoff for the entanglement of the modes in $\mathcal{W}_{1}$ and $\mathcal{W}_{2}$. 

In the main text, however, we require the split property for regions $\mathcal W_1$ and $\mathcal W_2$ that (1) have disjoint support separated by a finite spacelike distance and (2) are unbounded and extend to spatial infinity. The potential issue is property (2) since a general quantum theory in an unbounded region can, in principle, contain an infinite number of infrared modes. Furthermore, while the correlations between individual modes may be finite, their cumulative contributions need not converge. In this appendix we explain that this does not occur for the vacuum-at-infinity Hilbert space $\mathcal{H}$ constructed in sec.~\ref{subsec:Hilbreg}. For simplicity, we will restrict attention to the free massless scalar in four spacetime dimensions; however, as we explain in appendix~\ref{subsubsection:highdim}, there is no essential difficulty in generalizing our arguments to higher dimensions and other massive or massless fields. Indeed, the key physical mechanism is the angular momentum barrier which ensures that all modes of the quantum field must decay towards spatial infinity. This decay is sufficiently rapid even at low angular modes such that the Hilbert space cannot support a large amount of infrared entanglement. The purpose of this appendix is to make this intuition more precise.

\subsection{Geometry and outline of the proof}
\label{subsection:geometry}
We recall that the Hilbert space of quantum fields was constructed by deforming the Schwarzschild spacetime to be stationary to the future $M_{+}$ of a Cauchy surface $\Sigma_{+}$ as well as the past $M_{-}$ of a Cauchy surface $\Sigma_{-}$ (see Fig.~\ref{fig:bh}). The basic strategy will be to first prove the split property in the deformed region of the spacetime. Then, as we explain below, we obtain the desired split property in Schwarzschild spacetime by simply evolving the split algebras to the undeformed region. We choose, for definiteness, to prove the split property in $(M_{+},g_{+})$. Furthermore, for simplicity, we will restrict attention to four spacetime dimensions. The generalization of our arguments to $d\geq 4$ spacetime dimensions is provided  in sec.~\ref{subsubsection:highdim}. 

While the Hilbert space does not depend on the choice of spacetime $(M_{+},g_{+})$, it will be convenient for our arguments to make a particular choice of extension. We choose $(M_{+},g_{+})$ so that, in the deformed  regions of the spacetime, the metric is globally static, spherically symmetric and reflection-symmetric. In other words, the metric retains all of the symmetries of Schwarzschild spacetime but does not have a horizon. The resulting spacetime $M_{+}$ is a static wormhole geometry
\begin{equation}
g_{+,ab}dx^{a}dx^{b}= -N(s)^{2}dt^{2} + ds^{2}+r(s)^{2}d\Omega^{2}
\end{equation}
where $N(s)>0$ for all $s$ and $|s|$ is the proper radial distance from the wormhole throat  at $s=0$.  The region $s>0$ parametrizes the right asymptotically flat end and the region $s<0$ parametrizes the left asymptotically flat end. The function $r(s)$ is a smooth, even function which represents the areal radius of spheres and satisfies 
\begin{equation}
\label{eq:drds}
\bigg(\frac{dr}{ds}\bigg)^{2} = 1- \bigg(\frac{r_{\textrm{BH}}}{r}\bigg), \quad \quad r(0)=r_{\textrm{BH}}
\end{equation}
where $r_{\textrm{BH}}=2M$ is the radius of the black hole. Differentiating with respect to $s$ and using \eqref{eq:drds} yields 
\begin{equation}
\frac{d^{2}r}{ds^{2}} = \frac{r_{\textrm{BH}}}{2r^{2}}>0\,.
\end{equation}
So the throat is a minimal area sphere and the area of spheres increases as $|s|$ increases away from the throat.

The lapse $N(s)$ is also a smooth, even, strictly positive function. We obtain an additional restriction on the lapse from the fact that, in the deformation construction of sec.~\ref{subsec:Hilbreg}, the deformed metric is required to be isometric to Schwarzschild outside of a radius $R_{0}>r_{\textrm{BH}}$. Therefore, we additionally have that 
\begin{equation}
N^{2} = 1-\bigg(\frac{r_{\textrm{BH}}}{r}\bigg)\,,\qquad r\geq R_{0}>r_{\textrm{BH}}\,.
\end{equation}
So the constant $M$ is equivalent to the mass of the undeformed, black hole spacetime. In summary, $(M_{+},g_{+})$ is equivalent to the Schwarzschild geometry at large $r$ but replaces the interior by a regular, static wormhole. Choosing a Cauchy surface $\Sigma_{0}$ to be a constant $t$ slice of $(M_{+},g_{+})$ and constants $s_{a},s_{b}>0$, we define 
\begin{equation}
\label{eq:SigmaLR}
\Sigma_{\rm L} = \{(s,x^{A})\in \Sigma_{0}\,|\, s<-s_{a}\}\,, \quad \quad \Sigma_{\rm R} = \{(s,x^{A})\in \Sigma_{0}\,|\, s>s_{b}\}
\end{equation}
to be partial Cauchy surfaces for the left and right exteriors of the wormhole separated by a collar region from $-s_{a}<s<s_{b}$ that contains the throat
\begin{equation}
\Sigma_{\rm C} = \{(s,x^{A})\in \Sigma_{0}\,|\,-s_{a}\leq s\leq s_{b} \}\,.
\end{equation}
The spacetime regions $\mathcal{W}_{\rm L}$ and $\mathcal{W}_{\rm R}$ are domains of dependence of $\Sigma_{\rm L}$ and $\Sigma_{\rm R}$, respectively. 

Given a fixed choice of collar region, it will be convenient for the arguments in the latter part of this appendix to make a specific choice for the lapse function. To achieve this we choose constants 
\begin{equation}
0 < s_{a},s_{b} < s_{1} < s_{0}
\end{equation}
where $s_{0}$ satisfies $r(s_{0})=R_{0}$ and so for all $|s|>s_{0}$, the spacetime is isometric to Schwarzschild.  Our choice of lapse is given by 
\begin{equation}
N(s)^{2} = (1-\chi(s)) N_{0}^{2} + \chi(s)\bigg(1- \frac{r_{\textrm{BH}}}{r(s)}\bigg)
\end{equation}
where $N_{0}>0$ and $\chi(s)$ is a smooth, even function such that $\chi(s)$ vanishes for $|s|<s_{1}$ and is equal to $1$ for $|s|>s_{0}$. In summary, we choose $N(s)=N_0$ throughout the collar region and in a neighborhood extending to proper distance $s_{1}$ from the throat, after which it smoothly transitions to the lapse of a Schwarzschild metric and agrees with it for all $|s|>s_{0}$. 

\subsubsection*{Outline of the proof}

We now give a brief outline of our proof strategy. If the algebras $\mathcal{A}(\mathcal{W}_{\textrm{L}})$ and $\mathcal{A}(\mathcal{W}_{\textrm{R}})$ satisfy \eqref{eq:split2} for a Type I factor $\N$, then there is a unitary identification 
\begin{equation}
\mathcal{H}_{\textrm{out}} \cong \mathcal{H}_{\textrm{L}} \otimes \mathcal{H}_{\textrm{R}}
\end{equation}
for which $\N=\B(\mathcal{H}_{\rm L})$, $\A(W_{\rm L})\subset\B(\mathcal{H}_{\rm L})\;\overline{\otimes}\;\boldsymbol{1}$ and $\A(W_{\rm R})\subset\boldsymbol{1}\;\overline{\otimes}\;\B(\mathcal{H}_{\rm R})$. Equivalently, the joint algebra admits a normal product state $\ket{\Omega_{\textrm{split}}}\in \mathcal{H}_{\textrm{out}}$ which satisfies
\begin{equation}
\label{eq:Omegasplit}
\langle \Omega_{\textrm{split}} | a_{\textrm{L}}a_{\textrm{R}}|\Omega_{\textrm{split}}\rangle = \langle  \Omega_{\textrm{split}}|a_{\rm L}|  \Omega_{\textrm{split}}\rangle \langle \Omega_{\textrm{split}}|a_{\rm R}|  \Omega_{\textrm{split}}\rangle 
\end{equation}
for any $a_{\textrm{L}}\in \mathcal{A}(\mathcal{W}_{\textrm{L}})$ and $a_{\textrm{R}}\in \mathcal{A}(\mathcal{W}_{\textrm{R}})$.\footnote{Strictly speaking, a state is a positive normalized functional and need not be represented by a vector in the original vacuum Hilbert space. The quasiequivalence established below implies that $\Omega_{\mathrm{split}}$ is normal in the vacuum representation; in the standard-form realization of the joint algebra, it is therefore represented by a vector in the natural cone. We use $|\Omega_{\mathrm{split}}\rangle$ to denote this vector
representative.}

\begin{figure}
    \centering
    \includegraphics[width=0.9\linewidth]{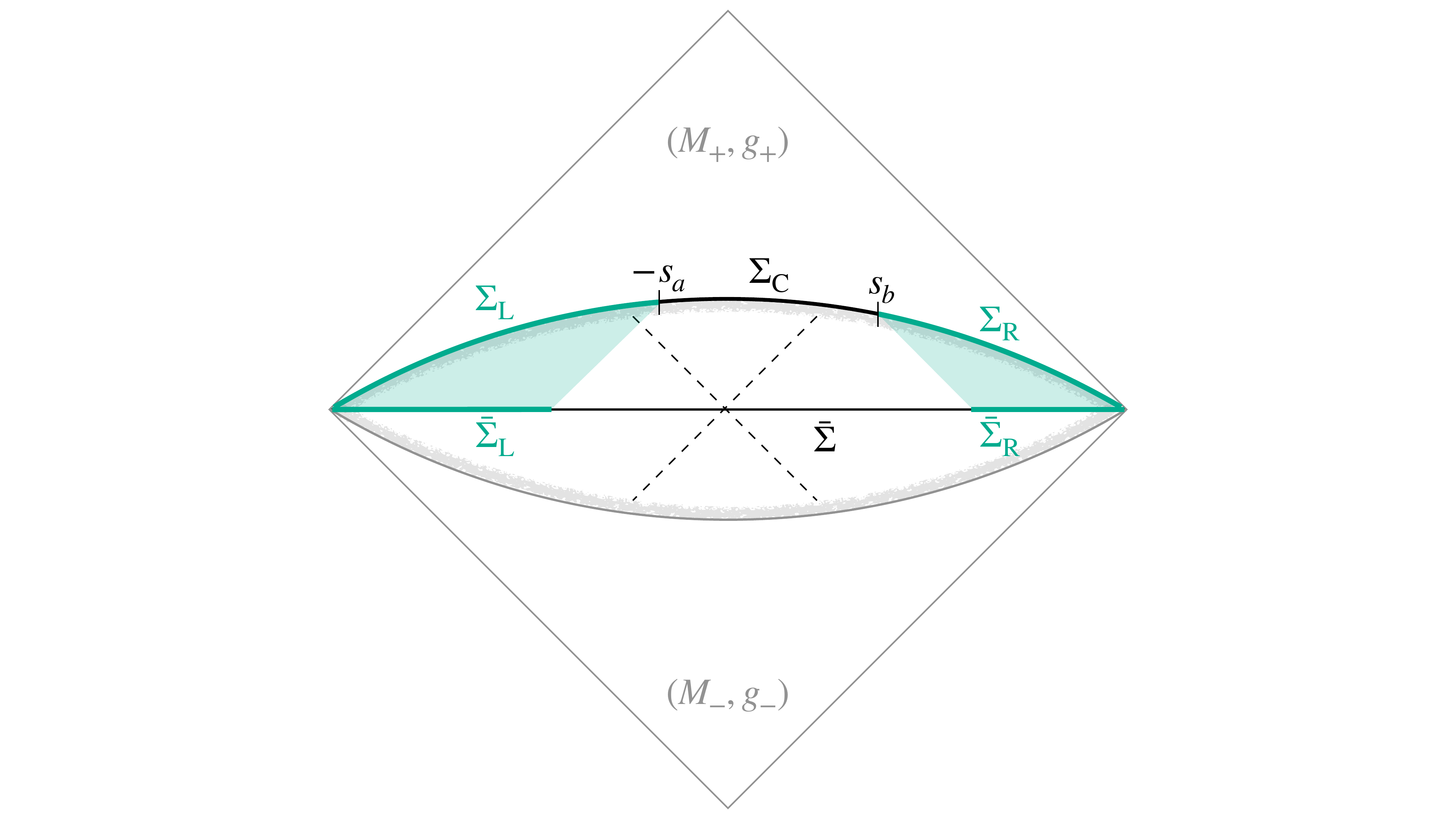}
    \vspace{-.3cm}
    \caption{ The upper static wormhole region $(M_+,g_+)$ with Cauchy surface $\Sigma_0=\Sigma_{\rm L}\cup\Sigma_{\rm C}\cup\Sigma_{\rm R}$, where $s=-s_a$ and $s=s_b$ bound the collar region $\Sigma_{\rm C}$. Unitary evolution of the split inclusion to a Cauchy surface $\bar\Sigma$ in the undeformed Schwarzschild spacetime preserves the split property for the evolved, disjoint subregions $\bar\Sigma_{\rm L}$ and $\bar\Sigma_{\rm R}$.}
    \label{fig:split}
\end{figure}

We then obtain, following \cite{Fewster:2015covariant}, the split property in the original, black hole spacetime by simply evolving the split algebras backwards in time to a  Cauchy surface $\overline{\Sigma}$ in the original, undeformed black hole spacetime (see Fig.~\ref{fig:split}). Since the regions $\Sigma_{\textrm{R}}$ and $\Sigma_{\textrm{L}}$ can be chosen so that the evolution to $\overline{\Sigma}$ preserves their disjoint spatial support and unitary time evolution preserves the split inclusion, we obtain the desired split property in black hole spacetimes for spacelike separated, infinitely extended regions $\overline{\mathcal{W}}_{\textrm{L}},\overline{\mathcal{W}}_{\textrm{R}}$ with disjoint support,
\begin{equation} \label{eq:undeformed-split}
\mathcal{A}(\overline{\mathcal{W}}_{\textrm{L}})  \subset \overline{\mathcal{N}}\subset \mathcal{A}(\overline{\mathcal{W}}_{\textrm{R}})^{\prime}\,
\end{equation}
where $\overline{\mathcal{N}}$ is the Type I factor obtained from unitary evolution of $\mathcal{N}$ from $\Sigma_{0}$ to $\overline{\Sigma}$. The main effort of this appendix will be to prove the existence of $\ket{\Omega_{\textrm{split}}}$ in $\mathcal{H}_{\textrm{out}}$. We essentially will show this by decomposing the state into modes and proving the angular momentum barrier leads to an exponential suppression in the correlations of these modes across the collar. For simplicity, we restrict attention to $d=4$ spacetime dimensions and subregions $\mathcal{W}_{\textrm{L}}$ and $\mathcal{W}_{\textrm{R}}$ on either side of the Einstein-Rosen bridge. However, as we explain in appendix~\ref{subsubsection:highdim}, our method of proof can be straightforwardly generalized to $d\geq 4$ dimensions and subregions where the collar is radially displaced into the left or right exterior. By the above arguments, if $\ket{\Omega_{\textrm{split}}}$ exists, we obtain the split property for a large class of infinitely extended regions in Schwarzschild spacetime. We will use this generalization in appendix~\ref{app:fns}.

\subsection{Quantization and quasiequivalence}
\label{eq:quantization}
To prove the split property we need to prove that $\ket{\Omega_{\textrm{split}}}$ is a normal state in $\mathcal{H}_{\textrm{out}}$. We will consider, as a prototypical example, a massless, minimally coupled scalar field in four spacetime dimensions. In this section we (1)  briefly review some key results and properties of the quantization of a scalar field on a static spacetime and (2) review  sufficient conditions for $\ket{\Omega_{\textrm{split}}}\in \mathcal{H}_{\textrm{out}}$. 

\subsubsection*{Quantization of a massless scalar field in a static spacetime}
\label{subsubsec:quantization}

In this section we briefly review the quantization of a massless, minimally coupled scalar field in the static spacetime $(M_{+},g_{+})$ described in sec.~\ref{subsection:geometry}. We refer the reader to \cite{Wald_1995,Fulling:1981cf,Hollands:2014eia} for more details. We first collect some results on classical scalar fields in static spacetimes. The scalar field satisfies 
\begin{equation}
\label{eq:EOM}
\Box_{g_{+}}\phi = 0 
\end{equation}
with conserved symplectic form 
\begin{equation}
\Omega^{\textrm{KG}}(\phi_{1},\phi_{2}) = \int_{\Sigma_{0}}N^{-1}\sqrt{h}d^{3}x~[\phi_{1}(s,x^{A})\partial_{t}\phi_{2}(s,x^{A}) - \phi_{2}(s,x^{A})\partial_{t}\phi_{1}(s,x^{A})]
\end{equation}
where, in our coordinates, $N^{-1}\sqrt{h}d^{3}x=N(s)^{-1}r(s)^{2}dsd\Omega$ where $\sqrt{h}d^{3}x$ is the volume measure of the induced metric $h_{ij}$ on $\Sigma_{0}$. In the coordinates  described in sec.~\ref{subsection:geometry}, the wave equation takes the form 
\begin{equation}
\partial_{t}^{2}\phi + H \phi = 0 
\end{equation}
where 
\begin{equation}
H = -\frac{N(s)}{r(s)^{2}}\frac{\partial }{\partial s}\bigg(N(s)r(s)^{2}\frac{\partial}{\partial s}\bigg) - \frac{N(s)^{2}}{r(s)^{2}}\Delta_{\mathbb{S}^{2}}
\end{equation}
where $\Delta_{\mathbb{S}^{2}}$ is the Laplacian on the unit $2$-sphere. 

The operator $H$ is initially defined on the space of smooth, compactly supported initial data on  $\Sigma_{0}$. On $C^{0}(\Sigma_{0})$, $H$ is non-negative with respect to the inner-product
\begin{equation}
\braket{\psi_{1},\psi_{2}}_{L^{2}(\Sigma_{0})} \equiv \int_{\Sigma_{0}}N^{-1}\sqrt{h}d^{3}x ~\overline{\psi_{1}}\psi_{2}
\end{equation}
 since, after integration by parts, one obtains the quadratic form
\begin{equation}
\label{eq:Hpos}
\braket{\psi,H\psi}_{L^{2}(\Sigma_{0})} =\int_{\Sigma_{0}} N \sqrt{h}d^{3}x~h^{ij}\overline{D_{i} \psi} D_{j}\psi =  \int_{\Sigma_{0}}N(s)\big[r(s)^{2}|\partial_{s}\psi|^{2} + |\nabla_{\mathbb{S}^{2}}\psi|^{2}\big]dsd\Omega \geq 0 
\end{equation}
for any $\psi\in C^{0}(\Sigma_{0},N^{-1}\sqrt{h}d^{3}x)$. We denote its Friedrichs extension as $H$. The operator $D_{i}$ is the covariant derivative compatible with $h_{ij}$ on $\Sigma_{0}$.  We use the same symbol $H$ for the positive, self-adjoint extension of this operator. We note that $H$ has no normalizable zero mode. Indeed, suppose there exists a $\psi$ in the domain of $H$ such that $H$ annihilates $\psi$. We have that 
\begin{equation}
H\psi = 0 \implies ||H^{1/2}\psi||^{2}_{L^{2}(\Sigma_{0})} = 0\,.
\end{equation}
Eq.~\eqref{eq:Hpos} then implies that $D_{i}\psi=0$ and, consequently, that $\psi$ cannot lie in $L^{2}(\Sigma_{0},N^{-1}\sqrt{h}d^{3}x)$. 

We now describe the quantization of the scalar field in terms of the initial data on $\Sigma_{0}$. The quantization algebra of the scalar field can be compactly  expressed in terms of the ``Weyl algebra'' 
\begin{equation}
\hat{W}(f) = e^{\ii\hat{\phi}(f)},   \quad \hat{W}(F_{1})\hat{W}(F_{2}) = e^{-\frac{\ii}{2}\Omega^{\textrm{KG}}(F_{1},F_{2})}\hat{W}(F_{1}+F_{2}),  \quad W(F)^{\ast}= W(-F)
\end{equation}
for any test functions $f$ in spacetime  and $F=Ef$ is the advanced-minus-retarded solution to \eqref{eq:EOM} with source $f$. A class of states of particular relevance to the problem at hand are (zero mean) Gaussian states which are uniquely specified by a choice of a positive, symmetric bilinear form $\mu_{\Psi}(F_{1},F_{2})$ on the space of solutions satisfying 
\begin{equation}
\label{eq:positivity}
|\Omega^{\textrm{KG}}(F_{1},F_{2})|\leq 2 \sqrt{\mu_{\Psi}(F_{1},F_{1})\mu_{\Psi}(F_{2},F_{2})}
\end{equation}
with which the Gaussian state $\ket{\Psi}$ is uniquely specified by the relation 
\begin{equation}
\label{eq:Gaussian}
\langle \Psi | \hat{W}(f)| \Psi \rangle  = e^{-\frac{1}{2}\mu_{\Psi}(F,F)}\,.
\end{equation}
In terms of the $n$-point correlation functions of $\hat{\phi}$, a zero mean Gaussian state is uniquely determined by specifying its $2$-point correlation function $\langle \Psi | \hat{\phi}(f_{1})\hat{\phi}(f_{2})|\Psi \rangle $. It is straightforward to show that the inner product $\mu_{\Psi}$ can be directly related to the symmetric part of $\langle \Psi | \hat{\phi}(f_{1})\hat{\phi}(f_{2})|\Psi \rangle $ whereas the state-independent, antisymmetric part can be related to $\Omega^{\textrm{KG}}$. The inequality \eqref{eq:positivity} ensures that the resulting state is positive.

An important example of a Gaussian state is the stationary vacuum state $\ket{\Omega_{\textrm{out}}}$ on the static spacetime $(M_{+},g_{+})$.
To describe the unique inner product $\mu_{0}$ which defines this state we note that any solution $F$ can be equivalently characterized by its initial data 
\begin{equation}
\label{eq:Fpq}
F = (q,p) \quad \quad \textrm{ where} \quad \quad q \equiv F\vert_{\Sigma_{0}} \textrm{ and } p\defn \partial_{t}F\vert_{\Sigma_{0}}\,.
\end{equation}
The inner product $\mu_{0}$ associated to $\ket{\Omega_{\textrm{out}}}$ is given by \cite{Fulling:1972md,Ashtekar:1975zn,Kay:1978yp}
\begin{equation}
\label{eq:mu0pq}
\mu_{0}(F_{1},F_{2}) =  \frac{1}{2}\big(\langle q_{1}, H^{1/2} q_{2} \rangle_{L^{2}(\Sigma_{0})}  + \langle p_{1}, H^{-1/2} p_{2} \rangle_{L^{2}(\Sigma_{0})} \big)\,.
\end{equation}
It is straightforward to show that $\mu_{0}$ satisfies \eqref{eq:positivity}.\footnote{Writing the symplectic product as  $\Omega^{\textrm{KG}}(F_{1},F_{2}) = \langle H^{1/4}q_{1}, H^{-1/4} p_{2} \rangle_{L^{2}(\Sigma_{0})}  - \langle H^{-1/4}  p_{1},H^{1/4} q_{2} \rangle_{L^{2}(\Sigma_{0})}$, \eqref{eq:positivity} follows from applying the Cauchy-Schwarz and triangle inequalities.}

The inner product $\mu_{0}$ plays an important structural role in the construction of the GNS Hilbert space. In particular, on the space of smooth, compactly supported initial data we consider
\begin{equation}
\langle F_{1}, F_{2}\rangle_{\mathcal{H}_{1}} \equiv  \mu_{0}(F_{1},F_{2}) + \frac{\ii}{2}\Omega^{\textrm{KG}}(F_{1},F_{2})\,.
\end{equation}
Quotienting by any null states and taking the completion yields a one-particle Hilbert space $\mathcal{H}_{1}$. The full GNS Hilbert space is the symmetric Fock space $\mathcal{H}_{\textrm{out}} \cong \mathcal{F}(\mathcal{H}_{1})\equiv  \mathbb{C} \oplus \bigoplus_{n=1}^{\infty}\mathcal{H}_{n} $
where $\mathcal{H}_{n}$ is the $n$-fold, symmetric tensor product of $\mathcal{H}_{1}$.

\subsubsection*{A sufficient condition for quasiequivalence }
\label{subsubsec:quasiequiv}
In the above analysis we considered the global algebra and Hilbert space associated to a Cauchy surface $\Sigma_{0}$. In this section we consider the restriction of the vacuum state to the local algebra $\mathcal{A}(\mathcal{W}_{\textrm{L}})\vee \mathcal{A}(\mathcal{W}_{\textrm{R}})$ which is obtained by quantizing the space of solutions with compactly supported data in $\Sigma_{\textrm{L}}\cup\Sigma_{\textrm{R}}$. Any such solution has a unique decomposition
\begin{equation}
F=F_{\textrm{L}}+F_{\textrm{R}}\,,
\end{equation}
where $F_{\textrm{L}}$ and $F_{\textrm{R}}$ are the solutions obtained by
evolving the corresponding left- and right-supported initial data. 
The symplectic form for initial data on these disjoint subregions of $\Sigma_{0}$ is 
\begin{equation}
    \Omega^{\textrm{KG}}(F_{1},F_{2}) = \Omega^{\textrm{KG}}(F_{1,\rm{L}},F_{2,\rm{L}}) + \Omega^{\textrm{KG}}(F_{1,\rm{R}},F_{2,\rm{R}})
\end{equation}
and so the data on $\Sigma_{\textrm{R}}$ and $\Sigma_{\textrm{L}}$ are symplectically orthogonal. By contrast, the decomposition of $\mu_{0}$ yields
\begin{equation}
\mu_{0}(F_{1},F_{2}) = \mu_{0}(F_{1,\rm{L}},F_{2,\rm{L}}) + \mu_{0}(F_{1,\rm{R}},F_{2,\rm{R}}) + \mu_{0}(F_{1,\rm{L}},F_{2,\rm{R}}) + \mu_{0}(F_{1,\rm{R}},F_{2,\rm{L}})
\end{equation}
where the cross-terms encode the fact that the vacuum state is correlated between $\Sigma_{\textrm{L}}$ and $\Sigma_{\textrm{R}}$. The split state is simply defined as a Gaussian state on $\mathcal{A}(\mathcal{W}_{\textrm{L}})\vee \mathcal{A}(\mathcal{W}_{\textrm{R}})$ with no correlations between $\mathcal{W}_{\textrm{L}}$ and $\mathcal{W}_{\textrm{R}}$. In other words, its inner product $\mu_{\textrm{split}}$ is given by 
\begin{equation}
\mu_{\textrm{split}}(F_{1},F_{2}) \equiv \mu_{0}(F_{1,\rm{L}},F_{2,\rm{L}}) + \mu_{0}(F_{1,\rm{R}},F_{2,\rm{R}})\,
\end{equation}
which can be shown to satisfy the positivity condition \eqref{eq:positivity}. Indeed, defining 
\begin{equation}
a_{\textrm{L/R}}\equiv \mu_{0}(F_{1,\rm{L/R}},F_{1,\rm{L/R}}) \quad \textrm{ and } b_{\textrm{L/R}}\equiv \mu_{0}(F_{2,\rm{L/R}},F_{2,\rm{L/R}})\,,
\end{equation}
we obtain
\begin{align}
\left|\Omega^{\textrm{KG}}(F_{1},F_{2})\right|
&=
\left|
\Omega^{\textrm{KG}}(F_{1,\rm{L}},F_{2,\rm{L}})
+
\Omega^{\textrm{KG}}(F_{1,\rm{R}},F_{2,\rm{R}})
\right|
\nonumber\\
&\leq
2\sqrt{
\mu_{0}(F_{1,\rm{L}},F_{1,\rm{L}})
+
\mu_{0}(F_{1,\rm{R}},F_{1,\rm{R}})
}
\sqrt{
\mu_{0}(F_{2,\rm{L}},F_{2,\rm{L}})
+
\mu_{0}(F_{2,\rm{R}},F_{2,\rm{R}})
}
\nonumber\\
&=
2\sqrt{
\mu_{\textrm{split}}(F_{1},F_{1})
\mu_{\textrm{split}}(F_{2},F_{2})
}\,,
\end{align}
where in the second line we used \eqref{eq:positivity} and the Cauchy-Schwarz inequality. By \eqref{eq:Gaussian}, $\mu_{\textrm{split}}$ uniquely defines a Gaussian state $\Omega_{\textrm{split}}$ on $\mathcal{A}(\mathcal{W}_{\textrm{L}})\vee \mathcal{A}(\mathcal{W}_{\textrm{R}})$. 

While this defines  $\Omega_{\textrm{split}}$ as an abstract, algebraic (mixed) state,
it remains to be determined whether it can be realized as a normal state in $\mathcal{H}_{\textrm{out}}$. We will actually prove a stronger statement that the GNS representations of both states are {\em quasiequivalent}: any normal state --- i.e., every state is representable as a density matrix --- in the GNS representation of $\Omega_{0}$ is also normal in the GNS representation of $\Omega_{\textrm{split}}$ and vice-versa.\footnote{The general theory of quasiequivalence of Gaussian states was developed
by Araki and Yamagami \cite{Araki-Yamagami}, who, building on the work of \cite{Powers:1970bu,Araki:1970zza,VanDaele:1971dh}, established necessary and sufficient criteria for quasiequivalence --- see also \cite{Caminiti:2026ewu,Caminiti:2026lbq,Satishchandran:2026ixo} for a discussion of the closely related notion of ``excitability''.} For the purposes of this appendix, we will not need the full necessary and sufficient conditions here since, following \cite{Buchholz:1973bk},  we will actually prove a stronger sufficient condition for quasiequivalence. We now explain these conditions. 

Since the states are (zero mean) Gaussian states, the sufficient criteria can be simply stated as relations between the inner products $\mu_{0}$ and $\mu_{\textrm{split}}$. The basic idea is to ensure that these inner products are not so different as they define Gaussian states that lie in disjoint Hilbert spaces. The first criterion is that the norms induced by $\mu_{0}$ and $\mu_{\textrm{split}}$ induce the same topology on the space of initial data on $\Sigma_{\textrm{L}}\cup \Sigma_{\textrm{R}}$. Therefore, we require  
\begin{equation}
\label{eq:topology}
c \mu_{\textrm{split}}(F,F) \leq \mu_{0}(F,F)\leq C \mu_{\textrm{split}}(F,F)
\end{equation}
for some $0 < c < C <\infty$ and for all $F$. This condition is precisely one of the quasiequivalence criteria of Araki and Yamagami. 

We now explain an additional condition that, together with \eqref{eq:topology}, is sufficient for quasiequivalence. Denote by $\mathcal{K}_{\textrm{split}}$ the Hilbert space obtained via the (semi) inner product $\mu_{\textrm{split}}$ --- i.e., we consider the space of smooth initial data of compact support on $\Sigma_{\textrm{L}}\cup \Sigma_{\textrm{R}}$, quotient by the null space, and complete with respect to $\mu_{\textrm{split}}$. Eq.~\eqref{eq:topology} implies that there exists a bounded, self-adjoint operator $T$ on $\mathcal{K}_{\textrm{split}}$ such that 
\begin{equation}
\label{eq:T}
\mu_{0}(F_{1},F_{2}) - \mu_{\textrm{split}}(F_{1},F_{2}) = \mu_{\textrm{split}}(F_{1},TF_{2})
\end{equation}
for all $F_{1},F_{2}\in \mathcal{K}_{\textrm{split}}$. The operator $T$ encodes the difference of the covariance of the Gaussian states $\Omega_{\textrm{out}}$ and $\Omega_{\textrm{split}}$. As noted by Buchholz (see Appendix B of  \cite{Buchholz:1973bk}), a sufficient condition for quasiequivalence is that the operator $T$ is a trace-class operator on $\mathcal{K}_{\textrm{split}}$
\begin{equation}
\label{eq:traceclass}
\textrm{Tr}_{\mathcal{K}_{\textrm{split}}}|T|<\infty
\end{equation}
where $|T|=(T^{\dagger}T)^{1/2}$.

In summary, the sufficient conditions for quasiequivalence are that (1) the norms $\mu_{0}$ and $\mu_{\textrm{split}}$ are equivalent in the sense of \eqref{eq:topology} and (2) the operator $T$ satisfies \eqref{eq:traceclass}.

\subsection{Proof of the existence of a normal product state}
In this subsection we prove the existence of a product state $|\Omega_{\textrm{split}}\rangle$ satisfying \eqref{eq:Omegasplit}.  In appendix~\ref{subsubsec:equivalenceofnorms} we prove that the norms satisfy \eqref{eq:topology}. In appendix~\ref{subsubsec:modedecomp} we obtain a mode decomposition of the state and the operator $T$ as well as bounds on various norms that arise in its computation. In appendix~\ref{subsubsec:prooftraceclass} we collect these bounds to prove that $T$ satisfies the trace-class condition \eqref{eq:traceclass}. 

\subsection*{Mathematical preliminaries}
The proof we will present is ``elementary'' in the sense that it makes repeated use of only a small number of standard, analytic estimates. These facts, while familiar in functional analysis and PDE theory, may be less familiar to a high-energy physics audience. We briefly collect them here for reference. 

If $(\Sigma,d\nu)$ is a measure space, we denote the $L^{p}$-norm of a measurable function $\psi$ by
\begin{equation}
\|\psi\|_{L^{p}(\Sigma,d\nu)}
\equiv
\left(
\int_{\Sigma}|\psi|^{p}\,d\nu
\right)^{1/p},
\qquad
1\leq p<\infty\,.
\end{equation}
$L^{p}$ spaces satisfy an analog of a Cauchy-Schwarz inequality known as Hölder's inequality. If $1<p,q<\infty$ satisfy $p^{-1}+q^{-1}=1$, then Hölder's inequality states that
\begin{equation} 
\label{eq:Holder-general}
\left|
\int_{\Sigma}\overline{\psi_{1}}\psi_{2}\,d\nu
\right|
\leq
\|\psi_{1}\|_{L^{p}(\Sigma)}
\|\psi_{2}\|_{L^{q}(\Sigma)}\,.
\end{equation}
For $p=q=2$, this is the Cauchy-Schwarz inequality. We will occasionally make use of the fact that $L^{p}$-norms on certain spaces $\Sigma$ satisfy a homogeneous Sobolev inequality, which, for a connected, two-ended asymptotically Euclidean $3$-manifold $\Sigma$, states that
\begin{equation}
\label{eq:sobolev-general}
\|\psi\|_{L^{3p/(3-p)}(\Sigma)}
\leq
C_{p}
\|D \psi\|_{L^{p}(\Sigma)}\,,
\qquad
1\leq p<3\,.
\end{equation}

Finally, we will make use of the interpolation theorem, which relates norms involving different powers of a non-negative, injective and self-adjoint operator $H$ on  a Hilbert
space $\mathcal{H}$. To state this theorem, we first define the common spectral core
\begin{equation}
\mathcal{D}_{\mathrm{sp}}
\equiv
\bigcup_{0<a<b<\infty}
\mathbf{1}_{[a,b]}(H)\mathcal{H}\,.
\end{equation}
For $s\in\mathbb{R}$, let $\dot{\mathcal{H}}_{s}$ be the completion of
$\mathcal{D}_{\mathrm{sp}}$ in the norm $\|\psi\|_{\dot{\mathcal{H}}_{s}}
\equiv
\|H^{s} \psi\|_{\mathcal{H}}$. Using the spectral theorem, we regard these spaces as compatible subspaces of
a common space of generalized vectors. Let $s_{0},s_{1}\in\mathbb{R}$, and suppose that a linear operator $A$,
initially defined on $\mathcal{D}_{\mathrm{sp}}$, has bounded extensions
\begin{equation}
A_{j}:\dot{\mathcal{H}}_{s_{j}}
\longrightarrow
\dot{\mathcal{H}}_{s_{j}}\,,
\qquad j=0,1\,,
\end{equation}
satisfying
\begin{equation}
\|A_{j}\psi\|_{\dot{\mathcal{H}}_{s_{j}}}
\leq
M_{j}\|\psi\|_{\dot{\mathcal{H}}_{s_{j}}}\,,
\qquad j=0,1
\end{equation}
for $M_{j}\geq 0$. We assume that the two endpoint extensions are compatible, meaning that
\begin{equation}
A_{0}\psi=A_{1}\psi\,,
\qquad
\psi\in
\dot{\mathcal{H}}_{s_{0}}
\cap
\dot{\mathcal{H}}_{s_{1}}\,.
\end{equation}
Then, for
\begin{equation}
s_{\kappa}
=
(1-\kappa)s_{0}+\kappa s_{1}\,,
\qquad
0\leq\kappa\leq1\,,
\end{equation}
the interpolation theorem implies that $A$ extends uniquely to a bounded
operator on $\dot{\mathcal{H}}_{s_{\kappa}}$ and satisfies
\begin{equation}
\label{eq:interpolation}
\|H^{s_{\kappa}}A\psi\|_{\mathcal{H}}
\leq
M_{0}^{1-\kappa}M_{1}^{\kappa}
\|H^{s_{\kappa}}\psi\|_{\mathcal{H}}\,.
\end{equation}
In the following we will make repeated use of Hölder's inequality, the Sobolev inequality and the interpolation theorem.  

\subsubsection{Proof of equivalence of norms}
\label{subsubsec:equivalenceofnorms}
We first prove that the norms $\mu_{0}$ and $\mu_{\textrm{split}}$ satisfy \eqref{eq:topology}. To show this we choose a smooth, real function $\eta$ on $\Sigma_{0}$ such that $0\leq \eta \leq 1$ where $\eta=1$ on $\Sigma_{\textrm{L}}$ and vanishes on $\Sigma_{\textrm{R}}$. Then given any compactly supported initial data $F$ on $\Sigma_{\textrm{L}}\cup \Sigma_{\textrm{R}}$, $\eta F$ corresponds to data compactly supported on $\Sigma_{\textrm{L}}$  and $(1-\eta)F$ is data compactly supported on $\Sigma_{\textrm{R}}$. Therefore, we have the identity 
\begin{equation}
\label{eq:musplitmu0}
\mu_{\textrm{split}}(F_{1},F_{2}) = \mu_{0}(\eta F_{1},\eta F_{2}) + \mu_{0}((1-\eta)F_{1},(1-\eta)F_{2})
\end{equation}
where, in the decomposition of \eqref{eq:Fpq}, multiplication by $\eta$ acts on both pieces of Cauchy data $\eta F = (\eta q,\eta p)$. Using this identity we separately prove the left- and right-hand side inequalities of \eqref{eq:topology}.

 We first prove that there exists $c>0$ such that $c\mu_{\textrm{split}}(F,F)\leq  \mu_{0}(F,F)$. Applying \eqref{eq:mu0pq}, this inequality amounts to proving that there exist positive constants $C^{\prime}$ and $C^{\prime\prime}$ such that 
\begin{equation}
\label{eq:qpineq}
||H^{1/4}(\eta q)||^{2}_{L^{2}(\Sigma_{0})} \leq C^{\prime} ||H^{1/4}q||^{2}_{L^{2}(\Sigma_{0})}\,, \quad \quad||H^{-1/4}(\eta p)||^{2}_{L^{2}(\Sigma_{0})} \leq C^{\prime\prime} ||H^{-1/4}p||^{2}_{L^{2}(\Sigma_{0})}
\end{equation}
where we used the fact that $H$ is self-adjoint and positive. We will also need the analogous inequalities where $\eta\to 1-\eta$, however the following arguments will yield these inequalities under this substitution.

To prove the first inequality we note that it suffices to prove the analogous bound involving  $H^{1/2}$
\begin{equation}
\label{eq:H12ineq}
||H^{1/2}(\eta q)||^{2}_{L^{2}(\Sigma_{0})} \leq C^{\prime\prime\prime} ||H^{1/2}q||^{2}_{L^{2}(\Sigma_{0})} \,.
\end{equation}
Indeed, if this inequality is satisfied then the square root of this inequality implies the multiplication by $\eta$ is bounded in the $H^{1/2}$ energy norm.  Applying the operator version of the interpolation theorem \eqref{eq:interpolation} with $s_{1}=1/2$ and $s_{0}=0$ and $\kappa=1/2$ so that $s_{\kappa}=1/4$ we obtain the inequality 
\begin{equation}
||H^{1/4}(\eta q)||_{L^{2}(\Sigma_{0})} \leq \sqrt{C^{\prime}}||H^{1/4}q||_{L^{2}(\Sigma_{0})}
\end{equation}
for some positive constant $C^{\prime}$. Squaring this inequality yields the desired result. 

We now prove \eqref{eq:H12ineq}. We use the fact that, by \eqref{eq:Hpos}, we have that
\begin{equation}
||H^{1/2}(\eta q)||^{2}_{L^{2}(\Sigma_{0})}=\braket{\eta q,H\eta q}_{L^{2}(\Sigma_{0})} =\int_{\Sigma_{0}}  \sqrt{h}d^{3}x~N h^{ij}D_{i}(\eta q) D_{j}(\eta q)\,.
\end{equation}
Applying the chain rule and using the triangle inequality we obtain
\begin{align}
h^{ij}D_{i}(\eta q) D_{j}(\eta q) 
\leq& ~2 \eta^{2}h^{ij}D_{i}q D_{j}q + 2 q^{2}h^{ij}D_{i}\eta D_{j}\eta\,, 
\end{align}
and so 
\begin{equation}
\label{eq:twoterms}
||H^{1/2}(\eta q)||^{2}_{L^{2}(\Sigma_{0})} \leq 2 \int_{\Sigma_{0}}\sqrt{h}d^{3}x ~N \eta^{2}h^{ij}D_{i}q D_{j}q + 2\int_{\Sigma_{0}}\sqrt{h}d^{3}x  ~Nq^{2}h^{ij}D_{i}\eta D_{j}\eta \,.
\end{equation}
Since $\eta$ is a bounded function, the first integral is bounded by the square of its supremum multiplied by the desired norm 
\begin{equation}
\label{eq:C1}
2\int_{\Sigma_{0}}\sqrt{h}d^{3}x ~N \eta^{2}h^{ij}D_{i}q D_{j}q \leq 2\int_{\Sigma_{0}}\sqrt{h}d^{3}x ~N h^{ij}D_{i}q D_{j}q=2 ||H^{1/2}q||^{2}_{L^{2}(\Sigma_{0})} \,.
\end{equation}
We now consider the second term in \eqref{eq:twoterms}. Since $\eta$ is constant on $\Sigma_{\textrm{L}}$ and $\Sigma_{\textrm{R}}$, the support of $D_{i}\eta$ is, at most, the compact collar region $\Sigma_{\textrm{C}}$. Therefore, we can replace the integral over the noncompact $\Sigma_{0}$ in the second term with an integral over $\Sigma_{\textrm{C}}$. Introducing a factor of $N^{-1}$ so the measure is $N^{-1}\sqrt{h}d^{3}x$ we obtain 
\begin{equation}
2\int_{\Sigma_{\textrm{C}}}N^{-1}\sqrt{h}d^{3}x  ~N^{2}q^{2}h^{ij}D_{i}\eta D_{j}\eta \leq C_{2}\int_{\Sigma_{\textrm{C}}}N^{-1}\sqrt{h}d^{3}x ~q^{2}
\end{equation}
where we again used the fact that $N^{2}h^{ij}D_{i}\eta D_{j}\eta$ is bounded so the integral can be bounded by its supremum on $\Sigma_{\rm C}$ multiplied by the integral of $q^{2}$. We now apply Hölder's inequality \eqref{eq:Holder-general} with $f=q^{2}$, $g=1$ and conjugate exponents $3$ and $3/2$, which yields 
\begin{equation}
\label{eq:pnorm}
\int_{\Sigma_{\textrm{C}}}N^{-1}\sqrt{h}d^{3}x ~q^{2} \leq \bigg(\int_{\Sigma_{\textrm{C}}}N^{-1}\sqrt{h}d^{3}x ~q^{6}\bigg)^{1/3}\bigg(\int_{\Sigma_{\textrm{C}}}N^{-1}\sqrt{h}d^{3}x ~1^{3/2}\bigg)^{2/3}\,.
\end{equation}
The second factor on the right is finite since $N^{-1}$ is smooth and positive and the integral is over a compact domain so we have that  
\begin{equation}\label{eq:sobloevnormq}
\int_{\Sigma_{\textrm{C}}}N^{-1}\sqrt{h}d^{3}x ~q^{2} \leq C^{\prime}_{2}\bigg(\int_{\Sigma_{\textrm{C}}}N^{-1}\sqrt{h}d^{3}x ~q^{6}\bigg)^{1/3}\,.
\end{equation}
Since $\Sigma_{\rm C}\subset \Sigma_0$, the integral over $q^{6}$ is no larger than the integral over all of $\Sigma_{0}$,
\begin{equation}
\label{eq:1/3}
\int_{\Sigma_{\textrm{C}}}N^{-1}\sqrt{h}d^{3}x ~q^{2} \leq C^{\prime\prime}_{2}\bigg(\int_{\Sigma_{0}}N^{-1}\sqrt{h}d^{3}x ~q^{6}\bigg)^{1/3}\,.
\end{equation}
The homogeneous Sobolev inequality \eqref{eq:sobolev-general} with $p=2$ and $d\nu=N^{-1}\sqrt{h}d^{3}x$ yields
\begin{equation}
\bigg(\int_{\Sigma_{0}}N^{-1}\sqrt{h}d^{3}x ~q^{6}\bigg)^{1/3} \leq C_{3}\int_{\Sigma_{0}}  \sqrt{h}d^{3}x~N^{-1} h^{ij}D_{i} q D_{j} q \,.
\end{equation}
Plugging this into \eqref{eq:1/3} and using the fact that $N^{-2}$ is bounded from above we can again bound the right-hand side by its supremum to obtain 
\begin{equation}
\label{eq:C4}
C_{3}\int_{\Sigma_{0}}  \sqrt{h}d^{3}x~N^{-1} h^{ij}D_{i} q D_{j} q  \leq C_{4} \int_{\Sigma_{0}}  \sqrt{h}d^{3}x~N h^{ij}D_{i} q D_{j} q = C_{4}||H^{1/2}q||^{2}_{L_{2}(\Sigma_{0})}
\end{equation}
where $C_{4}$ is $C_{3}$ times the supremum of $N^{-2}$. Combining \eqref{eq:twoterms}, \eqref{eq:C1} and \eqref{eq:C4} implies \eqref{eq:H12ineq} with $C'''=C_1+C_4$, and hence implies the first inequality in \eqref{eq:qpineq}. 

We now prove the second inequality in \eqref{eq:qpineq}. To prove this we note that since $H$ is positive and injective, the $L^{2}$ norm of $H^{-1/4}(\eta p)$ can be equivalently written as 
\begin{equation}
||H^{-1/4}(\eta p)||_{L^{2}(\Sigma_{0})} = \sup_{\psi\neq 0}\frac{|\braket{\psi,\eta p}_{L^{2}(\Sigma_{0})}|}{||H^{1/4}\psi||_{L^{2}(\Sigma_{0})}} = \sup_{\psi\neq 0}\frac{|\braket{\eta \psi, p}_{L^{2}(\Sigma_{0})}|}{||H^{1/4}\psi||_{L^{2}(\Sigma_{0})}}
\end{equation}
where we used the fact that $\eta$ is a real scalar function to move it to the first argument of the $L^{2}$ inner product. For a positive, self-adjoint operator $H$, we have
\begin{equation}
|\langle \eta \psi , p \rangle_{L^{2}(\Sigma_{0})} | = |\langle H^{1/4}(\eta \psi) , H^{-1/4}p \rangle_{L^{2}(\Sigma_{0})} | \leq \sqrt{C^{\prime}}||H^{1/4} \psi||_{L^{2}(\Sigma_{0})}||H^{-1/4}p||_{L^{2}(\Sigma_{0})}
\end{equation}
where we applied the Cauchy-Schwarz inequality as well as the inequality that we just proved $||H^{1/4}(\eta \psi)||^{2}_{L^{2}(\Sigma_{0})} \leq C^{\prime} ||H^{1/4}\psi||^{2}_{L^{2}(\Sigma_{0})}$. 
Dividing by $||H^{1/4}\psi||_{L^{2}(\Sigma_{0})}$ and taking the supremum we obtain 
\begin{equation}
||H^{-1/4}(\eta p)||_{L^{2}(\Sigma_{0})} \leq \sqrt{C^{\prime}}||H^{-1/4}p||_{L^{2}(\Sigma_{0})}\,.
\end{equation}
Squaring this results in the second inequality in \eqref{eq:qpineq}, and replacing $\eta \to 1-\eta$ in the above arguments yields the desired set of inequalities. All constants in the above expressions can be chosen to be positive.

To conclude the argument, we have shown by combining \eqref{eq:mu0pq} with \eqref{eq:qpineq} that there exist positive constants $C_{5},C_{6}>0$  such that 
\begin{equation}
\mu_{0}(\eta F,\eta F) \leq C_{5} \mu_{0}( F,F)\,, \quad \quad \mu_{0}((1-\eta) F,(1-\eta) F) \leq C_{6} \mu_{0}( F,F)\,.
\end{equation}
Then, using \eqref{eq:musplitmu0}, we have just proven that there exists a constant $c>0$ given by $c=(C_{5}+C_{6})^{-1}$ such that
\begin{equation}
\label{eq:ineqmusplimu0}
c\mu_{\textrm{split}}(F,F) \leq \mu_{0}( F,F)\,.
\end{equation}

It is far more straightforward to prove the right-hand side inequality of \eqref{eq:topology}. Writing $q$ and $p$ as 
\begin{equation}
q = \eta q + (1-\eta)q\,, \qquad p = \eta p + (1-\eta)p
\end{equation}
we then also have that $H^{1/4}q = H^{1/4}(\eta q) + H^{1/4}((1-\eta)q)$. Taking the norm, applying the triangle inequality and squaring the resulting expression yields 
\begin{equation}
||H^{1/4}q||_{L^{2}(\Sigma_{0})}^{2} \leq 2||H^{1/4}(\eta q)||_{L^{2}(\Sigma_{0})}^{2} + 2||H^{1/4}((1-\eta) q)||_{L^{2}(\Sigma_{0})}^{2}\,.
\end{equation}
The same argument yields an identical inequality with $q\to p$. Using \eqref{eq:mu0pq} and \eqref{eq:musplitmu0} we obtain 
\begin{equation}
\mu_{0}(F,F)\leq 2 \mu_{\textrm{split}}(F,F)\,.
\end{equation}
The inequality \eqref{eq:ineqmusplimu0} implies that there exists a positive constant $0<c_{0}<c$ which also satisfies \eqref{eq:ineqmusplimu0} and, choosing this constant so that $c_{0}<2$, yields the desired bound on $\mu_{0}$. 

\subsubsection{Mode decomposition and bounds on the trace}
\label{subsubsec:modedecomp}

\subsubsection*{Mode decomposition of the Hilbert space and operators}

In the remainder of this appendix we consider the trace-class condition \eqref{eq:traceclass}. The trace-class condition is a condition on $\mathcal{K}_{\textrm{split}}$ for the operator $T$ which, by \eqref{eq:T},  is the operator that relates the difference $\mu_{0}-\mu_{\textrm{split}}$ to $\mu_{\textrm{split}}$. To obtain a more explicit expression for $T$ we define the following inner product on  initial data $(q_{\textrm{L}},p_{\textrm{L}})$ supported  entirely on $\Sigma_{\textrm{L}}$,
\begin{equation}
\langle q_{1,\textrm{L}}, q_{2,\textrm{L}} \rangle_{q,\textrm{L}}  \equiv  \frac{1}{2}\langle H^{1/4}q_{1,\textrm{L}}, H^{1/4}q_{2,\textrm{L}} \rangle_{L^{2}(\Sigma_{0})} 
\end{equation}
and 
\begin{equation}
\langle p_{1,\textrm{L}}, p_{2,\textrm{L}} \rangle_{p,\textrm{L}}  \equiv  \frac{1}{2}\langle H^{-1/4}p_{1,\textrm{L}}, H^{-1/4}p_{2,\textrm{L}} \rangle_{L^{2}(\Sigma_{0})} \,.
\end{equation}
Let $\mathcal{K}_{q,\textrm{L}}$ and $\mathcal{K}_{p,\textrm{L}}$ be their  corresponding Hilbert space completions. We similarly define inner products for data $(q_{\textrm{R}},p_{\textrm{R}})$ supported entirely on $\Sigma_{\textrm{R}}$ given by the above expressions with the replacement $\textrm{L}\to \textrm{R}$. We denote their corresponding Hilbert spaces as $\mathcal{K}_{q,\textrm{R}}$ and $\mathcal{K}_{p,\textrm{R}}$. Using \eqref{eq:mu0pq} and \eqref{eq:musplitmu0}, the Hilbert space $\mathcal{K}_{\textrm{split}}$ can be decomposed as 
\begin{equation}
\label{eq:Ksplitdecomp}
\mathcal{K}_{\textrm{split}} \cong (\mathcal{K}_{q,\textrm{L}} \oplus \mathcal{K}_{p,\textrm{L}}) \oplus (\mathcal{K}_{q,\textrm{R}} \oplus \mathcal{K}_{p,\textrm{R}})\,.
\end{equation}
The operator $T$ encodes the correlations of the $q$ and $p$ between $\Sigma_{\textrm{L}}$ and $\Sigma_{\textrm{R}}$. To define this operator on the initial data function spaces we first observe that, by Cauchy-Schwarz,
\begin{equation}
\frac{1}{2}|\langle  H^{1/4}q_{\textrm{L}}, H^{1/4}q_{\textrm{R}} \rangle_{L^{2}(\Sigma_{0})}| \leq ||q_{\textrm{L}}||_{q,\textrm{L}}||q_{\textrm{R}}||_{q,\textrm{R}}
\end{equation}
and so the map $q_{\textrm{L}}\to \tfrac{1}{2} \langle  H^{1/4}q_{\textrm{L}}, H^{1/4}q_{\textrm{R}} \rangle_{L^{2}(\Sigma_{0})}$ is a bounded linear functional on $\mathcal{K}_{q,\textrm{L}}$. Therefore, there exists a  bounded operator 
\begin{equation}
T_{q}:\mathcal{K}_{q,\textrm{R}}\to \mathcal{K}_{q,\textrm{L}}
\end{equation}
which satisfies 
\begin{equation}
\label{eq:Tq}
\langle  q_{\textrm{L}}, T_{q}q_{\textrm{R}}\rangle_{q,\textrm{L}} = \frac{1}{2}\langle  q_{\textrm{L}}, H^{1/2}q_{\textrm{R}} \rangle_{L^{2}(\Sigma_{0})}
\end{equation}
for any $q_{\textrm{L}}\in \mathcal{K}_{q,\textrm{L}}$. Similarly, there exists a unique bounded operator $T_{p}$ such that 
\begin{equation}
\label{eq:Tp}
T_{p}:\mathcal{K}_{p,\textrm{R}}\to \mathcal{K}_{p,\textrm{L}}, \quad  \quad \langle  p_{\textrm{L}}, T_{p}p_{\textrm{R}} \rangle_{p,\textrm{L}} = \frac{1}{2}\langle  p_{\textrm{L}}, H^{-1/2}p_{\textrm{R}} \rangle_{L^{2}(\Sigma_{0})}\,.
\end{equation}
In the decomposition given by \eqref{eq:Ksplitdecomp} we can represent the initial data on $\Sigma_{\textrm{L}}\cup \Sigma_{\textrm{R}}$ by the $4$-tuple $(q_{\textrm{L}},p_{\textrm{L}},q_{\textrm{R}},p_{\textrm{R}})$. The full operator $T$ on  $\mathcal{K}_{\textrm{split}}$ has the block form 
\begin{equation}
T=
\begin{pmatrix}
0 & 0 & T_q & 0 \\
0 & 0 & 0 & T_p \\
T_q^{\dagger} & 0 & 0 & 0 \\
0 & T_p^{\dagger} & 0 & 0
\end{pmatrix},
\end{equation}
and so the trace of $|T|$ on $\mathcal{K}_{\textrm{split}}$ can be expressed as 
\begin{equation}
\label{eq:TclassTqTp}
\textrm{Tr}_{\mathcal{K}_{\textrm{split}}}|T| = 2 \big( \textrm{Tr}_{\mathcal{K}_{q,\textrm{R}}}|T_{q}|  +  \textrm{Tr}_{\mathcal{K}_{p,\textrm{R}}}|T_{p}|\big)
\end{equation}
where $|T_{q}|\equiv \sqrt{T_{q}^{\dagger}T_{q}}$ and $|T_{p}|\equiv \sqrt{T_{p}^{\dagger}T_{p}}$. 

To evaluate \eqref{eq:TclassTqTp} it will be useful to decompose the Hilbert space $L^{2}(\Sigma_{0},N^{-1}\sqrt{h}d^{3}x)$ with respect to modes. To achieve this, we define a new radial coordinate $z$ which satisfies 
\begin{equation}
\frac{dz}{ds} = \frac{1}{N(s)}
\end{equation}
with $z(0)=0$. 
If the left and right regions given by \eqref{eq:SigmaLR} end at $-s_{a}$ and $s_{b}$ respectively then we define constants $a<0$ and $b>0$ such that $a\equiv z(-s_{a})$ and  $b \equiv z(s_{b})$. In these coordinates,
\begin{equation}
\label{eq:SigmaLRz}
\Sigma_{\rm L} = \{(z,x^{A})\in \Sigma_{0}\,|\, z<a\} \quad \quad \Sigma_{\rm R} = \{(z,x^{A})\in \Sigma_{0}\,|\, z>b\}
\end{equation}
and the collar $\Sigma_{\textrm{C}}$ is the region $a\leq z \leq b$. This coordinate change achieves two simplifications. The first is that the measure in these coordinates is now $N^{-1}\sqrt{h}d^{3}x= r(z)^{2}dzd\Omega$. Secondly, the operator $H$ takes the form 
\begin{equation}
H = -\frac{1}{r^{2}}\frac{\partial}{\partial z}\bigg(r^{2}\frac{\partial}{\partial z}\bigg) - \frac{N^{2}}{r^{2}}\Delta_{\mathbb{S}^{2}}\,.
\end{equation}
In these coordinates, any function $\psi$ in $L^{2}(\Sigma, r^{2}dzd\Omega)$ admits the following expansion in terms of modes 
\begin{equation}
\label{eq:phimode}
\psi(z,x^{A}) = \frac{1}{r(z)}\sum_{\ell m}u_{\ell m}(z)Y_{\ell m}(x^{A})
\end{equation}
where the $Y_{\ell m}$ are the spherical harmonics on $\mathbb{S}^{2}$ and the inner product on $L^{2}(\Sigma, r^{2}dzd\Omega)$ yields the following inner product for the modes $u_{\ell m}$ and $v_{\ell^{\prime} m^{\prime}}$ 
\begin{equation}
\label{eq:umode}
 \langle u_{\ell m}, v_{\ell^{\prime} m^{\prime}}\rangle_{L^{2}(\mathbb{R})}\equiv \delta_{\ell \ell^{\prime}}\delta_{mm^{\prime}}\int_{\mathbb{R}}dz~ \overline{u_{\ell m}}(z) v_{\ell m}(z)\,.
\end{equation}

This decomposition implies the following decomposition of $H$,
\begin{equation}
H \simeq \bigoplus_{\ell =0}^{\infty}H_{\ell}\otimes 1_{2\ell +1}
\end{equation}
where $2\ell +1$ is the degeneracy of each spherical harmonic subspace of definite $\ell$ and 
\begin{equation}
H_{\ell} \equiv -\bigg(\frac{d}{d z}\bigg)^{2} + U_{\ell }(z) \quad \textrm{ and} \quad U_{\ell}(z) \equiv \frac{r^{\prime\prime}(z)}{r(z)} + \ell (\ell + 1)\frac{N(z)^{2}}{r(z)^{2}} \,.
\end{equation}
The initial data admit a similar decomposition as \eqref{eq:phimode} where the initial data on $\Sigma_{\textrm{L}}\cup \Sigma_{\textrm{R}}$ can be expressed in terms of the modes $(q_{\textrm{L},\ell m},p_{\textrm{L},\ell m},q_{\textrm{R},\ell m},p_{\textrm{R},\ell m})$. We denote the one-dimensional Hilbert space associated to the modes $q_{\textrm{L},\ell m}(z)$ and  $p_{\textrm{L},\ell m}(z)$ with norm
\begin{equation}
||q_{\textrm{L},\ell m}||^{2}\equiv \frac{1}{2}||H^{1/4}_{\ell}q_{\textrm{L},\ell m}||^{2}_{L^{2}(\mathbb{R})}\,, \qquad ||p_{\textrm{L},\ell m}||^{2}\equiv \frac{1}{2}||H^{-1/4}_{\ell}p_{\textrm{L},\ell m}||^{2}_{L^{2}(\mathbb{R})} 
\end{equation} 
as $\mathcal{K}_{q,\textrm{L},\ell}$ and $\mathcal{K}_{p,\textrm{L},\ell}$. The norm for the modes $q_{\textrm{R},\ell m}(z)$ and  $p_{\textrm{R},\ell m}(z)$ are defined in an identical manner and define the Hilbert spaces $\mathcal{K}_{q,\textrm{R},\ell}$ and $\mathcal{K}_{p,\textrm{R},\ell}$. The operators $T_{q}$ and $T_{p}$ are spherically symmetric and so their corresponding operator on each spherical harmonic subspace is 
\begin{equation}
T_{q,\ell}: \mathcal{K}_{q,\textrm{R},\ell} \to \mathcal{K}_{q,\textrm{L},\ell}\,, \qquad T_{p,\ell}:\mathcal{K}_{p,\textrm{R},\ell}\to \mathcal{K}_{p,\textrm{L},\ell}
\end{equation}
which have the obvious action on their respective Hilbert spaces descending from the action of \eqref{eq:Tq} and \eqref{eq:Tp}. In terms of these operators the trace of $|T|$ is given by 
\begin{equation}
\label{eq:TclassTqTplm}
\textrm{Tr}_{\mathcal{K}_{\textrm{split}}}|T| = 2\sum_{\ell =0}^{\infty}(2\ell+1) \big( \textrm{Tr}_{\mathcal{K}_{q,\textrm{R},\ell}}|T_{q,\ell}|  +  \textrm{Tr}_{\mathcal{K}_{p,\textrm{R},\ell}}|T_{p,\ell}|\big)\,.
\end{equation}

\subsubsection*{A Bound on the Trace of  $|T_{q,\ell}|$ and $|T_{p,\ell}|$}
To determine the trace of $|T_{q, \ell}|$ and $|T_{p, \ell}|$  we recall that this operator is defined by the matrix elements 
\begin{equation}
\label{eq:matrixelTq}
\left\langle q_{\textrm{L},\ell m}, T_{q,\ell} q_{\textrm{R},\ell m} \right\rangle = \frac{1}{2}\left\langle q_{\textrm{L},\ell m},H^{1/2}_{\ell}q_{\textrm{R},\ell m}\right\rangle_{L^{2}(\mathbb{R})}  
\end{equation}
and 
\begin{equation}
\label{eq:matrixelTp}
\left\langle p_{\textrm{L},\ell m}, T_{p,\ell} p_{\textrm{R},\ell m}\right\rangle = \frac{1}{2}\left\langle p_{\textrm{L},\ell m},H^{-1/2}_{\ell}p_{\textrm{R},\ell m}\right\rangle_{L^{2}(\mathbb{R})} \,.
\end{equation}
Therefore our immediate goal is to understand the matrix elements of $H_{\ell}^{1/2}$ and $H_{\ell}^{-1/2}$. However, unlike $H_{\ell}$ which acts locally on the radial modes, the fractional powers of $H_{\ell}$ act non-locally. To obtain a simple description of the matrix elements \eqref{eq:matrixelTq} and \eqref{eq:matrixelTp} we can rewrite them in terms of the ``resolvent'' $(H_{\ell}+\lambda^{2})^{-1}$ where $\lambda>0$. 

Since $H_{\ell}\geq 0$ and $H_{\ell}$ is injective, the operator $(H_{\ell}+\lambda^{2})^{-1}$ is bounded and positive on $L^{2}(\mathbb{R},dz)$. It is also non-local, but its kernel is the Green's function to a second order ODE. In the following, all resolvent formulas are to be understood at the level of matrix elements between smooth ``left'' and ``right'' supported data. In this sense, the operator $H^{-1/2}_{\ell}$ can be equivalently expressed as 
\begin{equation}
H^{-1/2}_{\ell} = \frac{2}{\pi} \int_{0}^{\infty}d\lambda ~(H_{\ell} + \lambda^{2})^{-1}\,.
\end{equation}
where the integral over $\lambda$ is also meant in the sense of matrix elements first defined for spectral cut-offs $\lambda\in [\epsilon,R]$ and then considering the limit as $\epsilon\to0$ and $R\to \infty$. 
Taking matrix elements yields 
\begin{equation}
\label{eq:matrixelpH}
\left\langle p_{\textrm{L},\ell m},H^{-1/2}_{\ell}p_{\textrm{R},\ell m}\right\rangle_{L^{2}(\mathbb{R})} = \frac{2}{\pi}\int_{0}^{\infty}d\lambda~\langle p_{\textrm{L},\ell m},(H_{\ell} + \lambda^{2})^{-1} p_{\textrm{R},\ell m}\rangle_{L^{2}(\mathbb{R})} \,.
\end{equation}
Similarly, $H^{1/2}_{\ell}$ can be expressed as 
\begin{equation}
\label{eq:H12}
H_{\ell}^{1/2} = \frac{2}{\pi}\int_{0}^{\infty}d\lambda~H_{\ell}(H_{\ell}+\lambda^{2})^{-1}\,.
\end{equation}
Since $H_{\ell}(H_{\ell}+\lambda^{2})^{-1}= 1 - \lambda^{2}(H_{\ell}+\lambda^{2})^{-1}$ and using the fact that $\langle q_{\textrm{L},\ell m}, q_{\textrm{R},\ell m}\rangle_{L^{2}(\mathbb{R})} =0$ we obtain 
\begin{equation}
\label{eq:Hellqmatrixel}
\langle q_{\textrm{L},\ell m},H^{1/2}_{\ell}q_{\textrm{R},\ell m}\rangle_{L^{2}(\mathbb{R})} = -\frac{2}{\pi}\int_{0}^{\infty}d\lambda~\lambda^{2}\langle q_{\textrm{L},\ell m},(H_{\ell} + \lambda^{2})^{-1} q_{\textrm{R},\ell m}\rangle_{L^{2}(\mathbb{R})} \,.
\end{equation}
These formulas 
for the matrix elements are useful because a one-dimensional, second-order Sturm-Liouville equation
\begin{equation}
\bigg(-\frac{d^{2}}{dz^{2}} + U_{\ell}(z) + \lambda^{2}\bigg)G_{\ell}(\lambda; z,z^{\prime}) = \delta(z-z^{\prime})
\end{equation}
has the unique Green's function 
\begin{equation}
\label{eq:Greensfunction}
G_{\ell}(\lambda;z,z^{\prime}) = \frac{u_{\ell,\lambda}^{-}(z_{<})u_{\ell,\lambda}^{+}(z_{>})}{W_{\ell, \lambda}}
\end{equation}
where $u_{\ell,\lambda}^{\pm}(z)$ are the strictly positive solutions to the homogeneous equation 
\begin{equation}
\label{eq:modeq}
\bigg(-\frac{d^{2}}{dz^{2}} + U_{\ell}(z) + \lambda^{2}\bigg)u=0
\end{equation}
which decay as  $ u_{\ell,\lambda}^{\pm}\sim e^{\mp\lambda z}$ as $z\to \pm \infty$. These functions are unique up to an overall constant and so we normalize them by requiring that, at the endpoints of the collar region, $u^{-}_{\ell \lambda}(a)=u^{+}_{\ell \lambda}(b)=1$.  The Wronskian $W_{\ell, \lambda}$ is independent of $z$ and with this normalization is given by 
\begin{equation}
W_{\ell, \lambda} =u^{-\prime}_{\ell \lambda}(z)u^{+}_{\ell \lambda}(z) - u^{-}_{\ell \lambda}(z)u^{+\prime}_{\ell \lambda}(z) =\frac{1}{G_{\ell}(\lambda;a,b)}\,.
\end{equation}
Finally $z_{<}=\min\{z,z^{\prime}\}$ and $z_{>}=\max\{z,z^{\prime}\}$. We will be particularly interested in this Green's function for $z\in \Sigma_{\textrm{L}}$ and $z^{\prime}\in\Sigma_{\textrm{R}}$. In this case we obtain the formula 
\begin{equation}
G_{\ell}(\lambda;z,z^{\prime}) = G_{\ell}(\lambda;a,b)u^{-}_{\ell , \lambda}(z)u^{+}_{\ell , \lambda}(z^{\prime})\,, \qquad \textrm{ $z<a<b<z^{\prime}$}\,.
\end{equation}
The key simplifying fact for the remainder of this proof is that the $G_{\ell}(\lambda;z,z^{\prime})$ factorizes into the product of two local functions when the points $z$ and $z^{\prime}$ remain on opposite sides of the collar. Therefore, we obtain the matrix element 
\begin{equation}
\begin{aligned}
\label{eq:matrixelGreens}
&\left\langle
q_{\textrm{L},\ell m},
(H_{\ell}+\lambda^{2})^{-1}q_{\textrm{R},\ell m}
\right\rangle_{L^{2}(\mathbb{R})} =
G_{\ell}(\lambda;a,b)
\left\langle
q_{\textrm{L},\ell m},
u^{-}_{\ell,\lambda}
\right\rangle_{L^{2}(\mathbb{R})}
\left\langle
u^{+}_{\ell,\lambda},
q_{\textrm{R},\ell m}
\right\rangle_{L^{2}(\mathbb{R})}\,.
\end{aligned}
\end{equation}
We, of course, obtain a similar formula for matrix elements of the resolvent with respect to $p_{\textrm{L},\ell m}$ and $p_{\textrm{R},\ell m}$. 

We note that there is a subtlety in the above formula that we now comment on. While  $u_{\ell,\lambda}^{-}(z)$ exponentially decays as $z\to -\infty$, it exponentially grows as $z\to +\infty$ for $\lambda>0$. The function $u_{\ell,\lambda}^{+}(z)$ has the opposite behavior where it decays as $z\to +\infty$ but exponentially grows as $z\to -\infty$ for $\lambda>0$.  While the inner products appearing in \eqref{eq:matrixelGreens} are well-defined due to the support properties of $q_{\textrm{L}}$ and $q_{\textrm{R}}$, the functions $u_{\ell,\lambda}^{+}(z)$ and $u_{\ell,\lambda}^{-}(z)$ are {\em not} in $L^{2}(\mathbb{R})$. Since only the restriction of $u_{\ell,\lambda}^{-}(z)$ to $(-\infty,a)$ and $u_{\ell,\lambda}^{+}(z)$ to $(b,\infty)$ enter into the above formula, we can obtain an equivalent expression for \eqref{eq:matrixelGreens} by replacing $u_{\ell,\lambda}^{-}$ and $u_{\ell,\lambda}^{+}$ with functions $v_{\ell,\lambda}^{-}$ and $v_{\ell,\lambda}^{+}$ which satisfy
\begin{equation} \label{eq:extension}
\begin{aligned}
v_{\ell,\lambda}^{-}(z)
&=
\begin{cases}
u_{\ell,\lambda}^{-}(z),
    & z\leq a, \\[2mm]
0,
    & z\geq b,
\end{cases}
&\qquad
v_{\ell,\lambda}^{+}(z)
&=
\begin{cases}
0,
    & z\leq a \\[2mm]
u_{\ell,\lambda}^{+}(z),
    & z\geq b
\end{cases}
\end{aligned}
\end{equation}
where $v_{\ell,\lambda}^{-},v_{\ell,\lambda}^{+}\in L^{2}(\mathbb{R},dz)$ for $\lambda>0$. For $\lambda,\ell=0$, the functions $u^{\pm}_{0,0}$ do not decay and so there is no $L^{2}$ extension. Therefore, unless otherwise stated we will restrict attention to the case of $\lambda>0$. The bounds we establish will control the limit of  \eqref{eq:matrixelGreens} as $\lambda\to 0$. To establish these bounds we will assume, for the moment, that the functions $v_{\ell,\lambda}^{\pm}$ can be chosen such that  $v_{\ell,\lambda}^{\pm}\in \textrm{Dom}(H^{1/2}_{\ell}) \cap \textrm{Dom}(H^{-1/2}_{\ell})$ for all $\ell\geq 0$ and $\lambda>0$ . After obtaining some estimates under this assumption, we will then prove that such a choice of $v_{\ell,\lambda}^{\pm}$ exists. With the choices of extension given by \eqref{eq:extension}, the matrix element is now 
\begin{equation}
\begin{aligned}
&\left\langle
q_{\textrm{L},\ell m},
(H_{\ell}+\lambda^{2})^{-1}q_{\textrm{R},\ell m}
\right\rangle_{L^{2}(\mathbb{R})} =
G_{\ell}(\lambda;a,b)
\left\langle
q_{\textrm{L},\ell m},
v^{-}_{\ell,\lambda}
\right\rangle_{L^{2}(\mathbb{R})}
\left\langle
v^{+}_{\ell,\lambda},
q_{\textrm{R},\ell m}
\right\rangle_{L^{2}(\mathbb{R})}\,.
\end{aligned}
\end{equation}

Using this formula together with the above provisional assumption on the regularity of $v_{\ell,\lambda}^{\pm}$, we now obtain bounds on the matrix elements of the resolvent for $\lambda>0$. All calculations below are initially performed on smooth, compactly supported left- and right-supported Cauchy data. Whenever an estimate is bounded in one of the local norms of $q$ or $p$, then the corresponding pairing or operator is subsequently extended by continuity to the relevant Hilbert-space completion. We continue to use the symbols $q$ and $p$ for elements of these completions, although such elements need not be smooth functions.

Writing $\left \langle v^{-}_{\ell,\lambda},q_{\textrm{L},\ell m} \right\rangle_{L^{2}(\mathbb{R})} = \left \langle H^{-1/4}_{\ell}v^{-}_{\ell,\lambda},H^{1/4}_{\ell}q_{\textrm{L},\ell m} \right\rangle_{L^{2}(\mathbb{R})} $ and applying Cauchy-Schwarz, we obtain 
\begin{align}
\label{eq:boundv-}
\left|
\left\langle
v^{-}_{\ell,\lambda},
q_{\textrm{L},\ell m}
\right\rangle_{L^{2}(\mathbb{R})}
\right|
&\leq
\left\|H_{\ell}^{-1/4}v^{-}_{\ell,\lambda}\right\|_{L^{2}(\mathbb{R})}
\left\|H_{\ell}^{1/4}q_{\textrm{L},\ell m}\right\|_{L^{2}(\mathbb{R})}
\nonumber\\
&=
\sqrt{2}\,
\left\|H_{\ell}^{-1/4}v^{-}_{\ell,\lambda}\right\|_{L^{2}(\mathbb{R})}
\left\|q_{\textrm{L},\ell m}\right\|.
\end{align}
Similarly,
\begin{equation}
\left|
\left\langle
v^{+}_{\ell,\lambda},
q_{\textrm{R},\ell m}
\right\rangle_{L^{2}(\mathbb{R})}
\right|
\leq
\sqrt{2}\,
\left\|H_{\ell}^{-1/4}v^{+}_{\ell,\lambda}\right\|_{L^{2}(\mathbb{R})}
\left\|q_{\textrm{R},\ell m}\right\|.
\end{equation}
For the inner product with $p$-data, we also obtain
\begin{align}
\left|
\left\langle
v^{-}_{\ell,\lambda},
p_{\textrm{L},\ell m}
\right\rangle_{L^{2}(\mathbb{R})}
\right|
&\leq
\sqrt{2}\,
\left\|H_{\ell}^{1/4}v^{-}_{\ell,\lambda}\right\|_{L^{2}(\mathbb{R})}
\left\|p_{\textrm{L},\ell m}\right\|\\
\left|
\left\langle
v^{+}_{\ell,\lambda},
p_{\textrm{R},\ell m}
\right\rangle_{L^{2}(\mathbb{R})}
\right|
&\leq
\sqrt{2}\,
\left\|H_{\ell}^{1/4}v^{+}_{\ell,\lambda}\right\|_{L^{2}(\mathbb{R})}
\left\|p_{\textrm{R},\ell m}\right\|\,.
\end{align}
For each fixed $\lambda>0$, the factorization of the resolvent kernel implies that its contribution to $T_{q,\ell}$ and $T_{p,\ell}$ is rank one. We combine the fact that the trace norm of a rank one operator factorizes into the product of the norms of its constituent factors together with the triangle inequality to bound the trace norm of the integral by the integral of the trace norm, yielding
\begin{equation}
\label{eq:TrTqell}
\textrm{Tr}_{\mathcal{K}_{q,\textrm{R},\ell}}|T_{q,\ell}| \leq \frac{2}{\pi}\int_{0}^{\infty} d\lambda ~\lambda^{2}G_{\ell}(\lambda;a,b)||H_{\ell}^{-1/4}v_{\ell,\lambda}^{-}||_{L^{2}(\mathbb{R})}||H_{\ell}^{-1/4}v_{\ell,\lambda}^{+}||_{L^{2}(\mathbb{R})}
\end{equation}
and 
\begin{equation}
\label{eq:TrTpell}
\operatorname{Tr}_{\mathcal{K}_{p,\textrm{R},\ell}}
|T_{p,\ell}|
\leq
\frac{2}{\pi}
\int_{0}^{\infty}
d\lambda\,
G_{\ell}(\lambda;a,b)
\left\|H_{\ell}^{1/4}v_{\ell,\lambda}^{-}\right\|_{L^{2}(\mathbb{R})}
\left\|H_{\ell}^{1/4}v_{\ell,\lambda}^{+}\right\|_{L^{2}(\mathbb{R})}.
\end{equation}

We comment on yet another subtlety in this expression. The integral from $(0,\infty)$ is defined in the limiting sense starting with the integral over a finite interval. The estimates below will show that the right-hand sides of \eqref{eq:TrTqell} and \eqref{eq:TrTpell} are finite. By the standard trace-class criterion for integrals of factorized kernels, this proves that the improper resolvent integrals converge in trace norm and define trace-class operators. Since their matrix elements agree on the dense space of smooth, compactly supported data, these operators are then equal to $T_{q,\ell}$ and $T_{p,\ell}$.

\subsubsection*{Bounds on the integrand of \eqref{eq:TrTqell} and \eqref{eq:TrTpell}}

The right-hand sides of \eqref{eq:TrTqell} and \eqref{eq:TrTpell} contain two different kinds of quantities. The first is the scalar Green's function coefficient $G_{\ell}(\lambda;a,b)$, which measures propagation from one side of the collar to the other. The remaining quantities are fractional $H_{\ell}$-norms of $v_{\ell,\lambda}^{\pm}$. We now establish bounds on both of these quantities. These bounds must not only be strong enough so the right-hand sides of \eqref{eq:TrTqell} and \eqref{eq:TrTpell} are integrable but also so that, ultimately, the sum over angular modes in \eqref{eq:TclassTqTplm} converges. 

We first apply the Cauchy-Schwarz inequalities, resulting in
\begin{equation}
\label{eq:H14bound}
||H_{\ell}^{1/4} v^{\pm}_{\ell,\lambda}||^{2}_{L^{2}(\mathbb{R})} \leq ||v^{\pm}_{\ell,\lambda}||_{L^{2}(\mathbb{R})} \cdot ||H^{1/2}_{\ell}v^{\pm}_{\ell,\lambda}||_{L^{2}(\mathbb{R})} 
\end{equation}
and 
\begin{equation}
\label{eq:H-14bound}
||H_{\ell}^{-1/4} v^{\pm}_{\ell,\lambda}||^{2}_{L^{2}(\mathbb{R})} \leq ||v^{\pm}_{\ell,\lambda}||_{L^{2}(\mathbb{R})} \cdot ||H^{-1/2}_{\ell}v^{\pm}_{\ell,\lambda}||_{L^{2}(\mathbb{R})} \,.
\end{equation}
Therefore, we need to bound $||v^{\pm}_{\ell,\lambda}||_{L^{2}}$, $||H^{1/2}_{\ell}v^{\pm}_{\ell,\lambda}||_{L^{2}(\mathbb{R})} $ and $||H^{-1/2}_{\ell}v^{\pm}_{\ell,\lambda}||_{L^{2}(\mathbb{R})} $. These norms explicitly depend on the choice of extension. The choice of extension we make is to define $v^{\pm}_{\ell,\lambda}(z)$ as continuous functions  which, in the collar region $[a,b]$, are given by  
\begin{equation}
\begin{aligned}
\label{eq:vpmcollar}
v_{\ell,\lambda}^{-}(z)
&=
\begin{cases}
1-\frac{z-a}{\delta_{\ell,\lambda}}\,,
    & a<z< a+\delta_{\ell,\lambda}\,, \\[2mm]
0\,,
    & a+\delta_{\ell,\lambda}\leq z\leq b\,,
\end{cases}
&\quad
v_{\ell,\lambda}^{+}(z)
&=
\begin{cases}
0\,,
    & a\leq z\leq b- \delta_{\ell,\lambda}\\[2mm]
\frac{z-b+\delta_{\ell,\lambda}}{\delta_{\ell,\lambda}}\,,
    & b-\delta_{\ell,\lambda}<z< b\,. \\[2mm]

\end{cases}
\end{aligned}
\end{equation}
In other words, each function enters the collar region and then decreases linearly over a distance $0<\delta_{\ell,\lambda}<\delta_{0}$ where we take $\delta_{0}<\tfrac{1}{2}|b-a|$. 

\paragraph*{\normalfont\underline{A bound on the norm of
$H^{1/2}v^{\pm}_{\ell,\lambda}$}}
To estimate $||H^{1/2}_{\ell}v^{\pm}_{\ell,\lambda}||_{L^{2}(\mathbb{R})} $ we note that the operator $H_{\ell}+\lambda^{2}$ naturally defines an ``energy'' on $L^{2}(\mathbb{R})$
\begin{equation}
E_{\ell,\lambda}[f] \equiv  \int_{-\infty}^{\infty}dz~(|Df|^{2}+\ell(\ell+1)W|f|^{2}+\lambda^{2}|f|^{2})
\end{equation}
where $D=\tfrac{d}{dz} - \tfrac{r^{\prime}}{r}$ and $W=(N/r)^{2}$. 
We denote $E^{-}_{\ell,\lambda}[f]$ as the energy integral restricted to the domain $(-\infty,a)$ and $E^{+}_{\ell,\lambda}[f]$ as the restriction to $(b,\infty)$. In terms of the energy, the norm we are interested in is $E_{\ell,0}[v^{\pm}_{\ell,\lambda}]=||H^{1/2}_{\ell}v^{\pm}_{\ell,\lambda}||^{2}_{L^{2}(\mathbb{R})}$. We focus on $v^{-}_{\ell,\lambda}$, and split the integral into the sum  
\begin{align}
\label{eq:twointegrals}
E_{\ell,0}[v^{-}_{\ell,\lambda}] =||H^{1/2}_{\ell}v^{-}_{\ell,\lambda}||^{2}_{L^{2}(\mathbb{R})}=& \int_{-\infty}^{a}dz~\big(|Du_{\ell,\lambda}^{-}|^{2} + \ell(\ell+1)W|u^{-}_{\ell,\lambda}|^{2} \big)  \nonumber \\
&+\int_{a}^{a+\delta_{\ell,\lambda}}dz~\big(|Dv_{\ell,\lambda}^{-}|^{2} + \ell(\ell+1)W|v^{-}_{\ell,\lambda}|^{2} \big)\,.
\end{align}
We first estimate the second integral of \eqref{eq:twointegrals}. On the transition interval, $\partial_{z}v^{-}_{\ell,\lambda}=-\delta_{\ell,\lambda}^{-1}$ and $\int_{a}^{a+\delta_{\ell,\lambda}}dz|v^{-}_{\ell,\lambda}|^{2}=\tfrac{\delta_{\ell,\lambda}}{3}$. Using the fact that $r^{\prime}/r$ and $W$ are bounded on this fixed compact neighborhood, there exists a constant $C>0$ such that 
\begin{equation}
\int_{a}^{a+\delta_{\ell,\lambda}}dz~\big(|Dv_{\ell,\lambda}^{-}|^{2} + \ell(\ell+1)W|v^{-}_{\ell,\lambda}|^{2} \big) \leq C \big(\delta^{-1}_{\ell,\lambda}+ (1+\ell(\ell+1))\delta_{\ell,\lambda}\big)\,.
\end{equation}
We now estimate the first integral of \eqref{eq:twointegrals}. To do so we note that
\begin{equation}
\label{eq:firstintegral}
\int_{-\infty}^{a}dz~\big(|Du_{\ell,\lambda}^{-}|^{2} + \ell(\ell+1)W|u^{-}_{\ell,\lambda}|^{2} \big)  \leq E_{\ell,\lambda}^{-}[u_{\ell,\lambda}^{-}]
\end{equation}
since their difference is manifestly positive. Therefore, it suffices to bound $E_{\ell,\lambda}^{-}[u_{\ell,\lambda}^{-}]$. To compute this we consider a trial function $f^{-}_{\textrm{trial}}$ with support from $[a-\delta_{\ell,\lambda},a]$
\begin{equation}
f_{\textrm{trial}}^{-}(z)
=
\begin{cases}
0,
    & z\leq a-\delta_{\ell,\lambda},\\[2mm]
1+\dfrac{z-a}{\delta_{\ell,\lambda}}\,,
    & a-\delta_{\ell,\lambda}<z\leq a\,.
\end{cases}
\end{equation}
The same estimates apply to $f_{\textrm{trial}}^{-}$ so we have that 
\begin{equation}
\label{eq:energyf}
E^{-}_{\ell,\lambda}[f_{\textrm{trial}}^{-}] \leq C [\delta^{-1}_{\ell,\lambda}+ \big(1+\ell(\ell+1)+\lambda^{2})\delta_{\ell,\lambda}\big)\,.
\end{equation}
To compare this with the energy of $u_{\ell,\lambda}^{-}$ we write $f_{\textrm{trial}}^{-}=u^{-}_{\ell,\lambda} + h$ where $h(a)=0$. Expanding the energy, we obtain $E^{-}_{\ell,\lambda}[f_{\textrm{trial}}^{-}]=E^{-}_{\ell,\lambda}[u^{-}_{\ell,\lambda}]+E^{-}_{\ell,\lambda}[h]+2I$ where 
\begin{equation}
\label{eq:I}
I = \int_{-\infty}^{a}dz~\bigg(\overline{D u^{-}_{\ell,\lambda}}Dh + [\ell(\ell+1)]W + \lambda^{2}]\overline{u^{-}_{\ell,\lambda}}h\bigg) = \int_{-\infty}^{a}dz h (H_{\ell}+\lambda^{2})\overline{u^{-}_{\ell,\lambda}}=0
\end{equation}
where in the second term we integrated by parts and used the fact that $h(a)=0$ and that $u_{\ell,\lambda}^{-}$ vanishes at $z\to-\infty$. Therefore, we have that $E^{-}_{\ell,\lambda}[f^{-}_{\textrm{trial}}] = E^{-}_{\ell,\lambda}[u^{-}_{\ell,\lambda}]+E^{-}_{\ell,\lambda}[h] + 2I\geq E^{-}_{\ell,\lambda}[u^{-}_{\ell,\lambda}]$. Combining this with \eqref{eq:firstintegral} and  \eqref{eq:energyf} we obtain 
\begin{equation}
E^{-}_{\ell,\lambda}[u^{-}_{\ell,\lambda}]\leq C \big(\delta^{-1}_{\ell,\lambda}+ (1+\ell(\ell+1)+\lambda^{2})\delta_{\ell,\lambda}\big)
\end{equation}
and so
\begin{equation}
    \label{eq:H12bound}
    ||H_{\ell}^{1/2}v^{-}_{\ell,\lambda}||^{2}_{L^{2}(\mathbb{R})} \leq C [\delta^{-1}_{\ell,\lambda}+ (1+\ell(\ell+1)+\lambda^{2})\delta_{\ell,\lambda}]\,.
\end{equation}
By an identical argument, $||H_{\ell}^{1/2}v^{+}_{\ell,\lambda}||^{2}_{L^{2}(\mathbb{R})}$ satisfies the same inequality. 

 We choose the width of the transition  $\delta_{\ell,\lambda}$ to be the minimum of $\delta_{0}$ and $(1+\ell(\ell+1)+\lambda^{2})^{-1/2}$. With this choice, the right-hand side of \eqref{eq:H12bound} is bounded from above by $C(1+\ell+\lambda)$. Therefore we obtain 
\begin{equation}
\label{eq:inequalities}
||H^{1/2}v^{\pm}_{\ell,\lambda}||^{2}_{L^{2}(\mathbb{R})}\leq C(1+\ell+\lambda) \,.
\end{equation}

\paragraph*{\normalfont\underline{A bound on the norm of $v^{\pm}_{\ell,\lambda}$}}
We will prove that the norm of $v^{\pm}_{\ell,\lambda}$ is bounded by
\begin{equation}
\label{eq:vpmbound}
||v^{\pm}_{\ell,\lambda}||_{L^{2}(\mathbb{R})}
\leq C(1+\lambda^{-1/2})
\end{equation}
where $C$ is a constant independent of $\ell$ and $\lambda$.
We first consider $v^{-}_{\ell,\lambda}$. We define the rescaled function
$f^{-}_{\ell,\lambda}=u^{-}_{\ell,\lambda}/r$, with which
$(H_{\ell}+\lambda^{2})u^{-}_{\ell,\lambda}=0$ becomes
\begin{equation}
\partial_{z}(r^{2}\partial_{z}f^{-}_{\ell,\lambda})
=
\big(\ell(\ell+1)N^{2}+\lambda^{2}r^{2}\big)f^{-}_{\ell,\lambda}\,.
\end{equation}
The right-hand side is non-negative and
$\lim_{z\to-\infty}r(z)^{2}\partial_{z}f^{-}_{\ell,\lambda}=0$, so
$f^{-}_{\ell,\lambda}$ is nondecreasing. Since
$u^{-}_{\ell,\lambda}(a)=1$, we have
$f^{-}_{\ell,\lambda}(a)=1/r(a)$ and hence
\begin{equation}
\label{eq:umonotonicity}
0<u^-_{\ell,\lambda}(z)\leq \frac{r(z)}{r(a)}
\quad (z\leq a)\,.
\end{equation}

We write
\begin{equation}
\label{eq:3integrals}
\begin{split}
||v^{-}_{\ell,\lambda}||_{L^{2}(\mathbb{R})}^{2}
=&
\int_{-\infty}^{z_{-}}dz~|u^{-}_{\ell,\lambda}(z)|^{2}
+\int_{z_{-}}^{a}dz~|u^{-}_{\ell,\lambda}(z)|^{2}\\
&+\int_{a}^{a+\delta_{\ell,\lambda}}dz~|v^{-}_{\ell,\lambda}(z)|^{2}
\end{split}
\end{equation}
where $z_{-}<a$ lies sufficiently far in the left asymptotic end.
To bound the first integral, we note that in the asymptotic region
$z\leq z_{-}$, the potential $U_{\ell}\geq0$, and so the mode
equation~\eqref{eq:modeq} implies that
$\partial_{z}^{2}u^{-}_{\ell,\lambda}
=(U_{\ell}+\lambda^{2})u^{-}_{\ell,\lambda}
\geq\lambda^{2}u^{-}_{\ell,\lambda}$.
Comparison with the solution of $y^{\prime\prime}=\lambda^{2}y$ that
decays as $z\to-\infty$, together with \eqref{eq:umonotonicity}, yields
\begin{equation}
0<u^{-}_{\ell,\lambda}(z)
\leq u^{-}_{\ell,\lambda}(z_{-})e^{-\lambda(z_{-}-z)}
\leq Ce^{-\lambda(z_{-}-z)}\,.
\end{equation}
Therefore, the first integral of~\eqref{eq:3integrals} is bounded by
\begin{equation}
\int_{-\infty}^{z_{-}}dz~|u^{-}_{\ell,\lambda}(z)|^{2}
\leq
C^{2}\int_{-\infty}^{z_{-}}dz~e^{-2\lambda(z_{-}-z)}
=\frac{C^{2}}{2\lambda}\,.
\end{equation}
By \eqref{eq:umonotonicity}, the second integral satisfies
\begin{equation}
\int_{z_{-}}^{a}dz~|u^{-}_{\ell,\lambda}(z)|^{2}
\leq
\frac{1}{r(a)^{2}}\int_{z_{-}}^{a}dz~r(z)^{2}
\leq C'\,.
\end{equation}
Explicitly evaluating the third integral using \eqref{eq:vpmcollar}, we
obtain
\begin{equation}
\int_{a}^{a+\delta_{\ell,\lambda}}
|v^{-}_{\ell,\lambda}(z)|^{2}dz
=\frac{\delta_{\ell,\lambda}}{3}
\leq\frac{\delta_{0}}{3}\,.
\end{equation}
Combining these estimates gives
\begin{equation}
||v^{-}_{\ell,\lambda}||^{2}_{L^{2}(\mathbb{R})}
\leq C''\bigg(1+\frac{1}{\lambda}\bigg)\,.
\end{equation}
The same argument applied to $v^{+}_{\ell,\lambda}$ implies that
$0<u^+_{\ell,\lambda}(z)\leq r(z)/r(b)$ for $z\geq b$ and gives
\begin{equation}
||v^{+}_{\ell,\lambda}||^{2}_{L^{2}(\mathbb{R})}
\leq C''\bigg(1+\frac{1}{\lambda}\bigg)\,.
\end{equation}
Taking the square root yields the desired bound.

\paragraph*{\normalfont\underline{A bound on the norm of $H_{\ell}^{-1/2}v^{\pm}_{\ell,\lambda}$}}
To estimate the norm of $H_{\ell}^{-1/2}v^{\pm}_{\ell,\lambda}$ we consider the full spatial function $\Phi_{\ell m\lambda}$ associated with $v^{\pm}_{\ell,\lambda}$ given by
\begin{equation}
\Phi^{\pm}_{\ell m\lambda}(z,x^{A})
=
\frac{v^{\pm}_{\ell,\lambda}(z)}{r(z)}Y_{\ell m}(x^{A})\,.
\end{equation}
Since the spherical harmonic decomposition is unitary and $H$ is spherically symmetric, we have
\begin{equation}
||H_{\ell}^{-1/2}v^{\pm}_{\ell,\lambda}||_{L^{2}(\mathbb{R})}
=
||H^{-1/2}\Phi^{\pm}_{\ell m\lambda}||_{L^{2}(\Sigma_{0})}\,.
\end{equation}
We also have that
\begin{align}
\|H^{-1/2}\Phi^{\pm}_{\ell m\lambda}\|_{L^{2}(\Sigma_{0})}
&=
\sup_{\psi\neq 0}
\frac{
|\langle\psi,\Phi^{\pm}_{\ell m\lambda}\rangle_{L^{2}(\Sigma_{0})}|
}{
\|H^{1/2}\psi\|_{L^{2}(\Sigma_{0})}
}
\nonumber\\
&\leq
\sup_{\psi\neq 0}
\frac{
\|\psi\|_{L^{6}(\Sigma_{0})}
\|\Phi^{\pm}_{\ell m\lambda}\|_{L^{6/5}(\Sigma_{0})}
}{
\|H^{1/2}\psi\|_{L^{2}(\Sigma_{0})}
}
\nonumber\\
&\leq
C\|\Phi^{\pm}_{\ell m\lambda}\|_{L^{6/5}(\Sigma_{0})}\,.
\end{align}
Here the supremum is over the domain of $H^{1/2}$. In the second line we used H\"older's inequality \eqref{eq:Holder-general} with conjugate exponents $6$ and $6/5$, and in the third line we used the inequality $||\psi||_{L^{6}(\Sigma_{0})}\leq C||H^{1/2}\psi||_{L^{2}(\Sigma_{0})}$ implied by the manipulations from \eqref{eq:sobloevnormq} to \eqref{eq:C4}. Therefore,
\begin{equation}
||H_{\ell}^{-1/2}v^{\pm}_{\ell,\lambda}||_{L^{2}(\mathbb{R})}
\leq
C||\Phi^{\pm}_{\ell m\lambda}||_{L^{6/5}(\Sigma_{0})}\,.
\end{equation}

We have thus reduced the problem to estimating the $L^{6/5}$-norm of $\Phi^{\pm}_{\ell m\lambda}$. Writing out this norm, we obtain
\begin{equation}
\label{eq:Phi6/5}
||\Phi^{\pm}_{\ell m\lambda}||^{6/5}_{L^{6/5}(\Sigma_{0})}
=
\bigg[\int_{\mathbb{R}}dz~|v^{\pm}_{\ell,\lambda}(z)|^{6/5}r(z)^{4/5}\bigg]
\bigg[\int_{\mathbb{S}^{2}}d\Omega~|Y_{\ell m}(x^{A})|^{6/5}\bigg]\,.
\end{equation}
Since $6/5<2$, applying H\"older's inequality \eqref{eq:Holder-general} yields
\begin{equation}
||Y_{\ell m}||_{L^{6/5}(\mathbb{S}^{2})}
\leq
(4\pi)^{1/3}||Y_{\ell m}||_{L^{2}(\mathbb{S}^{2})}
\leq
(4\pi)^{1/3}\,.
\end{equation}
The angular contribution to \eqref{eq:Phi6/5} is uniformly bounded, so we focus on the radial integral. By \eqref{eq:umonotonicity}, its right-hand analogue, and \eqref{eq:vpmcollar}, the radial integral over every fixed compact interval is uniformly bounded. It therefore remains to estimate the asymptotic tails. For $v^{-}_{\ell,\lambda}$, choose $z_{-}<a$ sufficiently far in the left asymptotic end and define $k\equiv z_{-}-z$. For $v^{+}_{\ell,\lambda}$, define $k\equiv z-z_{+}$, where $z_{+}>b$. We use the same letter $k$ since the resulting estimates are identical on the two ends.

The radial function satisfies $r(k)\asymp1+k$, and hence $r^{4/5}\leq C(1+k)^{4/5}$. Furthermore, the comparison estimate used above gives $|v^{\pm}_{\ell,\lambda}|\leq Ce^{-\lambda k}$ on the corresponding asymptotic end. Therefore,
\begin{equation}
|v^{\pm}_{\ell,\lambda}|^{6/5}r^{4/5}
\leq
Ce^{-6\lambda k/5}(1+k)^{4/5}
\end{equation}
and so
\begin{equation}
\label{eq:PhiIlambad}
||\Phi^{\pm}_{\ell m\lambda}||^{6/5}_{L^{6/5}(\Sigma_{0})}
\leq
C\big[1+I(\lambda)\big]
\end{equation}
where
\begin{equation}
\label{eq:Ilambda}
I(\lambda)
\equiv
\int_{0}^{\infty}dk~e^{-6\lambda k/5}(1+k)^{4/5}\,.
\end{equation}
We first consider $0<\lambda\leq1$. Changing variables to $y=\lambda k$ gives
\begin{equation}
I(\lambda)
=
\lambda^{-1}\int_{0}^{\infty}dy~e^{-6y/5}
\bigg(1+\frac{y}{\lambda}\bigg)^{4/5}
\qquad (0<\lambda\leq1)\,.
\end{equation}
For $\lambda\in(0,1]$, we have $(1+\tfrac{y}{\lambda})^{4/5}\leq\lambda^{-4/5}(1+y)^{4/5}$. Substituting this bound gives
\begin{equation}
I(\lambda)
\leq
\lambda^{-9/5}\int_{0}^{\infty}dy~e^{-6y/5}(1+y)^{4/5}
=
C_{\ast}\lambda^{-9/5}
\qquad (0<\lambda\leq1)\,.
\end{equation}
For $\lambda>1$, we have $e^{-6\lambda k/5}\leq e^{-6k/5}$, and so $I(\lambda)$ is uniformly bounded. Thus, for all $\lambda>0$,
\begin{equation}
I(\lambda)\leq C(1+\lambda^{-9/5})\,.
\end{equation}
Substituting this into \eqref{eq:PhiIlambad}, we find $||\Phi^{\pm}_{\ell m\lambda}||_{L^{6/5}(\Sigma_{0})}
\leq
C(1+\lambda^{-3/2})$
and hence
\begin{equation}
\label{eq:H-1/2ineq}
||H_{\ell}^{-1/2}v^{\pm}_{\ell,\lambda}||_{L^{2}(\mathbb{R})}
\leq
C(1+\lambda^{-3/2})\,.
\end{equation}
The key point is that this bound is uniform in $\ell$ and shows that, as $\lambda\to0$, the norm grows at most like $\lambda^{-3/2}$.

\paragraph*{\normalfont\underline{A bound on the coefficient $G_{\ell}(\lambda;a,b)$}}
We recall that, physically, $G_{\ell}(\lambda;a,b)$ is the propagator between the modes on the left and right of the collar. We will show that transmission through the potential barrier implies that 
\begin{equation}
\label{eq:Gellbound}
G_{\ell}(\lambda;a,b)\leq Ce^{-\gamma (\ell+\lambda)}\,.
\end{equation}
To prove this bound we first consider the case of $\ell=\lambda=0$. In this regime the solutions $u^{\pm}_{0,0}$ are of the form 
\begin{equation}
u^{-}_{0,0}(z)=\frac{r(z)}{C_{a}}\int_{-\infty}^{z}\frac{dy}{r(y)^{2}}\quad \quad u^{+}_{0,0}(z)=\frac{r(z)}{C_{b}}\int_{z}^{\infty}\frac{dy}{r(y)^{2}}
\end{equation}
where $C_{a}=r(a)\int_{-\infty}^{a}\tfrac{dy}{r(y)^{2}}$ and $C_{b}=r(b)\int_{b}^{\infty}\tfrac{dy}{r(y)^{2}}$. Since $r(z)\asymp1+|z|$ the integral converges and so the Wronskian $W_{0,0}=\frac{1}{C_{a}C_{b}}\int_{\mathbb{R}}\tfrac{dy}{r(y)^{2}}$ also converges. It follows that $G_{0}(0;z,z^{\prime})<\infty$.

We now obtain a bound on $G_{\ell}(\lambda;b,b)$. We note that the kernels $G_{\ell}$ given by \eqref{eq:Greensfunction} are manifestly nonnegative, $H_{\ell}-H_{0}\geq 0$ and we have the identity 
\begin{equation}
(H_{0}+\lambda^{2})^{-1} - (H_{\ell}+\lambda^{2})^{-1} = (H_{0}+\lambda^{2})^{-1}(H_{\ell}-H_{0})(H_{\ell}+\lambda^{2})^{-1}.
\end{equation}
The operator on the right has a non-negative kernel and hence $G_{\ell}(\lambda;z,z^{\prime})\leq G_{0}(\lambda;z,z^{\prime})$. Taking $\lambda_{\ast}\to 0$ 
\begin{equation}
    (H_{0}+\lambda_{\ast}^{2})^{-1}-(H_{0}+\lambda^{2})^{-1} = (\lambda^{2}-\lambda_{\ast}^{2})(H_{0}+\lambda_{\ast}^{2})^{-1}(H_{0}+\lambda^{2})^{-1}
\end{equation}
The kernel on the right-hand side is again non-negative. Taking $\lambda_{\ast}\to 0$ we obtain $G_{0}(\lambda;z,z^{\prime})\leq G_{0}(0,z,z^{\prime})$.  Combining these inequalities yields $0\leq G_{\ell}(\lambda;z,z^{\prime})\leq G_{0}(\lambda;z,z^{\prime})\leq G_{0}(0;z,z^{\prime})$. In particular, at the endpoint $b$ we obtain 
\begin{equation}
G_{\ell}(\lambda;b,b)\leq G_{\ell}(0;b,b)< C_{0}
\end{equation}
However, we will need the stronger bound of \eqref{eq:Gellbound} to prove that \eqref{eq:TrTqell} and \eqref{eq:TrTpell} converge. Furthermore, the exponential decay in $\ell$ will ensure that the sum over angular modes in the full trace of $|T|$ converges.  

To obtain this we now establish the stronger bound \eqref{eq:Gellbound} for large $\ell$ and $\lambda$. Using the definition of the Wronskian it is straightforward to show that
\begin{equation}
G_{\ell}(\lambda;a,b)=G_{\ell}(\lambda;b,b)\frac{u^{-}_{\ell,\lambda}(a)}{u^{-}_{\ell,\lambda}(b)}\,.
\end{equation}
Since $u^{-}_{\ell,\lambda}(a)=1$ and we already established that $G_{\ell}(\lambda;b,b)\leq C_{0}$ we just need to bound $u^{-}_{\ell,\lambda}(b)$. To obtain this bound we note that, in the collar $[a,b]$,  $r^{\prime \prime}/r>0$ and $W\geq C_{0}'>0$. Using these inequalities together with the fact that $(H_{\ell}+\lambda^{2})u^{-}_{\ell,\lambda}=0$ implies that 
\begin{equation}
\partial_{z}^{2}u^{-}_{\ell,\lambda}\geq k^{2}_{\ell,\lambda}u^{-}_{\ell,\lambda}
\end{equation}
where  $k_{\ell,\lambda}^{2}=\lambda^{2}+C_{0}\ell(\ell+1)$. Therefore we can obtain a bound on $u^{-}_{\ell \lambda}$ relative to solutions of the second order ODE $\partial_{z}^{2}y=k_{\ell,\lambda}^{2}y$. To obtain this bound we also need to bound the initial data $u_{\ell,\lambda}^{-}(a)$ and $\partial_{z}u_{\ell,\lambda}^{-}(a)$. From our normalization, $u_{\ell,\lambda}^{-}(a)=1$ and it follows from $E^{-}[u_{\ell,\lambda}^{-}]\geq 0$ and $(H_{\ell}+\lambda^{2})u_{\ell,\lambda}^{-}=0$ that there exists a constant $C_{1}$ such that  
\begin{equation}
\partial_{z}u_{\ell,\lambda}^{-}(a)=E^{-}[u_{\ell,\lambda}^{-}]+\frac{r^{\prime}(a)}{r(a)}\geq -C_{1}
\end{equation}
where $C_{1}$ is independent of $\ell,\lambda$. Comparing with the solution of $\partial_{z}^{2}y=k_{\ell,\lambda}^{2}y$ for $y(a)=1$ and $y^{\prime}(a)=-C_{1}$, we obtain 
\begin{equation}
u^{-}_{\ell \lambda}(b) \geq \cosh(k_{\ell, \lambda}(b-a)) - \frac{C_{1}}{k_{\ell \lambda}}\sinh(k_{\ell, \lambda}(b-a))\,.
\end{equation}
For $k_{\ell ,\lambda}>2C_{1}$ we find that there exists a constant $\gamma>0$ such that $u^{-}_{\ell,\lambda}(b)\geq C_{2}e^{\gamma k_{\ell,\lambda}(b-a)}$. 
This yields the desired bound \eqref{eq:Gellbound}. 

\subsubsection{Proof of trace-class condition}
\label{subsubsec:prooftraceclass}
The estimates we have obtained are strong enough to prove that $T$ is a trace-class operator. We first note that \eqref{eq:inequalities} and \eqref{eq:H-1/2ineq} imply that, as assumed, our choice of extension satisfies $v^{\pm}_{\ell,\lambda}\in \textrm{Dom}(H_{\ell}^{1/2})\cap \textrm{Dom}(H_{\ell}^{-1/2})$. Furthermore, using \eqref{eq:H14bound} and \eqref{eq:H-14bound}, these inequalities together with \eqref{eq:vpmbound} result in
\begin{equation}
||H_{\ell}^{1/4}v^{\pm}_{\ell,\lambda}||^{2}_{L^{2}(\mathbb{R})}\leq C_{1} (1+\ell+\lambda)^{1/2}(1+\lambda^{-1/2})\,, \quad \quad ||H_{\ell}^{-1/4}v^{\pm}_{\ell,\lambda}||^{2}_{L^{2}(\mathbb{R})}\leq C_{2} (1+\lambda^{-2})
\end{equation}
where $C_{1}$ and $C_{2}$ are independent of $\ell$ and $\lambda$. Plugging these estimates into the trace \eqref{eq:TrTqell} together with the bound on the propagator \eqref{eq:Gellbound} yields 
\begin{equation}
\textrm{Tr}|T_{q,\ell}| \leq Ce^{-\gamma \ell}\int_{0}^{\infty}d\lambda~e^{-\gamma \lambda} \lambda^{2}(1+\lambda^{-2}) \leq C_{q}e^{-\gamma \ell}
\end{equation}
where we note that the integral converges at small $\lambda$ due to the factor of $\lambda^{2}$ from the resolvent formula. Similarly, for \eqref{eq:TrTpell}, we obtain 
\begin{align}
\textrm{Tr}|T_{p,\ell}| \leq& Ce^{-\gamma \ell}\int_{0}^{\infty}d\lambda~e^{-\gamma \lambda}(1+\ell+\lambda)^{1/2}(1+\lambda^{-1/2})  \nonumber \\
\leq& C(1+\ell)^{1/2}e^{-\gamma \ell}\int_{0}^{\infty}d\lambda~e^{-\gamma \lambda}(1+\lambda)^{1/2}(1+\lambda^{-1/2}) \nonumber \\
\leq &C_{p}(1+\ell)^{1/2}e^{-\gamma \ell}\,.
\end{align}
We note that these estimates also justify taking the integration range to be $0\leq \lambda \leq \infty$ on the right-hand side of \eqref{eq:TrTqell} and \eqref{eq:TrTpell}. The exponential decay ensures that  $\textrm{Tr}_{\mathcal{K}_{\textrm{split}}}|T|$ given by \eqref{eq:TclassTqTplm} converges and so we finally obtain that $T$ is, indeed, a trace-class operator on $\mathcal{K}_{\textrm{split}}$. 

We have therefore established both sufficient conditions stated in~\ref{eq:quantization}: the covariances $\mu_0$ and $\mu_{\mathrm{split}}$ induce quasiequivalent norms by~\eqref{eq:topology}, and their relative covariance operator $T$ is trace class by~\eqref{eq:traceclass}. It follows that $\Omega_{\mathrm{split}}$ is a normal product state for $\A(W_{\rm L})$ and $\A(W_{\rm R})$ in the vacuum representation on $\mathcal H_{\mathrm{out}}$, as stated in~\eqref{eq:Omegasplit}. Then the pair $\A(W_{\rm L})$, $\A(W_{\rm R})$ is split, and there exists a Type I factor $\mathcal N$ such that
\begin{equation}
    \mathcal A(W_{\rm L}) \subset \mathcal N \subset \mathcal A(W_{\rm R})'\,.
\end{equation}
Finally as explained in~\ref{subsection:geometry}, unitary evolution to the undeformed Schwarzschild region then gives the split inclusion~\eqref{eq:undeformed-split}.

\subsection{Summary and generalizations}
\label{subsubsection:highdim}
While the preceding proof was necessarily technical in order to rigorously establish the required bounds on the trace, we can now re-examine the arguments in hindsight and attempt to distill the key ingredients. The crucial step was reducing the problem to a one-dimensional ODE and expressing the trace in terms of solutions of $(-\partial_{z}^{2}+U_{\ell}+\lambda^{2})u=0$ for $\lambda>0$. We write $U_{\ell}=U_{0}+\ell(\ell+1)W$ for $W=N^{2}/r^{2}$, where $W\geq C>0$ and $U_{0}\geq -C_{1}$ on the compact collar. Hence any mode with angular momentum $\ell>0$ encounters an effective potential barrier of height at least $\ell(\ell+1)C$. Physically, a mode with angular wavelength $r/\ell$ has transverse gradient energy of order $\ell^{2}/r^{2}$. For this mode to correlate on opposite sides of the collar, it must penetrate radially through the angular momentum barrier. For $\lambda>0$, the radial decay length of the modes $u_{\ell \lambda}^{\pm}$ is approximately $k_{\ell,\lambda}\sim \sqrt{\lambda^{2}+C\ell(\ell+1)}$ at large $\ell$ and $\lambda$. So propagating across the collar of width $b-a$ is exponentially suppressed $G_{\ell}(\lambda;a,b)\leq C e^{-(b-a)k_{\ell,\lambda}}\leq C e^{-\gamma(\lambda+\ell)}$ for some $\gamma>0$. The preceding proof merely makes this heuristic argument precise and controls the low $\ell$ and $\lambda$ behavior. 

While it was convenient in the proof to choose the collar $\Sigma_{\rm C}$ to straddle the throat, this was not essential in our arguments. Only two
collar properties were used: positive separation
of the two regions and, on each fixed compact collar, the geometric coefficients entering the radial operator admit bounds that are uniform in $\ell$ and $\lambda$. Indeed, while the particular constants entering the proof would change, the method can be straightforwardly applied when the collar is displaced by any finite amount into the right or left wedges. Therefore, the split property applies to any disjoint subregions $\Sigma_{\textrm{L}},\Sigma_{\textrm{R}}\cong (0,\infty)\times \mathbb{S}^{2} $ separated by any finite, spacelike collar $\Sigma_{\rm C}$. We will use this fact in appendix \ref{app:fns}. 

Additionally in the detailed proof of this appendix we have, for simplicity, restricted to four dimensions. However, there is no difficulty in generalizing our proof to higher dimensions. Indeed, even in the above heuristic argument, the only change is that modes now encounter a barrier height of $\ell(\ell+d-3)C$ and so their characteristic decay length is now of order  $k_{\ell,\lambda}\sim \sqrt{\lambda^{2}+C\ell(\ell+d-3)}$. This again yields exponential suppression of modes crossing the barrier. We claim that, similarly, the detailed proof in this appendix can be straightforwardly generalized to $d>4$, where the preceding estimates now depend on the spacetime dimension. Additionally, since the same physical ingredients apply equally to free Maxwell fields and linearized gravitons, we see no essential difficulty in adapting our methods of proof to these cases.

\section{The Hartle-Hawking Weight is Faithful, Normal, and Semifinite} \label{app:fns}
In this appendix we construct the Hartle-Hawking weight $\Psi_{\textrm{HH}}$ and show that it is a faithful, normal, semifinite weight on $\mathcal{A}(\mathcal{R})\subset\mathcal{B}(\mathcal{H})$, where $\H$ is the vacuum-at-infinity Hilbert space. We denote this representation of the algebra as $\pi_{0}$. As emphasized in sec.~\ref{subsec:HHUH}, the Hartle-Hawking state is not normal in the representation. However, it is normal in its own GNS representation $\pi_{\textrm{HH}}$ with corresponding von Neumann algebra $\mathcal{A}_{\textrm{HH}}(\mathcal{R})$. The reason that they lie in disjoint representations is because the Hartle-Hawking state looks like a thermal state near spatial infinity whereas $\mathcal{H}$ contains a dense set of states which approach the Minkowski vacuum state at infinity. On both of the algebras, the Schwarzschild time translations act as automorphisms of the algebra. We denote these automorphisms on their respective algebras as 
\begin{equation}
\alpha_{t}^{(0)}:\mathcal{A}(\mathcal{R}) \to \mathcal{A}(\mathcal{R})\,, \qquad \alpha_{t}^{(\textrm{HH})}:\mathcal{A}_{\textrm{HH}}(\mathcal{R}) \to \mathcal{A}_{\textrm{HH}}(\mathcal{R})  
\end{equation}
where $t\in \mathbb{R}$. In both cases, $\alpha_{t}^{(0)}$ and $\alpha_{t}^{(\textrm{HH})}$ evolve their respective algebras for a Schwarzschild time $t$, where we have set the inverse temperature $\beta=1$. In this appendix we will prove that there exists a faithful, normal, semifinite weight $\Psi_{\textrm{HH}}$ on $\mathcal{A}(\mathcal{R})$ such that the modular flow $\sigma_{t}^{\Psi_{\textrm{HH}}}$ induced by that weight is equivalent to Schwarzschild time evolution
\begin{equation}
\label{eq:HHmodflow}
\sigma_{t}^{\Psi_{\textrm{HH}}}=\alpha_{-t}^{(0)}\,, \quad \quad t\in\mathbb{R}
\end{equation}
and is unique up to a positive multiplicative constant. 

\subsection*{Outline of the proof}

Before providing the complete proof, we first present a sketch of our proof strategy and highlight some of the key steps. The basic idea is that it suffices to find a fixed faithful normal state $\rho$ on $\mathcal A(\mathcal R)$ such that, for every $t\in\mathbb R$,
\begin{equation}
\alpha_{-t}^{(0)}\circ\sigma_{-t}^{\rho}
\in\operatorname{Inn}\!\left(\mathcal A(\mathcal R)\right)\,.
\label{eq:pointwiseinner}
\end{equation}
Equivalently, for each $t$ there is a unitary
$w_t\in\mathcal A(\mathcal R)$ satisfying
$\alpha_{-t}^{(0)}
=\operatorname{Ad}(w_t)\circ\sigma_t^\rho$.
This initially gives the unitaries only pointwise in $t$. We must then choose them to form a strongly continuous $\sigma^\rho$-cocycle,
\begin{equation}
w_{t_1+t_2}=w_{t_1}\,\sigma_{t_1}^\rho(w_{t_2})\,.
\end{equation}
Connes' converse cocycle theorem then produces a faithful, normal, semifinite weight $\Psi_{\textrm{HH}}$ whose modular flow is $\sigma_t^{\Psi_{\textrm{HH}}}=\alpha_{-t}^{(0)}$.

We first establish \eqref{eq:pointwiseinner} separately for each fixed $t$. While the two representations $\pi_{0}$ and $\pi_{\textrm{HH}}$ disagree at infinity they are locally quasiequivalent.  In particular, they agree near the bifurcation surface and so admit a compatible identification there. We then use the split property to separate off these near-horizon degrees of freedom:
\begin{equation}
\label{eq:ARtensorprod}
\mathcal A(\mathcal R)
=
\mathcal C\;\overline\otimes\;
\mathcal B(\mathcal H_{\textrm R})\,,
\end{equation}
where $\mathcal{C}$ is a Type III$_{1}$ algebra localized in a finite collar near the bifurcation surface and $\mathcal B(\mathcal H_{\textrm R})$ is a Type I factor containing the remaining degrees of freedom in the right exterior. For each fixed $t$, we first compare the two flows near the bifurcation surface, where the representations are locally identified. If necessary, we enlarge the collar so that it contains both $\mathcal C$ and $\alpha_{-t}^{(0)}(\mathcal C)$. We then choose auxiliary reference states in the two representations whose modular flows agree on the identified collar.

We next use the fact that, in the Hartle-Hawking representation, Schwarzschild time evolution is already a modular flow, so the Connes' cocycle theorem allows us to relate it to the auxiliary modular flow by an inner automorphism. The local identification of algebras transfers this comparison to the collar algebra in $\mathcal A(\mathcal R)$. The split property ensures that any remaining difference acts only on a Type~I tensor factor and is therefore also inner. This proves \eqref{eq:pointwiseinner} for each $t$.

Although the reference states used in this construction may depend on $t$, they can all be compared with a single fixed faithful normal state $\rho$. We can then choose the implementing unitaries to form a strongly continuous $\sigma^\rho$-cocycle. Connes' converse cocycle theorem identifies this cocycle with a faithful, normal, semifinite weight whose modular flow is $\alpha_{-t}^{(0)}$.

\subsection*{Assumptions}

We state, for reference, the necessary ingredients of the proof as assumptions on the representations $\pi_{0}$ and $\pi_{\textrm{HH}}$ and their corresponding spacetime net of von Neumann algebras. 
\begin{enumerate}
\item \label{assump1} \underline{Time slice axiom:} If $\Sigma$ is a
Cauchy surface for a globally hyperbolic region $O$ and $O_{\Sigma}\subset O$
is any open neighborhood of $\Sigma$, then
\begin{equation}
\mathcal A(O_{\Sigma})=\mathcal A(O)\,,
\qquad
\mathcal A_{\mathrm{HH}}(O_{\Sigma})=\mathcal A_{\mathrm{HH}}(O)\,.
\end{equation}
We denote these algebras equivalently by $\mathcal A(\Sigma)$ and
$\mathcal A_{\mathrm{HH}}(\Sigma)$\,.

\item  \label{assump2} \underline{Factors and Haag duality:} $\mathcal{A}(\mathcal{R})$ and $\mathcal{A}_{\textrm{HH}}(\mathcal{R})$ are factors, and their commutants are the corresponding algebras of the left wedge 
\begin{equation}
\A(\mathcal L)=\mathcal{A}(\mathcal{R})'\,, \qquad \A_{\textrm{HH}}(\mathcal L)=\mathcal{A}_{\textrm{HH}}(\mathcal{R})^{\prime}\,.
\label{eq:HHexteriorduality}
\end{equation}
Choosing a reflection-symmetric Cauchy surface $\Sigma\cong \mathbb R\times \mathbb{S}^{d-2}$ and a proper radial coordinate $s$ with the bifurcation sphere at $s=0$ such that
\begin{equation}
\Sigma_{\textrm{C},a}=\{|s|\leq a\}\,,\qquad
\Sigma_{\textrm{L},a}=\{s<-a\}\,,\qquad
\Sigma_{\textrm{R},a}=\{s>a\}\,,
\end{equation}
we then additionally assume that 
\begin{equation}
\A(\Sigma_{\textrm{C},a})
=
\big(\A(\Sigma_{\textrm{L},a})\vee\A(\Sigma_{\textrm{R},a})\big)'\,.
\label{eq:HHcollarduality}
\end{equation}
\item  \label{assump3} \underline{Local quasiequivalence and wedge compatibility:} For every bounded
region $O$, the canonical identification extends to compatible, normal
$*$-isomorphisms $\theta_O:\A(O)\longrightarrow\A_{\mathrm{HH}}(O),$
which preserve wedge localization in the following sense 
\begin{equation}
\label{eq:wedgecompatibility}
\theta_O\!\left(\A(O)\cap\A(\mathcal{R})\right)
=
\A_{\mathrm{HH}}(O)\cap\A_{\mathrm{HH}}(\mathcal{R})\,.
\end{equation}
We also assume the analogous condition with $\mathcal R$ replaced by
$\mathcal L$.

\item \label{assump4} \underline{Two-sided split property:} For every sufficiently large $a$, there exists a constant $b>a$ and Type I$_{\infty}$ factors $\mathcal{N}_{\textrm{L}}$ and $\mathcal{N}_{\textrm{R}}$ such that the algebras admit the following split inclusion 
\begin{equation}
\label{eq:split}
\mathcal{A}(\Sigma_{\textrm{L},b}) \subset \mathcal{N}_{\textrm{L}}\subset \mathcal{A}(\Sigma_{\textrm{L},a})\,, \qquad \mathcal{A}(\Sigma_{\textrm{R},b}) \subset \mathcal{N}_{\textrm{R}}\subset \mathcal{A}(\Sigma_{\textrm{R},a})\,.
\end{equation}
\item  \label{assump5} \underline{Covariance of Schwarzschild evolution:} Schwarzschild evolution acts normally on the local algebras and acts as a continuous automorphism group $\alpha_{t}^{(0)}$ of $\mathcal{A}(\mathcal{R})$ and $\alpha_{t}^{(\textrm{HH})}$ of $\mathcal{A}_{\textrm{HH}}(\mathcal{R})$. The maps $\alpha_{t}^{(0)}$ and $\alpha_{t}^{(\textrm{HH})}$ carry every compact Cauchy neighborhood to a larger compact Cauchy neighborhood and satisfy an intertwining relation:

If $\tau_{t}$ is the Schwarzschild time evolution such that $\alpha_{t}^{(0)}(\mathcal{A}(\mathcal{O}))=\mathcal{A}(\tau_{t}(\mathcal{O}))$ and $\alpha_{t}^{(\textrm{HH})}(\mathcal{A}_{\textrm{HH}}(\mathcal{O}))=\mathcal{A}_{\textrm{HH}}(\tau_{t}(\mathcal{O}))$, then for every bounded subregion $O$ and $a\in \mathcal{A}(O)$ we have that 
\begin{equation}
\label{eq:intertwining}
\theta_{\tau_{t}O}(\alpha^{(0)}_{t}(a))=\alpha_{t}^{(\textrm{HH})}(\theta_{O}(a))\,.
\end{equation}

\item  \label{assump6} \underline{Hartle-Hawking state:} There is a faithful and normal Hartle-Hawking state $\omega_\mathrm{HH}$ on $\mathcal{A}_{\textrm{HH}}(\mathcal{R})$ whose modular flow is the Schwarzschild evolution $\alpha^{(\mathrm{HH})}_{-t}$.

\end{enumerate}

Assumptions (\ref{assump1}) -- (\ref{assump6}) will be sufficient 
to prove the existence of a Hartle-Hawking weight on $\mathcal A(\mathcal R)$. The time-slice axiom, factoriality, Haag duality, local quasiequivalence, and geometric covariance are standard structural properties of local quantum field theory \cite{Brunetti:2001dx,Verch:1992eg,Verch:1996wv}. They have, in particular, been rigorously established for bounded regions in free theory and the above assumptions reflect the expectation that these properties extend to the wedges considered here. The  assumptions which concern only the Hartle-Hawking representation have been proven in \cite{Kay:1985zs,Kay:1988mu,Sanders:2013vza,Kay:2025vpj}.  While we expect the
remaining assumptions involving $\pi_0$ and its compatibility with
$\pi_{\textrm{HH}}$ can be similarly proven,\footnote{We briefly sketch an argument for a free field: The time-slice axiom follows from well-posed
Cauchy evolution, while local quasiequivalence follows because the
deformation state is Hadamard \cite{Verch:1992eg}. Following \cite{Kay:1985zs,Kay:2025vpj} for $\pi_{0}$, wedge duality  \eqref{eq:HHexteriorduality} and that the wedge algebras are factors should follow from the ``one-particle'' quantization structures obtained in Appendix
\ref{app:split}. Wedge compatibility \eqref{eq:wedgecompatibility}
 follows from duality and the local quasiequivalence. The two-sided split property is proved in Appendix \ref{app:split}. Schwarzschild evolution preserves the $\pi_0$-folium by the unitary equivalence of the original and time-translated  representations in the deformed spacetime (see footnote \ref{foot:defindep}). Finally, the intertwining relation \eqref{eq:intertwining} follows since both representations carry the same geometric automorphism of the abstract Weyl net.} we have not attempted to rigorously show this here. Of the assumptions on $\pi_{0}$, the one that requires substantial additional justification is the split property which is known to fail for general, unbounded regions. In appendix \ref{app:split}, we present a proof of the split inclusion required in \eqref{eq:split}.

\subsection{Splitting off the asymptotic regions}

We first show the factorization \eqref{eq:ARtensorprod}.  Fix $a$ and $b$ as in assumption (\ref{assump4}) and define the Type I factor 
\begin{equation}
\mathcal{F} \equiv (\mathcal{N}_{\textrm{L}}\vee \mathcal{N}_{\textrm{R}})^{\prime} \subset \mathcal{B}(\mathcal{H})\,.
\end{equation}
By locality, $\mathcal{N}_{\textrm{L}}$ and $\mathcal{N}_{\textrm{R}}$ are commuting Type I factors, so their tensor product decompositions of $\mathcal{H}$ can be performed simultaneously. In particular, the Hilbert space decomposes as $\mathcal{H}=\mathcal{H}_{\textrm{L}}\otimes \mathcal{H}_{\textrm{C}}\otimes \mathcal{H}_{\textrm{R}}$ with $\mathcal{N}_{\textrm{L}}=\mathcal{B}(\mathcal{H}_{\textrm{L}})\otimes \mathbb{1}\otimes \mathbb{1}$, $\mathcal{N}_{\textrm{R}}= \mathbb{1}\otimes \mathbb{1}\otimes\mathcal{B}(\mathcal{H}_{\textrm{R}})$ and hence $\mathcal{F}=\mathbb{1}\otimes \mathcal{B}(\mathcal{H}_{\textrm{C}})\otimes \mathbb{1}$. Furthermore, $\mathcal{F}$ is localized within the collar. This follows from the fact that (1) the observables in $\Sigma_{\textrm{C},a}$ commute with $\mathcal{N}_{\textrm{L}}\vee \mathcal{N}_{\textrm{R}}$ and (2) Eq.~\eqref{eq:HHcollarduality} at radius $b$ yields $\mathcal{F}\subset (\mathcal{A}(\Sigma_{\textrm{L},b})\vee \mathcal{A}(\Sigma_{\textrm{R},b}))^{\prime}=\mathcal{A}(\Sigma_{\textrm{C},b})$. Therefore, we have 
\begin{equation}
\mathcal{A}(\Sigma_{\textrm{C},a})\subset \mathcal{F} \subset \mathcal{A}(\Sigma_{\textrm{C},b})\,.
\end{equation}

We note that the algebra $\mathcal{A}(\mathcal{R})$ contains $\mathcal{N}_{\textrm{R}}$ and commutes with $\mathcal{N}_{\textrm{L}}$ --- i.e., $\mathbb{1}\otimes \mathbb{1}\otimes \mathcal{B}(\mathcal{H}_{\textrm{R}})\subset \mathcal{A}(\mathcal{R}) \subset \mathbb{1}\otimes \mathcal{B}(\mathcal{H}_{\textrm{C}}\otimes \mathcal{H}_{\textrm{R}})$. The Type I tensor-factor decomposition implies\footnote{If $\one\otimes\B(\mathcal K)\subset M\subset\B(\H'\otimes\mathcal K)$, then $M'\subset\big(\one\otimes\B(\mathcal K)\big)'=\B(\H')\otimes\one$, so $M'=A\otimes\one$ for some von Neumann algebra $A$ on $\H'$, and taking commutants again gives $M=A'\,\overline{\otimes}\,\B(\mathcal K)$.}
\begin{equation}
\mathcal{A}(\mathcal{R}) = \one\otimes \mathcal{C}\;\overline{\otimes}\;\B(\H_{\mathrm R})\,,
\qquad
\mathcal{A}(\mathcal{R})' = \B(\H_{\mathrm L})\;\overline{\otimes}\;\tilde{\mathcal{C}}\otimes\one\,,
\qquad
\mathcal{C}\equiv \mathcal{A}(\mathcal{R})\cap \mathcal{F}\,,
\label{eq:M0split}
\end{equation}
where $\tilde{\mathcal{C}}\equiv \mathcal{C}'\cap \mathcal{F}$ is the relative commutant of $\mathcal{C}$ inside $\mathcal{F}$, and the second identity follows from the first by taking commutants. Since $\mathcal{A}(\mathcal{R})$ is a factor, so is $\mathcal{C}$. This leads to the desired factorization \eqref{eq:ARtensorprod}.

We now prove that an analogous factorization holds for $\mathcal{A}_{\textrm{HH}}(\mathcal{R})$. Let $\theta$ be the canonical identification on the bounded collar. Since $\mathcal{C}\subset \mathcal{F}\subset \mathcal{A}(\Sigma_{\textrm{C},b})$ and $\mathcal{C}\subset \mathcal{A}(\mathcal{R})$,  wedge compatibility implies that 
\begin{equation}
\theta(\mathcal{C}) \subset \mathcal{A}_{\textrm{HH}}(\Sigma_{\textrm{C},b})\cap \mathcal{A}_{\textrm{HH}}(\mathcal{R}) \subset \mathcal{A}_{\textrm{HH}}(\mathcal{R})\,.
\end{equation}
Moreover, since $\theta$ is a normal $\ast$-isomorphism and $\mathcal{F}$ is a Type I factor, it follows that $\theta(\mathcal{F})$ is also a Type I factor. We may therefore write
\begin{equation}
\label{eq:HHHilb}
\mathcal H_{\textrm{HH}}
=
\mathcal H_{\textrm C}\otimes\mathcal K_{\textrm{HH}}\,,
\qquad
\theta(\mathcal F)
=
\mathcal B(\mathcal H_{\textrm C})\otimes\one\,,
\end{equation}
where $\mathcal K_{\textrm{HH}}$ is an infinite-dimensional, separable Hilbert space. Since $\tilde{\mathcal C}=\mathcal C'\cap\mathcal F$ and $\theta$ is a normal $\ast$-isomorphism on $\mathcal F$, it preserves relative commutants. Therefore, $\theta(\tilde{\mathcal{C}})=\theta(\mathcal C)'\otimes\one$ where the commutant on the right-hand side is taken in
$\mathcal B(\mathcal H_{\textrm C})$. Taking the commutant in $\mathcal B(\mathcal H_{\textrm{HH}})$ then gives
\begin{equation}
\theta(\tilde{\mathcal C})'
=
\theta(\mathcal C)\;\overline\otimes\;
\mathcal B(\mathcal K_{\textrm{HH}})\,.
\label{eq:thetacommutant}
\end{equation}
On the other hand, wedge compatibility and wedge duality give $\theta(\tilde{\mathcal C}) \subset \mathcal A_{\textrm{HH}}(\mathcal L) = \mathcal A_{\textrm{HH}}(\mathcal R)'.$ Taking commutants reverses this inclusion, and hence $\mathcal A_{\textrm{HH}}(\mathcal R) \subset \theta(\tilde{\mathcal C})'$.  Combining this with $\theta(\mathcal C)\otimes\one \subset\mathcal A_{\textrm{HH}}(\mathcal R)$ and using \eqref{eq:thetacommutant}, we obtain
\begin{equation}
\theta(\mathcal C)\otimes\one
\subset
\mathcal A_{\textrm{HH}}(\mathcal R)
\subset
\theta(\mathcal C)\;\overline\otimes\;
\mathcal B(\mathcal K_{\textrm{HH}})\,.
\end{equation}
Since $\theta(\mathcal{C})$ is a factor, it follows by the tensor-splitting theorem of Ge-Kadison \cite{GeKadison1996} that there exists a von Neumann subalgebra $\mathcal{Q}_{\textrm{HH}}\subset \mathcal{B}(\mathcal{K}_{\textrm{HH}})$ such that 
\begin{equation}
\label{eq:HHsplit}
\mathcal{A}_{\textrm{HH}}(\mathcal{R})=\theta(\mathcal{C})\;\overline{\otimes}\;\mathcal{Q}_{\textrm{HH}}\,.
\end{equation}
We note that the split property for $\pi_{\textrm{HH}}$ was not invoked in the above arguments. We only required the split property for the $\pi_{0}$ representation of the algebra together with wedge compatibility and duality. In particular, $\mathcal{Q}_{\textrm{HH}}$ need not be Type I.

\subsection{Schwarzschild evolution as a modular flow}

We first fix the parameter $t$ in $\alpha_{-t}^{(0)}$. Let $\mathcal{F}$ be a collar factor as above with $\mathcal{C}=\mathcal{A}(\mathcal{R})\cap \mathcal{F}$. By assumption (\ref{assump5}), $\alpha_{-t}^{(0)}(\mathcal{F})$ for fixed $t$ is also localized in a bounded collar. We note that assumption (\ref{assump4}) holds for every sufficiently large collar radius, so the preceding collar factor construction can be repeated with an arbitrarily large collar. We may therefore choose  a constant $a_{D}$ sufficiently large and construct a Type I collar factor $\mathcal{F}_{D}$ such that 
\begin{equation}
\mathcal{F}\vee\alpha_{-t}^{(0)}(\mathcal{F})\subset \mathcal{A}(\Sigma_{\textrm{C},a_{D}}) \subset \mathcal{F}_{D}\,.
\end{equation}
Defining $\mathcal{D}\equiv \mathcal{A}(\mathcal{R})\cap \mathcal{F}_{D}$, we then have 
\begin{equation}
\mathcal{C}\subset \mathcal{D}\,, \quad \quad \textrm{and}\quad \quad \alpha_{-t}^{(0)}(\mathcal{C})\subset \mathcal{D}
\end{equation}
where the second inclusion uses the fact that $\alpha^{(0)}_{t}$ preserves $\mathcal{A}(\mathcal{R})$. Applying the same factorization argument from the previous subsection to $\mathcal{F}_{\textrm{D}}$ yields 
\begin{equation}
\mathcal{A}(\mathcal{R})=\mathcal{D}\;\overline{\otimes}\;\mathcal{B}(\mathcal{K}_{0})\quad \quad \textrm{and}\quad \quad \mathcal{A}_{\textrm{HH}}(\mathcal{R})=\theta(\mathcal{D})\;\overline{\otimes}\;\mathcal{Q}_{\textrm{HH}}
\end{equation}
where $\mathcal{K}_{0}$ is an infinite-dimensional, separable Hilbert space and $\theta$ will, from now on, denote the canonical isomorphism associated to the collar. These factors are allowed to depend on the fixed value of $t$. 

We choose faithful, normal states $\varphi$ on $\mathcal{D}$, $\chi_{0}$  on $\mathcal{B}(\mathcal{K}_{0})$ and $\chi_{\textrm{HH}}$ on $\mathcal{Q}_{\textrm{HH}}$. Such states exist because these von Neumann algebras act faithfully on separable Hilbert spaces. We then define
\begin{equation}
\psi_{0}\equiv \varphi \otimes \chi_{0}\,, \qquad \psi_{\textrm{HH}} \equiv (\varphi \circ \theta^{-1})\otimes \chi_{\textrm{HH}}
\end{equation}
so that $\psi_{0}$ and $\psi_{\textrm{HH}}$ are faithful normal states on $\mathcal A(\mathcal R)$ and
$\mathcal A_{\textrm{HH}}(\mathcal R)$, respectively.  Since the modular flow of a product state acts independently on each tensor factor, both flows preserve the shared collar factor and satisfy
\begin{equation}
\sigma_{t}^{\psi_{\textrm{HH}}}(\theta(d))=\theta(\sigma_{t}^{\psi_{0}}(d)) \quad \quad \textrm{ $\forall$ $d\in \mathcal{D}$, $t\in \mathbb{R}$}\,.
\end{equation}

Since both the Hartle-Hawking state $\omega_{\textrm{HH}}$ and the reference state $\psi_{\textrm{HH}}$ are faithful, normal states, the Connes' cocycle theorem implies that their modular flows are related by a strongly continuous family of unitaries $u_{t}$,
\begin{equation}
\alpha_{-t}^{\textrm{HH}} = \textrm{Ad}(u_{t})\circ \sigma_{t}^{\psi_{\textrm{HH}}}
\end{equation}
where $u_{t}\in \mathcal{A}_{\textrm{HH}}(\mathcal{R})=\theta(\mathcal{D})\;\overline{\otimes}\;\mathcal{Q}_{\textrm{HH}}$. 
We note that $u_{t}$ need not lie entirely in $\theta(\mathcal{D})$ so we cannot transport it into $\mathcal{A}(\mathcal{R})$ via $\theta^{-1}$. Therefore, we need to define a map from $\mathcal{Q}_{\textrm{HH}}$ to $\mathcal{B}(\mathcal{K}_{0})$. Since any infinite-dimensional, separable Hilbert spaces are isomorphic to each other, we choose an arbitrary unitary $V:\mathcal{K}_{0}\to \mathcal{K}_{\textrm{HH}}$ and define the faithful, normal unital representation $j: \mathcal{Q}_{\textrm{HH}}\to \mathcal{B}(\mathcal{K}_{0})$ by $j(q)=V^{\dagger}qV$ for all $q\in \mathcal{Q}_{\textrm{HH}}$. Together with the canonical identification $\theta$ on the bounded collar we obtain a normal unital embedding 
\begin{equation}
\iota\equiv\theta^{-1}\;\overline{\otimes}\; j\;:\;\mathcal{A}_{\textrm{HH}}(\mathcal{R})\longrightarrow \mathcal{A}(\mathcal{R})\,.
\label{eq:iotadef}
\end{equation}
This map is, of course, non-canonical but we will only need its action on the collar 
\begin{equation}
\label{eq:iotadef2}
\iota(\theta(d))=d \qquad \textrm{ for all $d\in \mathcal{D}$}\,.
\end{equation}

For every $c\in\mathcal C$, both $c$ and
$\alpha_{-t}^{(0)}(c)$ belong to $\mathcal D$ by construction. Since $\theta$ denotes the canonical identification on the enlarged collar, the intertwining relation \eqref{eq:intertwining} evaluated at $-t$, the Connes' cocycle relation, and the agreement of the reference modular flows on $\mathcal D$ give
\begin{align}
\theta\!\left(\alpha_{-t}^{(0)}(c)\right)
&=
\alpha_{-t}^{(\textrm{HH})}\!\left(\theta(c)\right) \\
&=
u_t\,\sigma_t^{\psi_{\textrm{HH}}}\!\left(\theta(c)\right)u_t^\dagger \\
&=
u_t\,\theta\!\left(\sigma_t^{\psi_0}(c)\right)u_t^\dagger \,.
\end{align}
Applying $\iota$ and using $\iota(\theta(d))=d$ for $d\in\mathcal D$,
we obtain
\begin{equation}
\alpha_{-t}^{(0)}(c)
=
\iota(u_t)\,\sigma_t^{\psi_0}(c)\,\iota(u_t)^\dagger\,,
\qquad c\in\mathcal C\,.
\label{eq:collarflow}
\end{equation}
Consequently, the automorphism
\begin{equation}
\gamma_t
\equiv
\sigma_{-t}^{\psi_0}
\circ\operatorname{Ad}\!\left(\iota(u_t)^\dagger\right)
\circ\alpha_{-t}^{(0)}
\end{equation}
fixes $\mathcal C$ pointwise:
\begin{equation}
\gamma_t(c)=c,
\qquad c\in\mathcal C\,.
\end{equation}

Since
\begin{equation}
\mathcal A(\mathcal R)
=
\mathcal C\;\overline\otimes\;\mathcal B(\mathcal H_{\textrm R})\,,
\qquad
\mathcal C'\cap\mathcal A(\mathcal R)
=
\one\otimes\mathcal B(\mathcal H_{\textrm R})\,,
\end{equation}
the automorphism $\gamma_t$ preserves
$\one\otimes\mathcal B(\mathcal H_{\textrm R})$. Its restriction to
this Type I factor is inner. Because $\gamma_t$ acts trivially on
$\mathcal C$, there is therefore a unitary
$z_t\in\one\otimes\mathcal B(\mathcal H_{\textrm R})$ such that $\gamma_t=\operatorname{Ad}(z_t)$
on all of $\mathcal A(\mathcal R)$. It follows that
\begin{equation}
\alpha_{-t}^{(0)}
=
\operatorname{Ad}(w_t)\circ\sigma_t^{\psi_0},
\qquad
w_t\equiv
\iota(u_t)\,\sigma_t^{\psi_0}(z_t)
\in\mathcal A(\mathcal R)\,.
\end{equation}

To apply the converse of Connes' cocycle theorem, we must compare the entire flow $\alpha_{-t}^{(0)}$ with the modular flow of a single fixed reference state. The preceding construction gives this comparison for each fixed $t$, but the auxiliary state $\psi_0$ may depend on $t$. We therefore fix a faithful normal state $\rho$ on $\mathcal A(\mathcal R)$ and show first that the difference between $\alpha_{-t}^{(0)}$ and $\sigma_t^\rho$ is inner for every $t$.

For each $t$, we apply Connes' cocycle theorem to the state $\psi_0$ chosen
at that value of $t$ and to $\rho$. This gives a unitary $v_t\in\mathcal A(\mathcal R)$ such that
\begin{equation}
\sigma_t^{\psi_0}
= \operatorname{Ad}(v_t)\circ\sigma_t^\rho\,.
\end{equation}
Combining this with $\alpha_{-t}^{(0)} =\operatorname{Ad}(w_t)\circ\sigma_t^{\psi_0}$ and defining $\widetilde w_t\equiv w_tv_t$, we obtain
\begin{equation}
\label{eq:pointwiseinner2}
\alpha_{-t}^{(0)}\circ\sigma_{-t}^\rho
=
\operatorname{Ad}(\widetilde w_t)
\in
\operatorname{Inn}\!\left(\mathcal A(\mathcal R)\right)\,,
\qquad t\in\mathbb R\,.
\end{equation}
This establishes pointwise innerness relative to the fixed state $\rho$. 

Although \eqref{eq:pointwiseinner2} holds for every $t$, the enlarged collar, the product states, and the embedding $\iota$ were allowed to depend on $t$. Consequently, the unitaries $\widetilde w_t$ have so far only been constructed pointwise in $t$. It remains to choose them so that they form a strongly continuous cocycle. To achieve this, we define $\beta_t\equiv\alpha_{-t}^{(0)}\circ\sigma_{-t}^{\rho}.$ By \eqref{eq:pointwiseinner2}, $\beta_t$ is inner for every $t$. Since $\mathcal A(\mathcal R)$ is a factor, a unitary implementing $\beta_t$ is unique up to a phase. Moreover, because $\mathcal A(\mathcal R)$ has separable predual, these implementing unitaries may be chosen measurably in $t$ \cite{Kechris1995,KuratowskiRyllNardzewski1965}. Thus there exists a measurable family of unitaries $\widetilde w_t\in\mathcal A(\mathcal R)$ such that
\begin{equation}
\alpha_{-t}^{(0)}
=
\operatorname{Ad}(\widetilde w_t)\circ\sigma_t^\rho\,,
\qquad t\in\mathbb R\,.
\end{equation}

We represent $\mathcal A(\mathcal R)$ in the GNS representation of
$\rho$ and denote its modular operator by $\Delta_\rho$. Define
$U_t\equiv\widetilde w_t\Delta_\rho^{\mathrm{i}t}$. The automorphism
$\operatorname{Ad}(U_t)$ restricts to $\alpha_{-t}^{(0)}$ on
$\mathcal A(\mathcal R)$ and to
$\operatorname{Ad}(\Delta_\rho^{\mathrm{i}t})$ on its commutant. Both
restrictions satisfy the group law. Since $\mathcal A(\mathcal R)$ is a
factor,
$\mathcal A(\mathcal R)\vee\mathcal A(\mathcal R)'
=\mathcal B(\mathcal H_\rho)$, and hence
\begin{equation}
U_{t_1}U_{t_2}=c(t_1,t_2)U_{t_1+t_2}
\end{equation}
for some phases $c(t_1,t_2)$. Thus $U_t$ is a measurable projective
one-parameter unitary group. By Bargmann's theorem \cite{bargmann}, its
phases may be chosen so that the resulting family
$\hat U_{t}$ is a strongly continuous unitary group. We now define
$\hat w_t\equiv\widehat U_t\Delta_\rho^{-\mathrm{i}t}$. Each
$\hat w_t$ differs from $\widetilde w_t$ only by a phase and
therefore lies in $\mathcal A(\mathcal R)$. The group law for
$\widehat U_t$ gives
\begin{equation}
\widehat w_{t_1+t_2}
=
\hat w_{t_1}\,\sigma_{t_1}^\rho(\widehat w_{t_2})\,,
\qquad
\alpha_{-t}^{(0)}
=
\operatorname{Ad}(\hat w_t)\circ\sigma_t^\rho \,.
\label{eq:HHcocycle}
\end{equation}
Thus $\hat w_t$ is the required strongly continuous
$\sigma^\rho$-cocycle.

The converse of the Connes' cocycle theorem
\cite[Thm.~1.2.4]{Connes:1973hg} states that every strongly continuous
unitary $\sigma^\rho$-cocycle arises from a faithful, normal, semifinite weight. It therefore gives a weight $\Psi_{\textrm{HH}}$ on $\mathcal A(\mathcal R)$ satisfying
\begin{equation}
\sigma_t^{\Psi_{\textrm{HH}}}
=
\operatorname{Ad}(\widehat w_t)\circ\sigma_t^\rho
=
\alpha_{-t}^{(0)}\,.
\end{equation}
This proves \eqref{eq:HHmodflow}. Finally, a standard consequence of
Connes' cocycle theorem is that two faithful, normal, semifinite
weights with the same modular flow differ by a positive operator
affiliated with the center. Since $\mathcal A(\mathcal R)$ is a factor,
this operator is a positive scalar. Hence, $\Psi_{\textrm{HH}}$ is unique
up to multiplication by a positive constant.

\bibliographystyle{JHEP}
\bibliography{ref}
\end{document}